\documentclass[useAMS,usenatbib]{mn2e}
\usepackage{epsfig}
\usepackage{float}
\usepackage{placeins}
\usepackage{graphicx}
\usepackage{epsfig}
\usepackage{bm}% bold math
\usepackage{amsfonts}
\usepackage{amsmath}
\usepackage{amssymb}
\usepackage{times}
\usepackage{natbib}
\usepackage{color}

\title[The core shift effect]{Core shift effect in blazars}

\author[Agarwal et al.]
{A. Agarwal$^{1}$\thanks{E-mail: aditiagarwal.phy@gmail.com}, 
P. Mohan$^{2}$\thanks{E-mail: pmohan@shao.ac.cn},
Alok C.\ Gupta$^{2,1}$\thanks{CAS PIFI Visiting Scientist},
A. Mangalam$^{3}$,
A.\ E.\ Volvach$^{4,5}$,
\newauthor M.\ F.\ Aller$^{6}$,
H.\ D.\ Aller$^{6}$,
M. F. Gu$^{2,7}$,
A.\ L{\"a}hteenm{\"a}ki$^{8,9}$,
M.\ Tornikoski$^{8}$, 
L.\ N.\ Volvach$^{4,5}$
\\ \\
$^{1}$Aryabhatta Research Institute of Observational Sciences (ARIES),
Manora Peak, Nainital - 263002, India\\
$^{2}$Shanghai Astronomical Observatory, Chinese Academy of Sciences, 80 Nandan Road, Shanghai 200030, China \\
$^{3}$Indian Institute of Astrophysics, Sarjapur Road, Koramangala, Bangalore - 560034, India \\
$^{4}$Radio Astronomy Laboratory of Crimean Astrophysical Observatory, Crimea \\
$^{5}$Taras Shevchenko National University of Kyiv, Kiev, Ukraine \\
$^{6}$Astronomy Department, University of Michigan, Ann Arbor, MI, U.S.A\\
$^{7}$ Key Laboratory for Research in Galaxies and Cosmology, Shanghai Astronomical Observatory, Chinese Academy of Sciences, 80 Nandan Road, \\ 
~~Shanghai 200030, China \\
$^{8}$Aalto University Mets{\"a}hovi Radio Observatory, Finland\\ 
$^{9}$Aalto University Department of Elecronics and Nanoengineering, Finland \\ 
}

\begin{document}
\newdimen\digitwidth
\setbox0=\hbox{2}
\digitwidth=\wd0
\catcode `#=\active
\def#{\kern\digitwidth}

\date{Accepted ....... Received  ......; in original form ......}

\pagerange{\pageref{firstpage}--\pageref{lastpage}} %\pubyear{2014}

\maketitle

\label{firstpage}

\begin{abstract}
We studied the pc-scale core shift effect using radio light curves for three blazars, S5 0716+714, 3C 279 and BL Lacertae, which were monitored at five frequencies ($\nu$) between 4.8 GHz and 36.8 GHz using the University of Michigan Radio Astronomical Observatory (UMRAO), the Crimean Astrophysical Observatory (CrAO), and Metsahovi Radio Observatory for over 40 years. Flares 
were Gaussian fitted to derive time delays between observed frequencies for each flare ($\Delta t$), peak amplitude ($A$), and their half width. Using $A \propto \nu^{\alpha}$ we infer $\alpha$ in the range $-$16.67 to 2.41 and using $\Delta t \propto \nu^{1/k_r}$, we infer $k_r \sim 1$, employed in the context of equipartition between magnetic and kinetic energy density for parameter estimation. From the estimated core position offset ($\Omega_{r \nu}$) and the core radius ($r_{\rm core}$), we infer that opacity model may not be valid in all cases. The mean magnetic field strength at 1 pc ($B_1$) and at the core ($B_{\rm core}$), are in agreement with previous estimates. %We use these estimates to test the magnetically arrested disk model which gives approximate but reasonable black hole spin for these sources. 
We apply the magnetically arrested disk model to estimate black hole spins in the range $0.15-0.9$ for these
blazars,
indicating that the model is consistent with expected accretion mode in such sources.
The power law shaped power spectral density has slopes $-$1.3 to $-$2.3 and is interpreted in terms of multiple shocks or magnetic instabilities.
 \end{abstract}
 

\section{Introduction}
\label{sec:Introduction}

Flat spectrum radio quasars (FSRQs) and BL Lacertae objects (BL Lac) together known as blazars are a subset of core dominated Active
Galactic Nuclei (AGNs) characterized by a luminous core, rapid variability over entire electro magnetic (EM) spectrum, high radio to optical
polarization, apparently superluminal jet components and non thermal emission from a Doppler boosted relativistic jet pointing $\leq$ 10$^{\circ}$ to the
observer line of sight (LOS).
Core-jet morphology is a common characteristic of most of these AGN in very long baseline interferometry (VLBI) images, where
core is the optically thick base of the jet (e.g. Blandford \& K\"onigl 1979).

The apparent systematic outward shift of the VLBI core position with decreasing observation frequency is the core shift effect, and can be
attributed to the synchrotron self absorption process (SSA) operating and causing the transition from optically thick to thin jet emission, the
optical depth given by,
\begin{equation}
\tau(r) \propto  r^{(1.5-\alpha)m+n-1} {\nu}^{2.5-\alpha}
\end{equation}
where $r$ is the radial distance from the center, $m$ is the power law index in the magnetic field decay ($B = B_{1} (r/1 \rm{pc})^{-m}$) while $n$ is the
power law index in the particle number density decay ($N = N_{1} (r/1 \rm{pc})^{-n}$) with distance. Here $N_{1}$ and $B_{1}$ refer to $N$ and $B$
at $r = 1$ pc where it is expected that the jet shape changes from conical to parabolic closer to the true jet origin (e.g.,  Beskin \& Nokhrina
2006; McKinney 2006; Krichbaum et al\ 2006; Nakamura \& Asada 2013). In the conical jet model, the VLBI core of blazars are compact, stationary and
bright features located at one end of the jet at sub-mas--mas scales. They are characterized by superluminal jet components which emerge and
separate and whose kinematics is governed by propagating disturbances or shock waves. 
The core is partially optically thick to SSA since the spectral slope is $\leq$ 1, which can be attributed to magnetic field geometry and the
relativistic electron density gradients. 
Cawthorne (2006) suggests that in some cases, the core could correspond to a stationary feature like a conical shock, inferred in some long term
studies of blazar jets  (e.g. Marscher 2009).

\begin{figure}
\epsfig{figure= 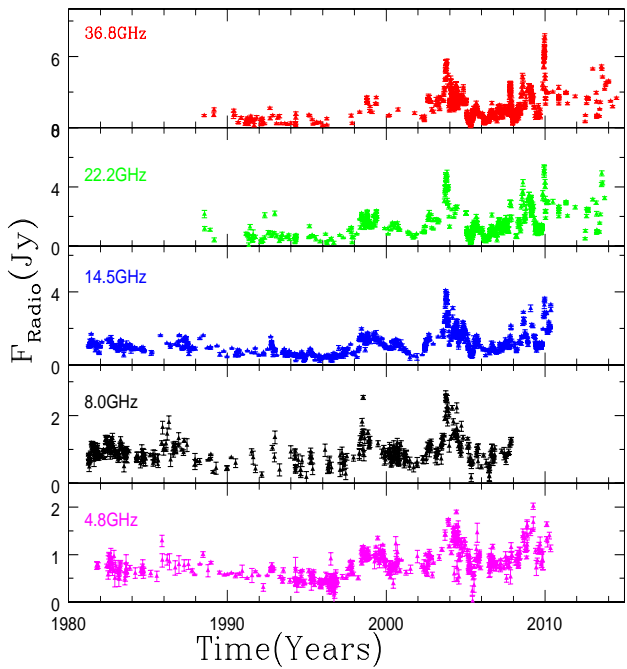,height=4.2in,width=3.6in,angle=0}
\caption{Long-term variability light curves of S5 0716+714 in the 4.8 GHz - 36.8 GHz frequency range.}
\end{figure}

Core shift measurements can provide information on the physical parameters of the VLBI jet including core magnetic field strength, distance from
the core to the base of the jet (e.g. Lobanov 1998; Hirotani 2005), spectral index, inputs to rotation measure maps, and clarify the nature of the
absorbing material in the SSA jet.
There are several ways to measure apparent core-shifts. One of them is the phase referencing technique where the telescope is switched between the
target source and a nearby phase calibrator, the switching time being shorter compared to the coherence time. However as this technique is complex,
resource intensive and relatively expensive in terms of the observing time, it has been applied only for some AGN e.g., 3C 390.1, 3C 395, 1038+528,
M81, 4C 39.25 (Guirado et al.\ 1995; Marcaide \& Shapiro 1984; Lara et al.\ 1994; Bietenholz, Bartel, \& Rupen 2004). Other methods require a
brightness distribution model where the core position is estimated by identifying individual optically thin jet features in the VLBI images which
do not change their position with frequency (e.g. Kovalev et al.\ 2008; Sokolovsky et al.\ 2011). Though, this method requires simultaneous
multiband VLBI observations. Another method involves calculation of the time delay between the lower and higher frequency radio emission which can
be attributed to the opacity effects. It is applied assuming that at a given frequency, a flare is associated with jet component ejection from the
core. The observation time of the flare and hence the shift increases with decreasing frequency (e.g. Kudryavtseva et al. 2011; Mohan et al. 2015). 
Magnitudes of core shifts are time dependent as they depend on the present state of an AGN and are resolution limited for the VLBI observations. Relative advantages of this method include the use of only single dish measurements with quasi-simultaneous
observations,
the ability to use VLBI core flux density measurements if available, and the available methodology for the timing
analysis
(Mohan et al. 2015). Shortcomings include a possible inability to capture a flare due to non-simultaneous observations,
the difficulty in identifying a flare in two frequencies as the same (opacity effect) or distinct, and the requirement of
more than two frequencies to confirm an increasing time delay and hence, the opacity based apparent core shift.
As multiple quasi-simultaneous observations are employed in our study, these shortcomings are met thus making it ideally suited.

We discuss the time delay based core shift estimation and the resulting jet parameters 
for three blazars, S5 0716+714, 3C 279, and BL Lac using single dish observations at five frequency bands from University of Michigan Radio Astronomical Observatory (UMRAO), Crimean Astrophysical Observatory (CrAO) and Mets\"ahovi Radio Observatory, observed between 1970 and 2015. %magnitude and other physical parameters related to jets.
The estimated parameters are compared with estimates from previous studies. We adopt a standard cold dark matter cosmological model with
Hubble constant $H_0 = 71~$km/s/Mpc, matter energy density $\Omega_{m} = 0.27$, and dark energy density $\Omega_\Lambda = 0.73$.
In Section 2, we briefly discuss the source properties relevant to the current study; in Section 3, the parameters that are calculated including
core position offset, magnetic field strength, and size of the emitting core are presented, the estimates of which can be employed in the
estimation of black hole spin in the context of the magnetically arrested disk (MAD) scenario;
 a description of observations and data reduction for our single dish observations
along with the analysis procedure is then presented in Section 4. This includes a piecewise Gaussian fitting and the variability analysis using the
Fourier periodogram technique.
In Section 5, we present the results of our analysis and their comparison with reports in literature. The implications and a brief summary of our
results is then presented in Section 6.

\section{Notes on Individual Sources}

\subsection{S5 0716+714}

S5 0716+714 is a bright, high declination ($\alpha_{2000.0}$ = 07h 21m 53.4s, $\delta_{2000.0}$ = $+71^{\circ} 20^{'} 36.4^{\prime \prime}$)
BL Lacertae object.
Based on the marginal detection of the host galaxy, a redshift of 0.31 $\pm$ 0.08 was reported by Nilsson et al.\ (2008). Recently,
Paiano et al.\ (2017) obtained a lower limit of z $>$ 0.10 during the high optical state of the source.
S5 0716+714 is a prominent blazar in optical bands and has shown variability on diverse
timescales from minutes through hours to days (e.g. Heidt \& Wagner 1996; Nesci et al.\ 1998; Giommi et al.\ 1999; Raiteri et al.\ 1999; Raiteri et
al.\ 2003; Gupta et al.\ 2008; Gupta et al.\ 2012; Agarwal et al.\ 2016) in radio to X ray bands (e.g. Wagner et al.\ 1996; Raiteri et al.\ 2003
and references therein). 
It was included in the S5 catalog after its discovery in the Bonn-NRAO Radio 5 GHz Survey (K{\"u}hr et al.\ 1981).

A lower limit on the apparent brightness temperature, $T_B$ $\geq$ 2 $\times$ 10$^{12}$ K was obtained using 5 GHz VLBI observations and suggests a
minimum equipartition Doppler factor of $\delta$ $\geq$ 4 (Bach et al.\ 2006). In addition, high bulk Lorentz factors of $\geq$ 16$-$21 were
estimated based on the inferred jet speeds (Jorstad et al.\ 2001). 
Radio maps from VLBI reveal a compact core-jet structure as has been suggested for BL Lacertae with different components moving at different velocities
(Bach et al.\ 2003). Jet speed $>$ 11-15 h$^{-1}$c were estimated from the component proper motions Jorstad et al.\ (2001), who also found quasi-periodic ejection of components every $\sim$
0.7 yr. An analysis of the above 26 VLBI observations in 5-22 GHz frequency band gives an apparent velocity in range 5-10 h$^{-1}$c. 
Radio flux density and linear polarization light curves obtained from UMRAO at 4.8, 8.0, and 14.5 GHz were presented by Aller et al.\ (1985) for the
1981-1984 period and those during 1985-1992 were studied by Wagner et al.\ (1996). Later, Teraesranta et al.\ (1998) presented this target's
radio data till 1999 which included 22 and 37 GHz observations from MRO.

\subsection{3C 279}

The FSRQ 3C 279 ($\alpha_{2000.0}$ = 12h 56m 11.17s, $\delta_{2000.0}$ = $-05^{\circ} 47^{'} 21.5^{\prime \prime}$) at a redshift of 0.536
(Burbidge \& Rosenburg 1965) was the first extragalactic radio source indicating superluminal motion (Cohen et al.\ 1971). It displays violent
multiwavelength flux variability and is well studied by various observation programs (e.g. Maraschi et al. 1994; Wehrle et al.\ 1998;
Lindfors et al.\ 2006, Collmar et al.\ 2007). Flux variability at X-rays, R band and 14.5 GHz frequency band have been found to be
significantly correlated, suggesting a possible co-spatial origin.
The radio morphology reveals a bright, stationary core with a jet extending to $\sim$ 5$\farcs$ along a position angle of 205$^{\circ}$ (de Pater \&
Perley 1983). Superluminal motion of the jet with apparent speed in range 5$c$ - 17$c$ was found to be associated with the target using very long
baseline array (VLBA) radio observations at 43 GHz between 1998 March and 2001 April (Jorstad et al.\ 2004). This source has been extensively
monitored with the high resolution VLBI revealing a one sided jet extending southwest on pc-scale characterized by bright knots ejected from the
core region, along with multiple projected apparent speeds and polarization angles (e.g. Jorstad et al.\ 2005; Unwin eta al.\ 1989; Chatterjee et
al.\ 2008; Wehrle et al.\ 2001).
Using VLBA polarimetry studies, electric field vector was found to be aligned with the jet direction on pc to kpc-scales implying that the magnetic
field is pre-dominantly perpendicular to the relativistic jets on above length scales. From VLBA radio observations at 43 GHz between 1998
March - 2001 April, a bulk Lorentz factor of 15.5 $\pm$ 2.5, viewing angle of 2$^{\circ}$.1 $\pm$ 1$^{\circ}$.1 and Doppler factor of 24.1 $\pm$
6.5 are obtained (Jorstad et al.\ 2004, 2005).

\begin{figure}
\epsfig{figure= 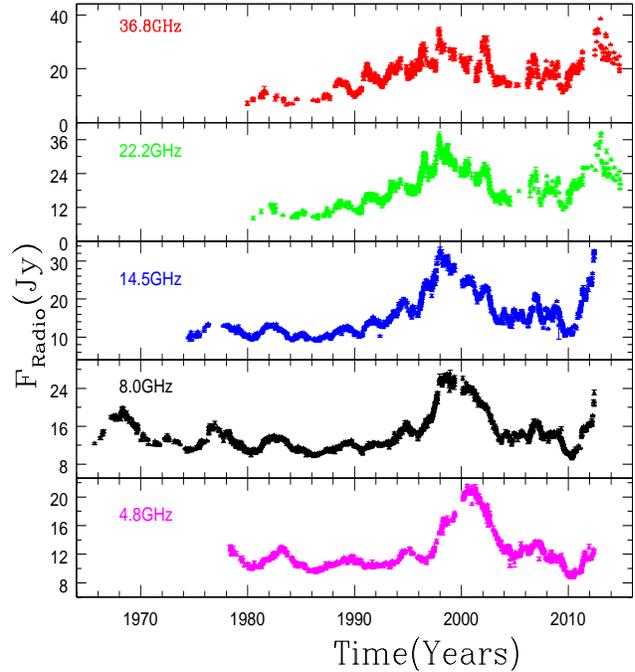,height=4.2in,width=3.6in,angle=0}
\caption{Long-term variability light curves of 3C 279 in the 4.8 GHz - 36.8 GHz frequency range.}
\end{figure}

\noindent
\begin{figure}
\epsfig{figure= 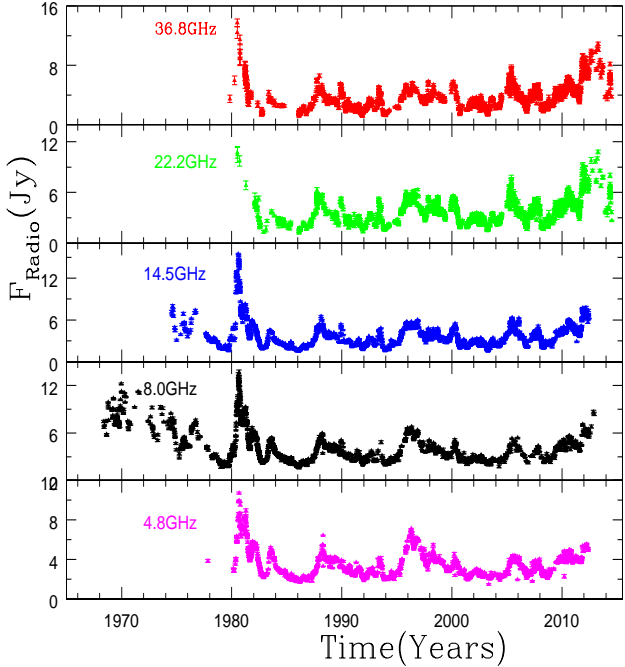,height=4.2in,width=3.6in,angle=0}
\caption{Long-term variability light curves of BL Lacertae in the 4.8 GHz - 36.8 GHz frequency range.}
\end{figure}

\subsection{BL Lacertae}
BL Lacertae ($\alpha_{2000.0}$ = 22h 02m 43.29s $\delta_{2000.0}$ = $+42^{\circ} 16^{'} 39.98^{\prime \prime}$) also known as IES 2200+420, is a
blazar archetype hosted by an elliptical galaxy at a redshift of $\sim$ 0.07
(Miller, French, \& Hawley 1978; Hyv{\"o}nen et al.\ 2007). Emission lines are either absent or extremely weak in BL Lacertae class of blazars. But broad H$_{\alpha}$ and H$_{\beta}$ lines have been found in the spectrum of this object rendering its classification in BL Lacertae class under delima (Vermeulen et al.\ 1995).
Since its association with radio source VRO 42.22.01  (Schmitt, 1968), it has been well studied across the entire EM spectrum.
It has been extensively observed in the optical band at diverse timescales i.e.
from minutes to years timescales (e.g. Villata et al.\ 2004; Agarwal \& Gupta 2015).

From $\sim$ 15 years of variability observations (June 1971 to January 1985), Webb et al.\ (1988) inferred recurrent variations after every 0.31,
0.60, and 0.88 yr using the Fourier periodogram. Marchenko et al.\ (1996) found a statistically significant long term component with $P$ = 7.8 $\pm$
0.2 yr in the 20 year duration light curve using the whitening of the time series method. Smith \& Nair (1995) found a period of 7.7 yr using a
20 yr duration light curve
using Fourier analysis thus, consistent with Marchenko et al.\ (1996). Later, Fan et al.\ (1998) obtained a long term period of 14 yr using the Jurkevich method on the
optical light curve.
The inclination angle for the Doppler boosted approaching jet of the BL Lacertae is $\sim$ 6$^{\circ}$-10$^{\circ}$ with a flow speed of 0.981-0.994 c
and a bulk Lorentz factor of 7.0 $\pm$ 1.8 (Jorstad et al.\ 2005).

\section{Frequency-dependent core shifts and tests of the magnetically arrested disk (MAD) scenario}

\subsection{Core shift and jet parameters}
The optically thick radio core is not completely resolved with VLBI and thus the emission from the core region is mainly from the $\tau$ = 1
surface at a particular frequency. Due to its frequency dependence, the radio core position follows $r\propto \nu^{-1/k_{r}}$ (e.g. Konigl 1981;
O'Sullivan \& Gabuzda 2009) where $r$ is the distance from the central region, $\alpha$ is the spectral index in the power law flux density
dependence $S_\nu \propto \nu^{\alpha}$, and the index $k_{r}$ is
\begin{equation}
k_{r} = \frac{(3 - 2\alpha)m + 2n - 2}{5 - 2\alpha}
\label{k_r}
\end{equation}
assuming that the ambient medium pressure on the jet is negligible and external pressure is non-negligible with non-zero gradient along the jet (Lobanov 1998).

The quantity $k_{r}$ becomes independent of $\alpha$ if there is equipartition between the jet particle density and magnetic field energy density
for which $m = 1$ and $n = 2$ (e.g. Hutter \& Mufson 1986; Lobanov 1998). The above combination of $m$ and $n$ is valid for synchrotron emission
from compact VLBI cores (Konigl 1981) for a constant jet speed and half opening angle. With $k_{r} = 1$, the core shift between two frequencies $\nu_1$ and $\nu_2$ is
\begin{equation}
\Delta r = r_{\nu_1} - r_{\nu_2} = \frac{(\nu_1 - \nu_2)}{\nu_1 \nu_2}.
\label{k_r}
\end{equation}
In the models proposed by \citet{1998A&A...330...79L} and \citet{2005ApJ...619...73H}, magnetic field at some given distance along the jet frame is
derived without using the information about the radiative flux of the jet. The magnetic field calculated then would depend on the normalization of
electron distribution. These works assumed equipartition between the relativistic electrons and the magnetic field. 
For S5 0716+714, 3C 279, and BL Lacertae we use the modified equations following \citet{Z15} which improves on the method
previously followed in the application to 3C 454.3 in (\citet{2015MNRAS.452.2004M}; Paper 1). The angular core separation $\Delta \theta$ (mas) for
a mean component proper motion $\mu$ (mas/y) is
\begin{equation}
\Delta \theta = \mu \Delta t.
\end{equation}
The core offset parameter $\Omega_{r \nu}$ (pc~GHz$^{1/k_r}$) which is the core offset distance per unit observation frequency $\nu^{1/k_r}$ (GHz) interval is
\begin{equation}
\Omega_{r \nu} = (1 {\rm mas}) \frac{D_L \Delta \theta}{(1+z)^2 \left(\nu^{-1/k_r}-\nu^{-1/k_r}_0\right)},
\label{om}
\end{equation}
where $D_L$ is the luminosity distance. The core offset distance $r_{\rm core}$ (pc) is the projected distance of the actual emitting core (at a
frequency $\nu_0 \to \infty$) at the observation frequency $\nu$ (GHz) and for a jet inclination angle $i$ is
\begin{equation}
r_{\rm core} = \frac{\Omega_{r \nu}}{\sin i} \nu^{-1/k_r}.
\label{rc}
\end{equation}
The magnetic field strength $B_\beta$ (G) at a distance $h_1 = h/(1 {\rm pc})$ is
\begin{eqnarray}
\tiny
B_\beta = \left(\frac{(h_P (1+z))^\frac{p+4}{p+2}}{\delta \sin i}\frac{B_{cr}}{m_e c^2} \left(\frac{8 \alpha_F}{(1 {\rm pc})~C_2 (p)~ \sigma_T \tan \theta_0}\right)^\frac{2}{p+2} \right)^\frac{p+2}{p+6}\\ \nonumber
 \frac{1}{h_1 \beta^\frac{2}{p+6}} \left(\frac{\Omega_{r \nu}}{10^{-9/k_r}}\right)^\frac{p+4}{p+6},
\label{bb}
\end{eqnarray}
where $p = 2 \alpha+1$ is the particle energy distribution index, $h_P$ is Planck's constant, $m_e$ is the electron mass, $e$ is its
charge, $B_{cr} = 2 \pi m^2_e c^3/(e h_P)$, $\alpha_F = 1/137$ is the fine structure constant, $\beta$ is the ratio between the particle kinetic
energy density and the magnetic field energy density and the constants $C_1$, $C_2$ and $C_3$ are defined in terms of $p$ (Zdziarski et al. 2012).
The magnetic field strength $B_F$ (G) at the distance $h_1$, accounting for an optically thin jet based flux $F_\nu$ (includes jet and core emission) is,
\begin{eqnarray}
B_F = \frac{\delta (1+z)^7}{D^4_L \sin^3 i} \frac{B^5_{cr} h^7_P}{c^{12}} \left(\frac{\alpha_F C_1 (p) C_3 (p)~ \tan \theta_0}{(1 {\rm Jy})~ 24 \pi^3 C_2 (p)}\right)^2\\ \nonumber
\frac{1}{h_1 F^2_\nu} \left(\frac{\Omega_{r \nu}}{10^{-9/k_r}}\right)^{5}.
\end{eqnarray}
The magnetic field strength $B_{\rm core}$ (G) at the emitting core at the distance $r_{\rm core}$ (pc) is
\begin{equation}
B_{\rm core} = B ~ r^{-1}_{\rm core},
\label{bcore}
\end{equation}
where $B$ can be calculated by determining $B_\beta$, $B_F$ and the ratio $B_\beta/B_F$ from their means, used as a normalizing multiplicative factor.

\subsection{Application of the MAD model}

Magnetic fields are expected to play a prominent role in structuring the accretion flow, the disk-jet connection, and the collimated outflow at the sub--pc to pc-scales and in possible helical signatures at the kpc-scales. Further, using the  derived field strength, the bolometric and jet luminosity one can explore electrodynamical and hybrid jet models and place constraints on the spin and mass of the black hole. 

\citet{Z14} considered samples of blazars and radio galaxies and obtained core shift and magnetic field strengths. 
They considered a model of jet formation from black hole (BH) spin-energy extraction \citep{BZ77}. \citet{Z15} considered models with the accretion
being magnetically arrested (MAD; \citet{Narayan2003}). Such flows have dragged so much magnetic flux to the BH that it becomes dynamically
important owing to obstruction of accretion, and is given by $\Phi_{BH} \simeq 50 (\dot{M} c)^{1/2} = 10^{-4} M_8^{3/2} \dot{m}^{1/2}$ pc$^2$
\citep{Tchek2011, McKinney2012} where the accretion rate is in units of the Eddington rate. The surface value of magnetic flux depending on $h$ is (Tchekhovskoy et al.\ 2009),
\begin{equation}
B_{\rm pj}={Bc\Gamma_{\rm j} \over \Omega_{\rm f} r_{\rm j}}\left(2\sigma\ \over 1+\sigma \right)^{1/2}
\label{bp}
\end{equation}
where $B$ is the transverse average magnetic field strength. To compare with the observations, we use the average value of $B'_{\phi}$ (Zdiarski et al. 2015),
\begin{equation}
B =B'_{\phi{\rm j}}\left(1+\sigma\over 2\sigma\right)^{1/2},
\end{equation}
where $\sigma$ is the ratio of the magnetic to particle energy and the magnetic flux in terms of the mean toroidal field $B$ is
\begin{equation}
\Phi_{\rm j}\equiv 2\upi \int_0^{r_{\rm j}}{\rm d}r\, r B_{\rm p}(r)={2 \over 2-\alpha}\upi B_{\rm pj}r_{\rm j}^2.
\end{equation}
Using $\alpha=2/(1+\sigma)$ where $\alpha$ is the radial index of the variation of the poloidal field,
\begin{equation}
\Phi_{\rm j}={\upi B_{\rm pj}r_{\rm j}^2 \over \sigma}.
\end{equation}
The ratio of the angular frequency of the field lines to the BH angular frequency is defined by a parameter $l$ as

\begin{equation}
l = {\Omega_{\rm f} \over \Omega_{\rm H}}
\label{l}
\end{equation}
where $r_{\rm H}=[1+(1-a^2)^{1/2}]r_{\rm g}$ is the BH horizon radius. Now using (\ref{bp}) and (\ref{l}) we find

\begin{equation}
\Phi_{\rm j}={\upi Bc r_{\rm j} \Gamma_{\rm j} \over l \Omega_{\rm H}\sigma} \left(2\sigma\ \over 1+\sigma \right)^{1/2}
\end{equation}

and using 
\begin{equation}
\Omega_{\rm H} = {ac \over 2r_{\rm H}},
\end{equation}

we obtain

\begin{equation}
\Phi_{\rm j}={\upi 2^{3/2}B r_{\rm j} r_{\rm H} \Gamma_{\rm j} \over \sigma l a} \left(\sigma \over 1+\sigma \right)^{1/2},
\label{phij}
\end{equation}
where
\begin{equation}
r_{\rm j} = h\theta_{\rm j}, ~~{\rm and} ~~\theta_{\rm j}\Gamma_{\rm j}=s\sigma^{1/2}.
\label{rj}
\end{equation}

Applying eqn (\ref{rj}) in (\ref{phij}), we derive

\begin{equation}
\Phi_{\rm j}={\upi 2^{3/2}hB r_{\rm H}s \over l a} \left(1 \over 1+\sigma \right)^{1/2}
\label{phij1}
\end{equation}

where $a$ is the dimensionless spin parameter. Now,

\begin{equation}
{B \over B'_{\phi{\rm j}}} = \left(1+\sigma \over 2\sigma \right)^{1/2}
\label{b}
\end{equation}

and inserting (\ref{b}) in (\ref{bp}) we get,

\begin{equation}
B_{\rm pj}={B'_{\phi{\rm j}}  c\Gamma_{\rm j} \over \Omega_{\rm f} r_{\rm j}}.
\end{equation}

According to Narayan et al.\ (2003) the poloidal flux treading the BH on one hemisphere is given as:

\begin{equation}
\Phi_{\rm BH}=\phi_{\rm BH}(\dot M c)^{1/2} r_{\rm g},
\end{equation}

If we equate $\Phi_{\rm j}$ with $\Phi_{\rm BH}$ applying the relation $r_{\rm H}=[1+(1-a^2)^{1/2}]r_{\rm g}$, we arrive at

\begin{equation}
{\upi 2^{3/2}hB s \over l a} \left(1 \over 1+\sigma \right)^{1/2} r_{\rm g}[1+(1-a^2)^{1/2}] = \phi_{\rm BH}(\dot M c)^{1/2} r_{\rm g}.
\end{equation}

Putting  $L = \epsilon \dot M c^2$, we find

\begin{equation}
[{1+(1-a^2)^{1/2} \over a}]^2 h^{2}B^{2}c \left(1 \over 1+\sigma \right) = q L,
\label{q}
\end{equation}

where 

\begin{equation}
q = {l^{2}\phi_{\rm BH}^{2} (1+ \sigma) \over s^{2} \epsilon 2^{3} \upi^{2}}
\end{equation}

is a dimensionless parameter and from equation (\ref{q}),

\begin{equation}
w(a)= {q L \over h^{2}B^{2}c}= q  L_{46} ({{0.2 \rm G pc} \over h B})^{-2},
\label{wqL}
\end{equation}

where,  the quantity 
\begin{equation}
w(a)= [{1+(1-a^2)^{1/2} \over a}]^2,
\label{wa}
\end{equation}
and the bolometric luminosity $L_{46}$ is in units of $10^{46}$ erg/s.
For typical values of the parameters, $s=0.5, l=0.5, \epsilon=0.4, \phi_{\rm BH}= 50, \sigma=0.5$, we find that
$q \simeq 120 $. Inverting the eqn (\ref{wa}), we obtain
\begin{equation}
a(w) = \displaystyle {2 \sqrt{w} \over w^2 +1}. 
\label{eqnaw}
\end{equation}
The function $a(w)$ is plotted in Fig. \ref{aw}. It predicts that the values of $0.296 < w < 1$ are disallowed since they yield spin values $a > 1$. The relation given in Figure 2 of \citet{Z14} and \citet{Z15} also validate a variant of eqn. (\ref{wqL}) with the choice of
$\phi_{BH}=50$ and $\epsilon=0.4$. We have adopted these values and computed $w$ and $q$ to obtain approximate values of spin
consistent with known luminosity and variability in these sources.
For each source, we have calculated a linear fit of log(B$_{core}$) against log(r$_{core}$). The corresponding fit values of the intercept
were used to calculate the spin values for each source.
The luminosity is expected to be peaked at an intermediate $a$ due to the
fact that in the BZ model,
the horizon traps smaller flux for larger spin as the horizon shrinks with $a$ but at lower spin, rate of change of flux which is proportional
to the spin reduces. The spin in combination with the varying $Bh$ in the sources, can explain the jet luminosity within the context of the
MAD model. We test the model predictions using the core shift measurements for the three sources in section 5.2.

\section{Observations and analysis}

The long term light curves of blazars are obtained from the University of Michigan Radio Astronomical Observatory (UMRAO) which has monitored
compact variable radio sources spanning over $\sim$ 40 years with a 26 m paraboloid dish and provides multifrequency (4.8, 8.0 and 14.5 GHz), high
time resolution (better than a month) light curves (Aller et al.\ 1985, 1999). Observations at 4.8 GHz started in 1978 on regular basis, at 8.0 GHz
in 1965 and at 14.5 GHz in 1974. Monitoring at 22.2 GHz and 36.8 GHz were carried out with the 22-m radio telescope (RT-22) of the Crimean
Astrophysical Observatory (CrAO; Volvach 2006). We use modulated radiometers in combination with the registration regime ``ON-ON" for collecting
data from the telescope (Nesterov, Volvach, \& Strepka 2000). The 14 m radio telescope of Aalto University Mets\"ahovi Radio Observatory in Finland
was used for observations at 37.0 GHz. Data obtained at Mets\"ahovi and RT-22 were combined in a single array to supplement each other. A detailed
description of the data reduction and analysis of Mets\"ahovi data is presented in Teraesranta et al.\ (1998).

Fig.\ 1 shows the total flux density light curves for S5 0716+714 at five frequencies from 4.8 GHz to 36.8 GHz while Fig.\ 2 displays the same for
3C 279 and Fig.\ 3 for BL Lacertae, spanning more than four decades of observations.

The method employed here was applied to the FSRQ, 3C 454.3, in paper 1 with the results being consistent with previous studies. The radio light
curves of S5 0716+714, 3C 279 and BL Lacertae are piecewise Gaussian fitted as described in paper 1 using 

\begin{equation}
y = A e^{[-(t-\overline{m})^2/(2 \sigma^2)]},
\label{Gaussian}
\end{equation}
where $y$ is the typical flare profile, $A$ is the flare amplitude, $\overline{m}$ is the peak position and $\sigma$ is the flare width.
Following the procedure described in Paper 1. Initially pre-processing was performed to locate the maxima and minima of the flaring regions.
Gaussian filtering was applied from the baseline at the minimum amplitude of the LC with some initial values of maxima value, maxima position and
half the difference between consecutive minima ($A_0$, $\overline{m}_0$ and $\sigma_0$) respectively. Trial values of $A$, $\overline{m}$ and
$\sigma$ were generated from the regions around the initial values. These trial values account for possible values of the flare peak amplitude,
position, and difference in the original LC. Single Gaussian fit procedure described in Paper 1 is cycled through the entire LC covering all flares
till the best $\chi^2$ fit value was obtained.

\begin{figure*}

{\includegraphics[scale=0.33]{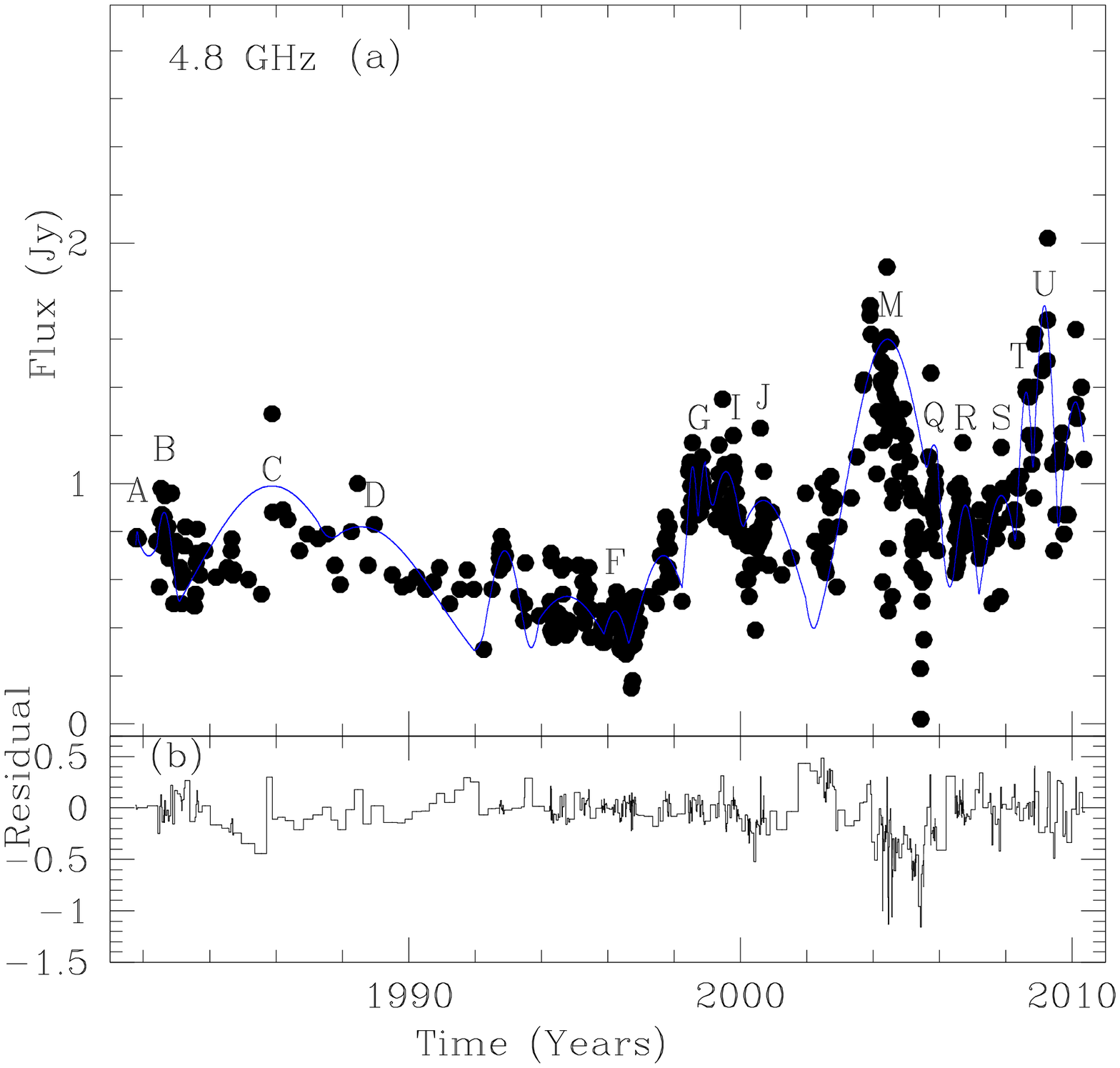}}
\hspace*{0.3in}{\includegraphics[scale=0.33]{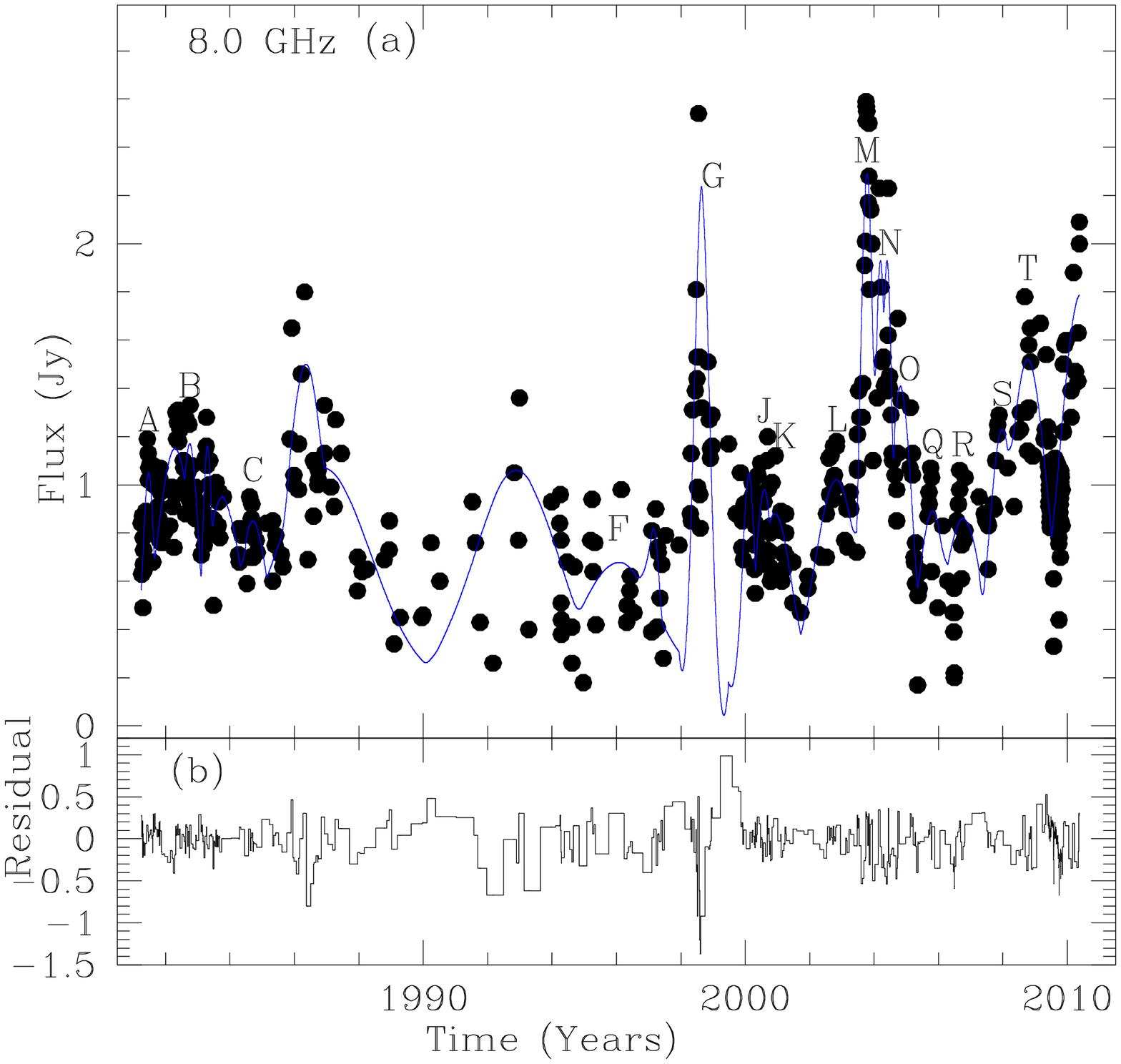}}
{\includegraphics[scale=0.33]{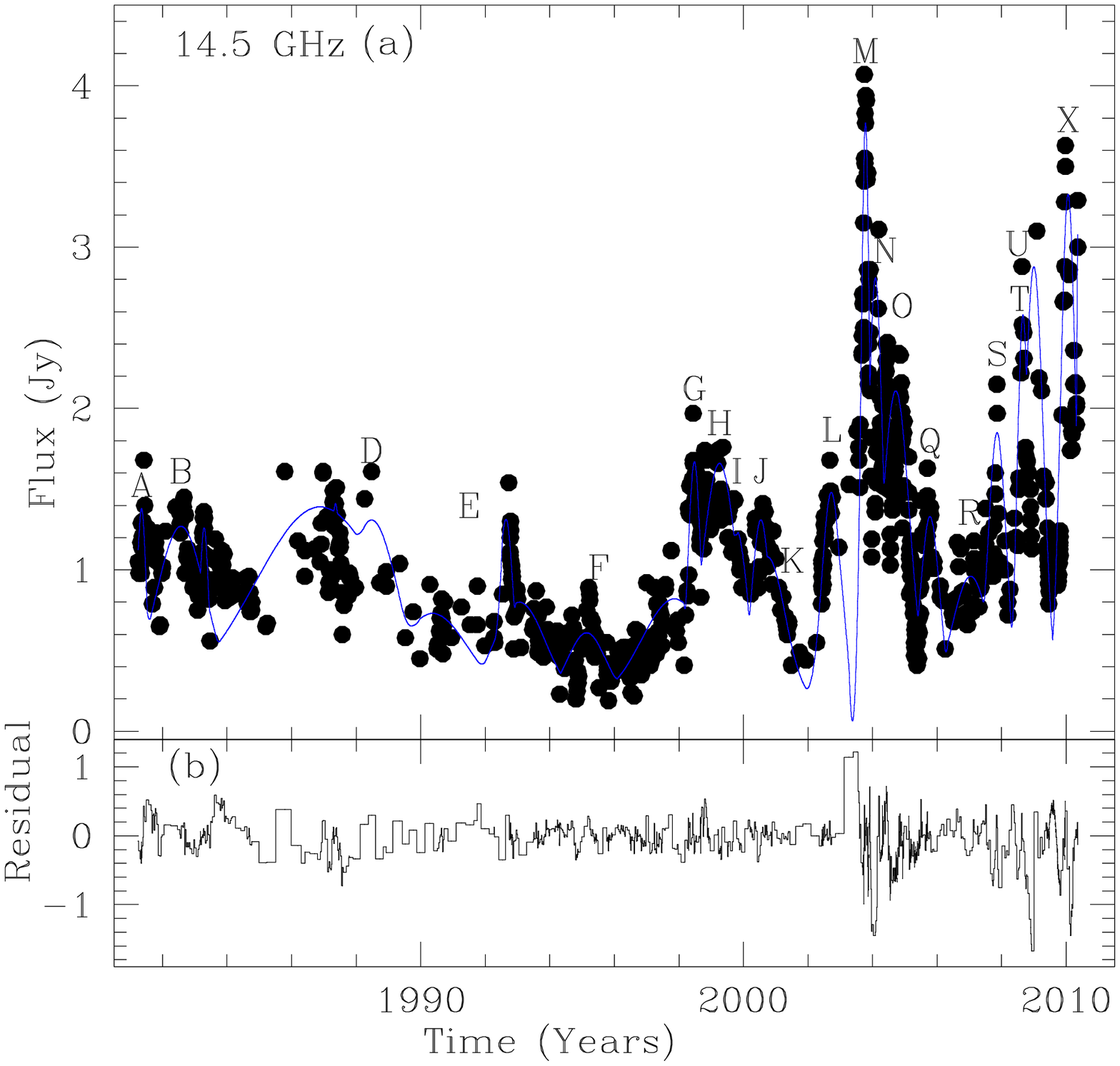}}
\hspace*{0.3in}{\includegraphics[scale=0.33]{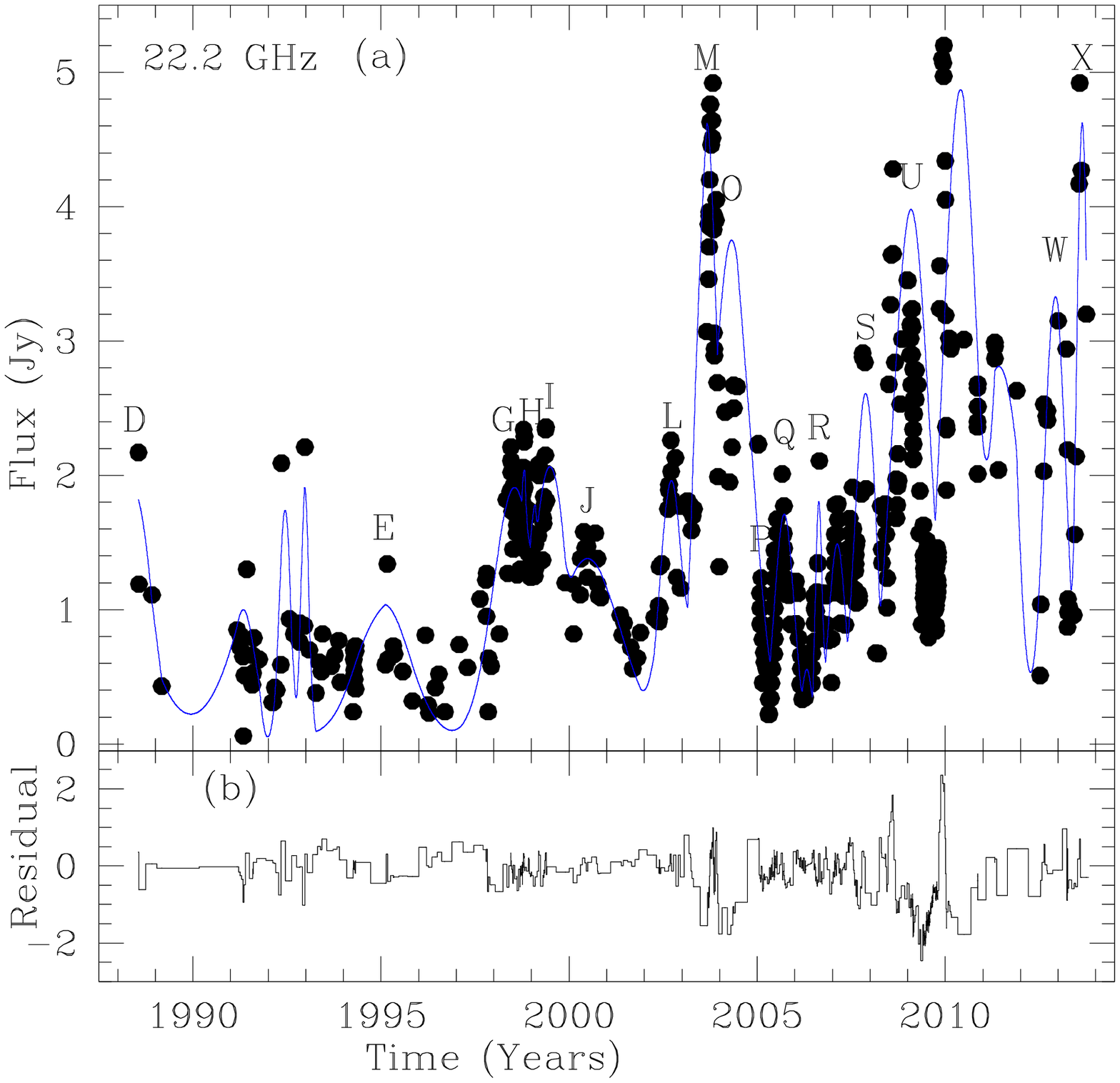}}
{\includegraphics[scale=0.33]{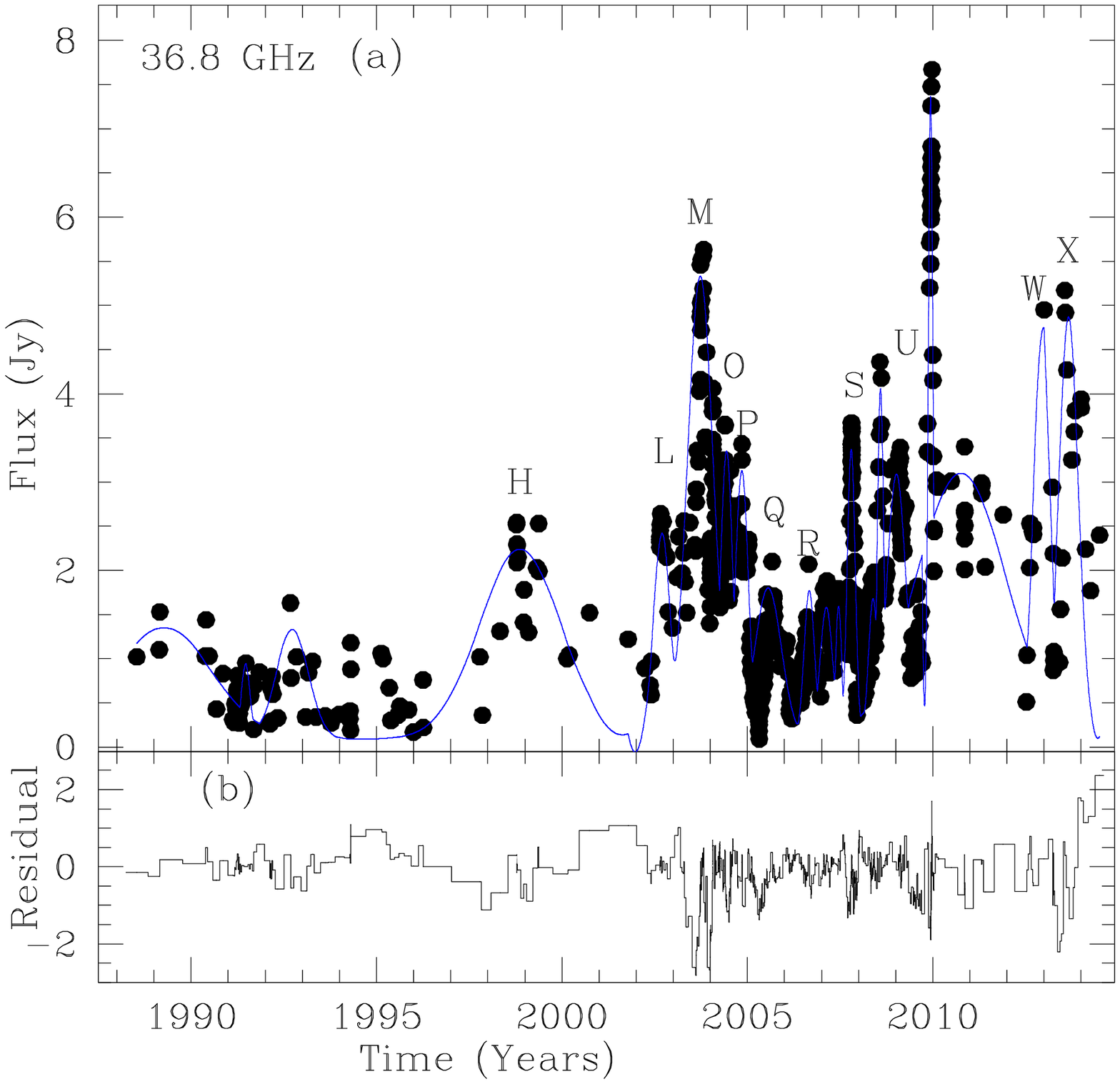}}
\caption{4.8 GHz, 8.0 GHz, 14.5 GHz, 22.2 GHz and 36.8 GHz light curve flares fit for S5 0716+714 with a piecewise Gaussian
function. The residual in the lower panels are calculated as $[y(A_i,\overline{m}_i,\sigma_i)-x_i(t_i)]/{\rm Standard ~Deviation(x_i (t_i))}$.}
\label{s5}
\end{figure*}

\begin{table}
{\bf Table 1.} Gaussian fit based parameters, time lags and spectral index for the light curves of S5 0716+714. 
\centering
\scalebox{.73}{
\begin{tabular}{|l|l|l|l|l|l|l|}
\hline
Flare & Frequency & Amplitude & Position & Width & Time lag & Spectral index \\
      & (GHz)     & $A$ (Jy)  & $\overline{m}$ (year) & $\sigma$ (years) & $\Delta t$ (years) & $\alpha$ \\
\hline

A     & 14.5      & 1.19$\pm$0.00 & 1981.35$\pm$0.17 & 0.14$\pm$0.00 & 0.00 & 0.40\\
      & 8.0       & 0.88$\pm$0.06 & 1981.47$\pm$0.07 & 0.18$\pm$0.07 & 0.12$\pm$0.18 & \\
      & 4.8       & 0.78$\pm$0.28 & 1981.82$\pm$0.02 & 0.11$\pm$0.01 & 0.47$\pm$0.17 & \\ \hline      
B     & 14.5      & 1.08$\pm$0.38 & 1982.57$\pm$0.14 & 0.77$\pm$0.03 & 0.00 & 0.20\\
      & 8.0      & 1.00$\pm$0.32 & 198.75$\pm$0.05 & 0.27$\pm$0.08 & 0.18$\pm$0.15 & \\
      & 4.8       & 0.86$\pm$0.16 & 1982.63$\pm$0.04 & 0.37$\pm$0.12 & 0.06$\pm$0.15 & \\\hline
C     & 8.0      & 0.68$\pm$0.08 & 1984.71$\pm$0.18 & 0.44$\pm$0.19 & 0.00 & $-$0.69\\ 
      & 4.8       & 0.97$\pm$0.28 & 1985.88$\pm$0.07 & 2.39$\pm$1.10 & 0.17$\pm$0.19 & \\ \hline
D     & 22.2      & 1.81$\pm$0.58 & 1988.45$\pm$0.09 & 0.42$\pm$0.15 & 0.00 & 0.59 \\
      & 14.5      & 1.12$\pm$0.06 & 1988.48$\pm$0.03 & 0.82$\pm$0.34 & 0.03$\pm$0.09 & \\
      & 4.8       & 0.80$\pm$0.34 & 1988.56$\pm$0.05 & 2.38$\pm$0.78 & 0.11$\pm$0.10 & \\ \hline
E     & 22.2      & 0.98$\pm$0.44 & 1995.07$\pm$0.00 & 0.69$\pm$0.26 & 0.00 & $-$16.46\\
      & 14.5      & 0.42$\pm$0.26 & 1995.14$\pm$0.02 & 0.60$\pm$0.26 & 0.07$\pm$0.02 & \\ \hline
F     & 14.5      & 0.32$\pm$0.14 & 1995.51$\pm$0.06 & 0.4$\pm$0.17 & 0.00 & $-$15.43 \\
      & 8.0       & 0.51$\pm$0.12 & 1996.05$\pm$0.01 & 1.17$\pm$0.76 & 0.54$\pm$0.06 & \\
      & 4.8       & 0.45$\pm$0.28 & 1996.22$\pm$0.01 & 0.48$\pm$0.22 & 0.71$\pm$0.06 & \\ \hline
G     & 22.2      & 1.85$\pm$0.14 & 1998.55$\pm$0.13 & 0.56$\pm$0.42 & 0.00 & 0.19 \\
      & 14.5      & 1.48$\pm$0.46 & 1998.48$\pm$0.05 & 0.20$\pm$0.07 & $-$0.07$\pm$0.14 & \\
      & 8.0       & 2.07$\pm$0.10 & 1998.65$\pm$0.19 & 0.22$\pm$0.77 & 0.10$\pm$0.23 & \\
      & 4.8       & 1.07$\pm$0.12 & 1998.95$\pm$0.19 & 0.31$\pm$0.09 & 0.40$\pm$0.23 & \\ \hline
H     & 36.8      & 2.15$\pm$0.24 & 1998.87$\pm$0.19 & 1.09$\pm$0.58 & 0.00 & 0.41 \\ 
      & 22.2      & 1.72$\pm$0.22 & 1999.10$\pm$0.19 & 0.21$\pm$0.01 & 0.23$\pm$0.27 & \\ 
      & 14.5      & 1.47$\pm$0.20 & 1999.26$\pm$0.10 & 0.53$\pm$0.13 & 0.39$\pm$0.21 & \\\hline      
I     & 22.2      & 2.00$\pm$0.24 & 1999.48$\pm$0.18 & 0.45$\pm$0.12 & 0.00 & 0.48 \\
      & 14.5      & 1.05$\pm$0.42 & 1999.80$\pm$0.06 & 0.32$\pm$0.06 & 0.32$\pm$0.19 & \\
      & 4.8       & 1.03$\pm$0.48 & 1999.56$\pm$0.10 & 0.65$\pm$0.18 & 0.08$\pm$0.21 & \\ \hline
J     & 22.2      & 1.32$\pm$0.02 & 2000.49$\pm$0.15 & 0.87$\pm$0.28 & 0.00 & 0.31\\
      & 14.5      & 1.12$\pm$0.30 & 2000.54$\pm$0.15 & 0.29$\pm$0.03 & 0.05$\pm$0.21 & \\
      & 8.0       & 0.81$\pm$0.08 & 2000.59$\pm$0.18 & 0.27$\pm$0.05 & 0.10$\pm$0.23 & \\
      & 4.8       & 0.91$\pm$0.40 & 2000.69$\pm$0.05 & 1.17$\pm$0.00 & 0.20$\pm$0.16 & \\ \hline      
K     & 14.5      & 0.89$\pm$0.26 & 2000.75$\pm$0.19 & 0.54$\pm$0.29 & 0.00 & $-$15.54\\
      & 8.0       & 0.71$\pm$0.32 & 2000.92$\pm$0.01 & 0.51$\pm$0.26 & 0.17$\pm$0.19 & \\ \hline 
L     & 36.8      & 2.33$\pm$0.36 & 2002.70$\pm$0.04 & 0.22$\pm$0.02 & 0.00 & 0.63 \\ 
      & 22.2      & 1.90$\pm$0.46 & 2002.72$\pm$0.09 & 0.30$\pm$0.05 & 0.02$\pm$0.10 & \\ 
      & 14.5      & 1.29$\pm$0.02 & 2002.72$\pm$0.04 & 0.29$\pm$0.28 & 0.02$\pm$0.06 & \\           
      & 8.0       & 0.85$\pm$0.00 & 2002.83$\pm$0.06 & 0.66$\pm$0.26 & 0.13$\pm$0.07 & \\\hline 
M     & 36.8      & 5.24$\pm$0.38 & 2003.73$\pm$0.03 & 0.35$\pm$0.10 & 0.00 & 0.55 \\ 
      & 22.2      & 4.56$\pm$0.38 & 2003.66$\pm$0.16 & 0.29$\pm$0.12 & $-$0.07$\pm$0.16 & \\ 
      & 14.5      & 3.58$\pm$0.12 & 2003.78$\pm$0.05 & 0.13$\pm$0.17 & 0.05$\pm$0.06 & \\  
      & 8.0       & 2.12$\pm$0.50 & 2003.78$\pm$0.03 & 0.21$\pm$0.02 & 0.05$\pm$0.04 & \\      
      & 4.8       & 1.58$\pm$0.28 & 2004.43$\pm$0.04 & 1.28$\pm$0.61 & 0.70$\pm$0.05 & \\\hline       
N     & 14.5      & 2.62$\pm$0.52 & 2004.09$\pm$0.13 & 0.22$\pm$0.00 & 0.00 & 0.70 \\
      & 8.0       & 1.76$\pm$0.46 & 2004.20$\pm$0.10 & 0.18$\pm$0.00 & 0.11$\pm$0.16 & \\ \hline 
O     & 36.8      & 3.26$\pm$0.54 & 2004.44$\pm$0.04 & 0.16$\pm$0.01 & 0.00 & 0.56 \\ 
      & 22.2      & 3.69$\pm$0.32 & 2004.32$\pm$0.11 & 0.52$\pm$0.05 & $-$0.12$\pm$0.12 & \\ 
      & 14.5      & 1.92$\pm$0.36 & 2004.72$\pm$0.13 & 0.42$\pm$0.07 & 0.28$\pm$0.14 & \\           
      & 8.0       & 1.24$\pm$0.44 & 2004.83$\pm$0.03 & 0.35$\pm$0.12 & 0.39$\pm$0.05 & \\\hline       
P     & 36.8      & 3.04$\pm$0.02 & 2004.85$\pm$0.04 & 0.17$\pm$0.05 & 0.00 & 2.41 \\ 
      & 22.2      & 0.88$\pm$0.38 & 2005.10$\pm$0.04 & 0.11$\pm$0.06 & 0.25$\pm$0.06 & \\  \hline     
Q     & 36.8      & 1.71$\pm$0.38 & 2005.56$\pm$0.04 & 0.36$\pm$0.11 & 0.00 & 0.34 \\ 
      & 22.2      & 1.65$\pm$0.26 & 2005.72$\pm$0.02 & 0.26$\pm$0.07 & 0.16$\pm$0.04 & \\ 
      & 14.5      & 1.14$\pm$0.40 & 2005.78$\pm$0.03 & 0.29$\pm$0.01 & 0.22$\pm$0.05 & \\  
      & 8.0       & 0.72$\pm$0.20 & 2005.82$\pm$0.13 & 0.43$\pm$0.02 & 0.26$\pm$0.14 & \\      
      & 4.8       & 1.14$\pm$0.44 & 2005.83$\pm$0.14 & 0.46$\pm$0.03 & 0.27$\pm$0.15 & \\\hline   

\end{tabular}}
\label{s5t}
\end{table}

\begin{table}
{\bf Table 1.} for S5 0716+714 continued.
\centering
\scalebox{.73}{
\begin{tabular}{|l|l|l|l|l|l|l|}
\hline
Flare & Frequency & Amplitude & Position & Width & Time lag & Spectral index \\
      & (GHz)     & $A$ (Jy)  & $\overline{m}$ (year) & $\sigma$ (years) & $\Delta t$ (years) & $\alpha$ \\
\hline

R     & 36.8      & 1.68$\pm$0.58 & 2006.67$\pm$0.08 & 0.14$\pm$0.01 & 0.00 & 0.50\\ 
      & 22.2      & 1.75$\pm$0.00 & 2006.64$\pm$0.02 & 0.09$\pm$0.10 & $-$0.03$\pm$0.08 & \\ 
      & 14.5      & 0.77$\pm$0.36 & 2007.04$\pm$0.02 & 0.57$\pm$0.21 & 0.37$\pm$0.08 & \\  
      & 8.0       & 0.69$\pm$0.34 & 2006.76$\pm$0.03 & 0.58$\pm$0.03 & 0.09$\pm$0.08 & \\      
      & 4.8       & 0.89$\pm$0.02 & 2006.78$\pm$0.09 & 0.40$\pm$0.27 & 0.11$\pm$0.12 & \\  \hline 
S     & 36.8      & 3.28$\pm$0.08 & 2007.80$\pm$0.05 & 0.11$\pm$0.06 & 0.00 & 0.68 \\ 
      & 22.2      & 2.55$\pm$0.10 & 2007.88$\pm$0.04 & 0.29$\pm$0.14 & 0.08$\pm$0.06 & \\ 
      & 14.5      & 1.66$\pm$0.28 & 2007.86$\pm$0.09 & 0.28$\pm$0.11 & 0.06$\pm$0.10 & \\  
      & 8.0       & 1.06$\pm$0.22 & 2007.98$\pm$0.19 & 0.40$\pm$0.04 & 0.18$\pm$0.20 & \\      
      & 4.8       & 0.93$\pm$0.06 & 2007.86$\pm$0.09 & 0.62$\pm$0.27 & 0.06$\pm$0.10 & \\  \hline      
T     & 36.8      & 3.97$\pm$0.22 & 2008.59$\pm$0.05 & 0.08$\pm$0.00 & 0.00 & 0.59 \\ 
      & 14.5      & 2.39$\pm$0.00 & 2008.66$\pm$0.02 & 0.18$\pm$0.18 & 0.07$\pm$0.05 & \\  
      & 8.0       & 1.35$\pm$0.36 & 2008.77$\pm$0.09 & 0.60$\pm$0.18 & 0.18$\pm$0.10 & \\      
      & 4.8       & 1.36$\pm$0.22 & 2008.61$\pm$0.10 & 0.28$\pm$0.07 & 0.02$\pm$0.11 & \\  \hline   
U     & 36.8      & 3.00$\pm$0.44 & 2009.02$\pm$0.01 & 0.26$\pm$0.14 & 0.00 & 0.27 \\ 
      & 22.2      & 3.92$\pm$0.48 & 2009.09$\pm$0.02 & 0.48$\pm$0.01 & 0.07$\pm$0.02 & \\ 
      & 14.5      & 2.69$\pm$0.18 & 2009.00$\pm$0.01 & 0.29$\pm$0.09 & $-$0.02$\pm$0.01 & \\  
      & 4.8       & 1.72$\pm$0.10 & 2009.16$\pm$0.17 & 0.34$\pm$0.05 & 0.14$\pm$0.17 & \\  \hline       
V     & 36.8      & 7.28$\pm$0.24 & 2009.94$\pm$0.05 & 0.06$\pm$0.07 & 0.00 & 0.90 \\ 
      & 14.5      & 3.14$\pm$0.38 & 2010.06$\pm$0.13 & 0.23$\pm$0.05 & 0.12$\pm$0.14 & \\  \hline    
W     & 36.8      & 4.84$\pm$0.16 & 2012.94$\pm$0.10 & 0.22$\pm$0.14 & 0.00 & 0.73 \\ 
      & 22.2      & 3.33$\pm$0.48 & 2012.92$\pm$0.14 & 0.28$\pm$0.06 & $-$0.02$\pm$0.17 & \\  \hline   
X     & 36.8      & 4.78$\pm$0.24 & 2013.66$\pm$0.08 & 0.26$\pm$0.05 & 0.00 &  0.09\\ 
      & 22.2      & 4.56$\pm$0.42 & 2013.64$\pm$0.12 & 0.15$\pm$0.03 & $-$0.02$\pm$0.14 & \\    
 
\hline
\end{tabular}}
\label{s5t2}
\end{table}

The light curves, best fit Gaussians and the residuals for S5 0716+714 are displayed in Figure 4, while for BL Lacertae they are presented in Fig 7.
For 3C 279, we split the light curve into two segments 
due to the difference in the flaring amplitudes. Segment 1 is from the beginning of observations till 1997 covering low amplitude flares compared
to larger amplitude flares in Segment 2 from 1998 to the end of the observation. Due to flaring behaviour being different during these epochs,
applying the Gaussian fits to the entire light curve can lead to loss in accounting for prominent features (even though low amplitude) from the
Segment 1 thus causing incorrect results. For the light curves of Segment 1 of 3C 279, the Gaussian fitting and the residuals
are presented in Fig.\ 5. The Gaussian fits along with the LC and residual (in
the bottom panel) for the second segment of 3C 279 are shown in Figure 6 for five frequency
bands. In all these figures, black points show the data points while the solid blue curve is the sum of all fitted Gaussians, covering the main
features in the light curve. Residuals plotted in the lower panel of these figures are calculated as $[y(A_i,\overline{m}_i,\sigma_i)-x_i(t_i)]/{\rm Standard~ Deviation(x_i (t_i))}$.

The parameters derived from the Gaussian fit procedure are given in Tables 1$-$4. Columns 1--7 give the flare nomenclature, observing frequency,
maximum amplitude of the flare ($A$), epoch of maximum flux, full width at half maximum (FWHM) from the Gaussian fit, time delays and spectral
index calculated using the convention
$A \propto \nu^\alpha$ 
for those flares present in at least three frequencies.

The periodogram analysis employed here is similar to that carried out for the analysis of X-ray, radio and optical light curves from Seyfert
galaxies and blazars in (Mohan \& Mangalam 2014; Mohan et al.\ 2015; Mohan et al.\ 2016). It is employed in the determination of the power spectral
density which is emergent from various physical process causing the variability in the light curve (jet and accretion disk based) and effects due
to the sampling process during the accumulation of the light curve. It is also used to infer any possible statistically significant quasi-periodic
features in the light curves originating from the above physical processes. Here, the 4.8 - 36.8 GHz light curves were interpolated and re-sampled
at regular intervals of $\Delta t$ = 0.1 d for the time series analysis with the normalized Fourier periodogram, 
\begin{equation}
P(f_j)=\frac{2 \Delta t}{\mu^2 N} |F(f_j)|^2
\end{equation}
where $\Delta t$ is the sampling time step for the evenly sampled light curve $x(t_n) = x(n \Delta t)$, and $|F(f_j)|$ is its
discrete Fourier transform evaluated at frequencies $f_j = j/(N \Delta t)$ with  $j = 1, 2,..,(N/2-1)$. The analysis procedure includes the periodogram estimation, parametric model fits to determine the PSD shape, model selection and statistical significance testing and is discussed in
more detail in the above mentioned work and references therein.

\begin{table}
{\bf Table 2.} {Gaussian fit based parameters, time lags and spectral index for segment 1 light curves of 3C 279.}
\centering
\scalebox{.73}{
\begin{tabular}{|l|l|l|l|l|l|l|}
\hline
Flare & Frequency & Amplitude & Position & Width & Time lag & Spectral index \\
      & (GHz)     & $A$ (Jy)  & $\overline{m}$ (year) & $\sigma$ (years) & $\Delta t$ (years) & $\alpha$ \\
\hline

A     & 14.5      & 1.37$\pm$0.10 & 1980.63$\pm$0.07 & 0.14$\pm$0.03 & 0.00 & 0.05 \\
      & 8.0       & 1.33$\pm$0.52 & 1980.59$\pm$0.11 & 0.20$\pm$0.01 & $-$0.04$\pm$0.13 & \\\hline
B     & 36.8      & 5.38$\pm$0.34 & 1981.64$\pm$0.07 & 0.37$\pm$0.55 & 0.00 & 0.17 \\
      & 22.2      & 4.71$\pm$0.54 & 1982.02$\pm$0.18 & 0.66$\pm$0.04 & 0.38$\pm$0.19 & \\
      & 14.5      & 4.17$\pm$0.04 & 1981.86$\pm$0.07 & 1.26$\pm$0.34 & 0.22$\pm$0.07 & \\
      & 8.0       & 4.26$\pm$0.28 & 1982.34$\pm$0.10 & 0.46$\pm$0.15 & 0.18$\pm$0.12 & \\\hline
C     & 8.0       & 3.99$\pm$0.34 & 1983.03$\pm$0.01 & 0.85$\pm$0.34 & 0.00 & 0.37 \\ 
      & 4.8       & 3.31$\pm$0.20 & 1983.08$\pm$0.06 & 1.20$\pm$0.54 & 0.50$\pm$0.06 & \\ \hline
D     & 36.8      & 3.67$\pm$0.58 & 1987.26$\pm$0.08 & 0.65$\pm$0.08 & 0.00 & 0.52 \\
      & 22.2      & 3.39$\pm$0.26 & 1987.42$\pm$0.07 & 0.17$\pm$0.10 & 0.16$\pm$0.11 & \\
      & 14.5      & 1.54$\pm$0.58 & 1987.37$\pm$0.06 & 0.37$\pm$0.15 & 0.11$\pm$0.10 & \\
      & 8.0       & 1.38$\pm$0.20 & 1987.64$\pm$0.00 & 0.68$\pm$0.28 & 0.38$\pm$0.08 & \\
      & 4.8       & 1.88$\pm$0.04 & 1988.98$\pm$0.11 & 1.50$\pm$0.17 & 1.72$\pm$0.14 & \\ \hline
E     & 36.8      & 9.16$\pm$0.58 & 1988.59$\pm$0.06 & 0.47$\pm$0.11 & 0.00 & 0.80 \\
      & 22.2      & 5.77$\pm$0.58 & 1988.78$\pm$0.10 & 1.21$\pm$0.11 & 0.19$\pm$0.12 & \\
      & 14.5      & 3.60$\pm$0.38 & 1989.32$\pm$0.15 & 0.97$\pm$0.18 & 0.73$\pm$0.16 & \\
      & 8.0       & 3.13$\pm$0.28 & 1989.60$\pm$0.10 & 1.07$\pm$0.27 & 1.01$\pm$0.12 & \\ 
      & 4.8       & 2.10$\pm$0.12 & 1989.71$\pm$0.07 & 0.38$\pm$0.12 & 1.12$\pm$0.09 & \\ \hline      
F     & 36.0      & 14.84$\pm$0.08 & 1991.28$\pm$0.20 & 0.73$\pm$0.08 & 0.00 & 1.08 \\
      & 22.2      & 8.41$\pm$0.42  & 1991.47$\pm$0.04 & 0.74$\pm$0.28 & 0.19$\pm$0.20 & \\ 
      & 14.5      & 6.10$\pm$0.36 & 1991.82$\pm$0.06 & 0.76$\pm$0.21 & 0.54$\pm$0.21 & \\
      & 8.0       & 2.66$\pm$0.56 & 1991.95$\pm$0.0 & 1.00$\pm$0.33 & 0.67$\pm$0.21 & \\
      & 4.8       & 1.14$\pm$0.34 & 1991.37$\pm$0.18 & 0.76$\pm$0.23 & 0.82$\pm$0.30 & \\ \hline
G     & 36.8      & 16.96$\pm$0.30 & 1994.39$\pm$0.10 & 0.61$\pm$0.18 & 0.00 & 0.64 \\
      & 22.2      & 13.45$\pm$0.16 & 1994.37$\pm$0.00 & 1.33$\pm$0.29 & $-$0.02$\pm$0.10 & \\
      & 14.5      & 10.72$\pm$0.14 & 1994.68$\pm$0.15 & 0.90$\pm$0.10 & 0.29$\pm$0.18 & \\
      & 8.0       & 6.91$\pm$0.00 & 1994.83$\pm$0.16 & 0.96$\pm$0.31 & 0.44$\pm$0.19 & \\
      & 4.8       & 3.07$\pm$0.04 & 1995.17$\pm$0.07 & 1.19$\pm$0.51 & 0.78$\pm$0.12 & \\ \hline
H     & 36.8      & 22.05$\pm$0.44 & 1996.53$\pm$0.13 & 0.80$\pm$0.22 & 0.00 & 0.02 \\ 
      & 22.2      & 21.83$\pm$0.42 & 1996.62$\pm$0.17 & 0.51$\pm$0.20 & 0.09$\pm$0.21 & \\
\hline
\end{tabular}}
\label{3C279p}
\end{table}

\begin{table}
{\bf Table 3.} Gaussian fit based parameters, time lags and spectral index for segment 2 light curves for 3C 279.
\centering
\scalebox{.73}{
\begin{tabular}{|l|l|l|l|l|l|l|}
\hline
Flare & Frequency & Amplitude & Position & Width & Time lag & Spectral index \\
      & (GHz)     & $A$ (Jy)  & $\overline{m}$ (year) & $\sigma$ (years) & $\Delta t$ (years) & $\alpha$ \\
\hline
I     & 14.5      & 14.34$\pm$0.18 & 1996.80$\pm$0.12 & 0.63$\pm$0.17 & 0.00 & 0.80\\
      & 8.0       & 8.89$\pm$0.20 & 1996.86$\pm$0.14 & 0.39$\pm$0.09 & 0.06$\pm$0.18 & \\\hline
J     & 22.2      & 14.90$\pm$0.22 & 1997.30$\pm$0.07 & 0.11$\pm$0.02 & 0.00 & 0.00 \\
      & 14.5      & 22.31$\pm$0.54 & 1997.95$\pm$0.01 & 0.58$\pm$0.0 & 0.65$\pm$0.07 & \\
      & 4.8       & 16.84$\pm$0.38 & 1998.04$\pm$0.05 & 0.59$\pm$0.18 & 0.74$\pm$0.09 & \\ \hline
K     & 36.8      & 23.17$\pm$0.52 & 1997.82$\pm$0.05 & 1.50$\pm$0.41 & 0.00 & 0.35 \\ 
      & 22.2      & 26.30$\pm$0.58 & 1998.00$\pm$0.16 & 0.49$\pm$0.00 & 0.18$\pm$0.17 & \\
      & 14.5      & 20.70$\pm$0.42 & 1998.79$\pm$0.07 & 1.01$\pm$0.31 & 0.97$\pm$0.09 & \\
      & 8.0       & 18.28$\pm$0.16 & 1998.74$\pm$0.13 & 1.4$\pm$0.41 & 0.92$\pm$0.14 & \\
      & 4.8       & 8.02$\pm$0.04 & 1998.72$\pm$0.05 & 1.1$\pm$0.32 & 0.90$\pm$0.05 & \\ \hline
L     & 36.8      & 12.48$\pm$0.50 & 2000.73$\pm$0.10 & 0.53$\pm$0.01 & $-$0.00 & 0.04 \\
      & 22.2      & 16.87$\pm$0.06 & 2000.77$\pm$0.50 & 0.65$\pm$0.21 & 0.04$\pm$0.51 & \\
      & 14.5      & 15.42$\pm$0.32 & 2000.68$\pm$0.05 & 0.75$\pm$0.24 & $-$0.05$\pm$0.11 & \\
      & 8.0       & 14.42$\pm$0.30 & 2000.93$\pm$0.04 & 1.66$\pm$0.51 & 0.20$\pm$0.12 & \\
      & 4.8       & 12.30$\pm$0.10 & 2000.22$\pm$0.15 & 1.60$\pm$0.50 & 0.49$\pm$0.18 & \\ \hline
M     & 22.2      & 12.12$\pm$0.46 & 2001.56$\pm$0.03 & 0.11$\pm$0.04 & 0.00 &  0.27 \\
      & 14.5      & 10.81$\pm$0.58 & 2001.63$\pm$0.05 & 0.14$\pm$0.05 & 0.07$\pm$0.06 & \\\hline
N     & 36.0      & 20.1$\pm$0.44 & 2002.14$\pm$0.02 & 0.43$\pm$0.18 & 0.00 & 0.48 \\
      & 22.2      & 15.72$\pm$0.32  & 2002.07$\pm$0.17 & 0.60$\pm$0.18 & $-$0.07$\pm$0.17 & \\ 
      & 14.5      & 12.82$\pm$0.58 & 2002.19$\pm$0.06 & 0.70$\pm$0.07 & 0.05$\pm$0.06 & \\\hline
O     & 36.8      & 5.56$\pm$0.52 & 2003.85$\pm$0.05 & 0.44$\pm$0.16 & 0.00 & 0.12 \\
      & 22.2      & 6.14$\pm$0.12 & 2003.86$\pm$0.11 & 0.44$\pm$0.08 & 0.01$\pm$0.12 & \\
      & 14.5      & 6.75$\pm$0.42 & 2003.97$\pm$0.03 & 0.42$\pm$0.1 & 0.12$\pm$0.06 & \\
      & 8.0       & 5.47$\pm$0.36 & 2003.18$\pm$0.15 & 0.47$\pm$0.02 & 0.33$\pm$0.16 & \\
      & 4.8       & 4.1$\pm$0.54 & 2003.51$\pm$0.11 & 0.48$\pm$0.13 & 0.66$\pm$0.12 & \\ \hline
P     & 14.5      & 7.50$\pm$0.42 & 2004.64$\pm$0.06 & 0.08$\pm$0.00 & 0.00 & 0.49 \\ 
      & 8.0       & 5.41$\pm$0.18 & 2005.01$\pm$0.11 & 0.31$\pm$0.10 & 0.37$\pm$0.12 & \\
      & 4.8       & 4.44$\pm$0.22 & 2005.67$\pm$0.14 & 0.61$\pm$0.15 & 1.03$\pm$0.15 & \\ \hline
Q     & 14.5      & 6.15$\pm$0.40 & 2005.51$\pm$0.08 & 0.51$\pm$0.21 & 0.00 & 0.13 \\
      & 8.0       & 5.70$\pm$0.104 & 2005.71$\pm$0.17 & 0.54$\pm$0.12 & 0.20$\pm$0.19 & \\\hline
R     & 36.8      & 13.72$\pm$0.46 & 2006.70$\pm$0.08 & 0.20$\pm$0.03 & 0.00 &  0.44 \\
      & 22.2      & 12.96$\pm$0.44 & 2006.86$\pm$0.13 & 0.35$\pm$0.29 & 0.16$\pm$0.15 & \\
      & 14.5      & 10.67$\pm$0.06 & 2006.94$\pm$0.06 & 0.42$\pm$0.11 & 0.24$\pm$0.10 & \\
      & 8.0       & 7.72$\pm$0.26 & 2007.00$\pm$0.11 & 0.56$\pm$0.14 & 0.30$\pm$0.14 & \\
      & 4.8       & 4.72$\pm$0.34 & 2007.36$\pm$0.01 & 0.47$\pm$0.15 & 0.66$\pm$0.08 & \\ \hline
S     & 8.0      & 6.76$\pm$0.22 & 2006.57$\pm$0.17 & 0.20$\pm$0.08 & 0.00 & 0.57 \\
      & 4.8       & 5.06$\pm$0.28 & 2006.58$\pm$0.01 & 0.55$\pm$0.15 & 0.31$\pm$0.17 & \\ \hline
T     & 36.8      & 11.55$\pm$0.50 & 2007.96$\pm$0.09 & 0.19$\pm$0.08 & 0.00 & 0.55 \\
      & 22.2      & 11.69$\pm$0.14 & 2008.08$\pm$0.04 & 0.35$\pm$0.24 & 0.12$\pm$0.10 & \\
      & 14.5      & 6.92$\pm$0.00 & 2008.03$\pm$0.12 & 0.48$\pm$0.13 & 0.07$\pm$0.15 & \\ 
      & 4.8       & 3.29$\pm$0.38 & 2008.42$\pm$0.05 & 0.46$\pm$0.17 & 0.46$\pm$0.11 & \\ \hline
U     & 36.8      & 11.09$\pm$0.50 & 2008.53$\pm$0.01 & 0.20$\pm$0.04 & 0.00 & 0.49 \\
      & 22.2      & 11.55$\pm$0.40 & 2008.98$\pm$0.03 & 0.14$\pm$0.06 & 0.45$\pm$0.03 & \\
      & 14.5      & 8.75$\pm$0.48 & 2008.78$\pm$0.11 & 0.40$\pm$0.01 & 0.25$\pm$0.11 & \\
      & 8.0       & 6.03$\pm$0.10 & 2009.03$\pm$0.05 & 0.63$\pm$0.45 & 0.50$\pm$0.05 & \\
      & 4.8       & 2.84$\pm$0.18 & 2009.11$\pm$0.03 & 0.47$\pm$0.14 & 0.58$\pm$0.03 & \\ \hline
V     & 36.8      & 7.44$\pm$0.08 & 2010.00$\pm$0.07 & 0.09$\pm$0.23 & 0.00 & 1.04 \\
      & 22.2      & 6.86$\pm$0.06 & 2010.05$\pm$0.06 & 0.09$\pm$0.09 & 0.05$\pm$0.09 & \\
      & 14.5      & 1.57$\pm$0.28 & 2010.08$\pm$0.16 & 0.19$\pm$0.13 & 0.08$\pm$0.17 & \\
      & 4.8       & 0.47$\pm$0.26 & 2010.29$\pm$0.04 & 0.08$\pm$0.03 & 0.29$\pm$0.08 & \\ \hline  
W     & 36.8      & 10.05$\pm$0.12 & 2010.45$\pm$0.09 & 0.27$\pm$0.02 & 0.00 & 0.35 \\
      & 22.2      & 9.94$\pm$0.08 & 2010.51$\pm$0.03 & 0.28$\pm$0.10 & 0.06$\pm$0.09 & \\
      & 14.5      & 12.33$\pm$0.42 & 2011.45$\pm$0.10 & 0.38$\pm$0.04 & 1.00$\pm$0.13 & \\
      & 8.0       & 6.50$\pm$0.30 & 2011.53$\pm$0.18 & 0.50$\pm$0.05 & 1.08$\pm$0.20 & \\
      & 4.8       & 3.19$\pm$0.04 & 2011.57$\pm$0.16 & 0.24$\pm$0.07 & 1.12$\pm$0.18 & \\    
\hline
\end{tabular}}
\label{3C279p2}
\end{table}

\begin{table}
{\bf Table 4.} Gaussian fit based parameters, time lags and spectral index for the light curves of BL Lacertae.
\centering
\scalebox{.73}{
\begin{tabular}{|l|l|l|l|l|l|l|}
\hline
Flare & Frequency & Amplitude & Position & Width & Time lag & Spectral index \\
      & (GHz)     & $A$ (Jy)  & $\overline{m}$ (year) & $\sigma$ (years) & $\Delta t$ (years) & $\alpha$ \\
\hline

1     & 14.5      & 6.20$\pm$0.44 & 1974.61$\pm$0.16 & 0.20$\pm$0.74 & 0.00 & 0.92 \\
      & 8.0       & 3.59$\pm$0.26 & 1975.67$\pm$0.06 & 0.30$\pm$0.07 & 0.06$\pm$0.17 & \\\hline      
2     & 14.5      & 5.56$\pm$0.52 & 1976.62$\pm$0.00 & 0.90$\pm$0.29 & 0.00 & $-$0.05 \\
      & 8.0       & 5.73$\pm$0.44 & 1976.81$\pm$0.08 & 0.70$\pm$0.12 & 0.19$\pm$0.08 & \\\hline
      & 4.8       & 2.20$\pm$0.08 & 1977.93$\pm$0.11 & 0.52$\pm$0.10 & 0.31$\pm$0.11 & \\     
3     & 14.5      & 1.72$\pm$0.06 & 1978.34$\pm$0.02 & 0.41$\pm$0.13 & 0.00 & $-$0.24\\ 
      & 8.0       & 1.58$\pm$0.32 & 1978.45$\pm$0.16 & 0.34$\pm$0.01 & 0.11$\pm$0.16 & \\ \hline
4     & 36.8      & 12.34$\pm$0.30 & 1980.59$\pm$0.04 & 0.52$\pm$0.17 & 0.00 &  0.08 \\
      & 22.2      & 9.17$\pm$0.10  & 1980.50$\pm$0.04 & 1.12$\pm$1.03 & $-$0.09$\pm$0.04 & \\
      & 14.5      & 13.66$\pm$0.22 & 1980.60$\pm$0.01 & 0.19$\pm$0.17 & 0.01$\pm$0.01 & \\
      & 8.0       & 11.75$\pm$0.16 & 1980.66$\pm$0.02 & 0.20$\pm$0.19 & 0.07$\pm$0.04 & \\
      & 4.8       & 8.74$\pm$0.48  & 1980.65$\pm$0.09 & 0.16$\pm$0.10 & 0.06$\pm$0.10 & \\ \hline
5     & 14.5      & 6.46$\pm$0.38 & 1981.15$\pm$0.15 & 0.29$\pm$0.14 & 0.00 & $-$0.07 \\
      & 8.0       & 7.38$\pm$0.00 & 1981.18$\pm$0.12 & 0.28$\pm$0.01 & 0.03$\pm$0.19 & \\ 
      & 4.8       & 6.97$\pm$0.32 & 1981.19$\pm$0.11 & 0.27$\pm$0.03 & 0.04$\pm$0.19 & \\ \hline 
6     & 14.5      & 4.18$\pm$0.38 & 1981.94$\pm$0.16 & 0.34$\pm$0.06 & 0.00 & 0.05 \\
      & 8.0       & 4.75$\pm$0.50 & 1981.96$\pm$0.03 & 0.38$\pm$0.09 & 0.02$\pm$0.16 & \\
      & 4.8       & 3.87$\pm$0.24 & 1982.01$\pm$0.17 & 0.39$\pm$0.24 & 0.07$\pm$0.23 & \\ \hline
7     & 36.8      & 2.97$\pm$0.26 & 1983.47$\pm$0.16 & 0.45$\pm$0.28 & 0.00 & $-$0.12 \\
      & 22.2      & 2.26$\pm$0.22 & 1983.71$\pm$0.00 & 0.80$\pm$0.19 & 0.24$\pm$0.16 & \\
      & 14.5      & 2.70$\pm$0.52 & 1983.53$\pm$0.03 & 0.49$\pm$0.12 & 0.06$\pm$0.16 & \\
      & 8.0       & 3.51$\pm$0.04 & 1983.71$\pm$0.14 & 0.49$\pm$0.35 & 0.24$\pm$0.21 & \\
      & 4.8       & 3.22$\pm$0.16 & 1983.64$\pm$0.18 & 0.39$\pm$0.02 & 0.17$\pm$0.24 & \\ \hline
8     & 36.8      & 1.41$\pm$0.22 & 1986.43$\pm$0.07 & 0.0$\pm$0.01 & 0.00 & $-$16.67 \\ 
      & 14.5      & 0.82$\pm$0.02 & 1986.50$\pm$0.03 & 0.19$\pm$0.17 & 0.07$\pm$0.08 & \\
      & 8.0       & 0.77$\pm$0.10 & 1986.50$\pm$0.08 & 0.23$\pm$0.03 & 0.07$\pm$0.11 & \\
      & 4.8       & 0.52$\pm$0.14 & 1986.55$\pm$0.00 & 0.20$\pm$0.05 & 0.12$\pm$0.07 & \\ \hline
9     & 8.0      & 1.09$\pm$0.14 & 1987.02$\pm$0.03 & 0.16$\pm$0.06 & 0.00 & 1.20 \\ 
      & 4.8      & 0.59$\pm$0.18 & 1987.06$\pm$0.06 & 0.24$\pm$0.03 & 0.04$\pm$0.07 & \\ \hline      
10    & 36.8      & 5.18$\pm$0.50 & 1987.98$\pm$0.01 & 0.41$\pm$0.13 & 0.00 & 0.10 \\
      & 22.2      & 4.75$\pm$0.28 & 1988.12$\pm$0.11 & 0.57$\pm$0.34 & 0.14$\pm$0.11 & \\
      & 14.5      & 4.71$\pm$0.32 & 1988.07$\pm$0.11 & 0.37$\pm$0.19 & 0.09$\pm$0.11 & \\
      & 8.0       & 4.03$\pm$0.02 & 1988.14$\pm$0.16 & 0.45$\pm$0.05 & 0.16$\pm$0.16 & \\
      & 4.8       & 4.44$\pm$0.56 & 1988.20$\pm$0.03 & 0.29$\pm$0.11 & 0.22$\pm$0.03 & \\ \hline
11    & 36.8      & 2.54$\pm$0.30 & 1988.95$\pm$0.13 & 0.57$\pm$0.16 & 0.00 & $-$0.03 \\
      & 14.5      & 2.70$\pm$0.22 & 1988.72$\pm$0.03 & 0.23$\pm$0.11 & $-$0.23$\pm$0.13 & \\
      & 8.0       & 2.85$\pm$0.34 & 1988.80$\pm$0.02 & 0.12$\pm$0.05 & $-$0.15$\pm$0.13 & \\
      & 4.8       & 2.66$\pm$0.32 & 1988.65$\pm$0.11 & 0.22$\pm$0.03 & $-$0.30$\pm$0.17 & \\ \hline
12    & 14.5      & 2.42$\pm$0.36 & 1989.26$\pm$0.11 & 0.46$\pm$0.02 & 0.00 & $-$0.01 \\
      & 8.0       & 3.19$\pm$0.20 & 1989.19$\pm$0.16 & 0.19$\pm$0.41 & $-$0.07$\pm$0.19 & \\ 
      & 4.8       & 2.43$\pm$0.46 & 1989.34$\pm$0.07 & 0.52$\pm$0.10 & 0.08$\pm$0.13 & \\ \hline 
13    & 36.8      & 4.10$\pm$0.52 & 1989.79$\pm$0.18 & 0.34$\pm$0.20 & 0.00 & 0.32 \\
      & 22.2      & 3.76$\pm$0.22 & 1990.01$\pm$0.18 & 0.36$\pm$0.13 & 0.22$\pm$0.18 & \\
      & 14.5      & 3.56$\pm$0.14 & 1990.01$\pm$0.18 & 0.17$\pm$0.17 & 0.22$\pm$0.18 & \\
      & 4.8       & 1.95$\pm$0.56 & 1990.01$\pm$0.03 & 0.30$\pm$0.03 & 0.22$\pm$0.18 & \\ \hline
14    & 36.8      & 1.96$\pm$0.42 & 1990.59$\pm$0.01 & 0.21$\pm$0.04 & 0.00 & 0.12 \\
      & 14.5      & 1.60$\pm$0.30 & 1990.62$\pm$0.11 & 0.28$\pm$0.01 & 0.03$\pm$0.11 & \\
      & 8.0       & 1.70$\pm$0.24 & 1990.66$\pm$0.00 & 0.32$\pm$0.13 & 0.07$\pm$0.01 & \\
      & 4.8       & 1.47$\pm$0.44 & 1990.68$\pm$0.00 & 0.35$\pm$0.12 & 0.09$\pm$0.01 & \\ \hline
15    & 36.8      & 1.30$\pm$0.12 & 1991.04$\pm$0.11 & 0.18$\pm$0.03 & 0.00 & $-$0.10 \\
      & 22.2      & 1.88$\pm$0.24 & 1991.16$\pm$0.01 & 0.51$\pm$0.16 & 0.12$\pm$0.02 & \\
      & 14.5      & 1.52$\pm$0.04 & 1991.45$\pm$0.01 & 0.29$\pm$0.15 & 0.41$\pm$0.02 & \\
      & 8.0       & 1.93$\pm$0.20 & 1991.48$\pm$0.03 & 0.24$\pm$0.15 & 0.44$\pm$0.03 & \\
      & 4.8       & 1.72$\pm$0.30 & 1991.41$\pm$0.12 & 0.31$\pm$0.03 & 0.37$\pm$0.16 & \\ \hline
16    & 36.8      & 2.20$\pm$0.30 & 1992.56$\pm$0.06 & 0.30$\pm$0.09 & 0.00 & 0.22 \\
      & 22.2      & 1.93$\pm$0.26 & 1992.61$\pm$0.16 & 0.37$\pm$0.07 & 0.05$\pm$0.17 & \\
      & 14.5      & 1.83$\pm$0.42 & 1992.55$\pm$0.10 & 0.36$\pm$0.04 & $-$0.01$\pm$0.12 & \\
      & 8.0       & 1.85$\pm$0.22 & 1992.60$\pm$0.13 & 0.36$\pm$0.07 & 0.04$\pm$0.14 & \\
      & 4.8       & 1.22$\pm$0.42 & 1992.73$\pm$0.02 & 0.44$\pm$0.19 & 0.17$\pm$0.06 & \\ \hline      
17    & 36.8      & 3.66$\pm$0.34 & 1993.43$\pm$0.02 & 0.17$\pm$0.09 & 0.00 & $-$0.23 \\
      & 22.2      & 3.72$\pm$0.48 & 1993.43$\pm$0.09 & 0.16$\pm$0.03 & 0.00$\pm$0.09 & \\
      & 14.5      & 3.03$\pm$0.20 & 1993.46$\pm$0.03 & 0.18$\pm$0.19 & 0.03$\pm$0.04 & \\
      & 8.0       & 3.24$\pm$0.30 & 1993.47$\pm$0.15 & 0.18$\pm$0.06 & 0.04$\pm$0.15 & \\
      & 4.8       & 2.16$\pm$0.12 & 1993.47$\pm$0.04 & 0.20$\pm$0.16 & 0.04$\pm$0.04 & \\ \hline 
18    & 36.8      & 2.64$\pm$0.48 & 1995.50$\pm$0.04 & 0.27$\pm$0.02 & 0.00 & $-$0.12 \\
      & 22.2      & 4.54$\pm$0.02 & 1995.91$\pm$0.07 & 0.48$\pm$0.27 & 0.41$\pm$0.08 & \\
      & 14.5      & 3.89$\pm$0.60 & 1995.97$\pm$0.15 & 0.64$\pm$0.11 & 0.47$\pm$0.15 & \\
      & 8.0       & 4.71$\pm$0.36 & 1996.06$\pm$0.06 & 0.66$\pm$0.28 & 0.51$\pm$0.07 & \\
      & 4.8       & 5.12$\pm$0.58 & 1996.21$\pm$0.04 & 0.54$\pm$0.08 & 0.71$\pm$0.04 & \\ \hline       
19    & 36.8      & 3.98$\pm$0.54 & 1995.91$\pm$0.09 & 0.32$\pm$0.09 & 0.00 & $-$0.01 \\
      & 22.2      & 4.83$\pm$0.34 & 1996.03$\pm$0.09 & 0.14$\pm$0.02 & 0.12$\pm$0.09 & \\
      & 14.5      & 4.26$\pm$0.02 & 1996.36$\pm$0.06 & 0.21$\pm$0.07 & 0.45$\pm$0.11 & \\
      & 8.0       & 5.07$\pm$0.14 & 1996.29$\pm$0.14 & 0.5$\pm$0.01 & 0.38$\pm$0.17 & \\ \hline     

\end{tabular}}
\label{BLLAC}
\end{table}

\begin{table}
{\bf Table 4.} for BL Lacertae continued.
\centering
\scalebox{.73}{
\begin{tabular}{|l|l|l|l|l|l|l|}
\hline
Flare & Frequency & Amplitude & Position & Width & Time lag & Spectral index \\
      & (GHz)     & $A$ (Jy)  & $\overline{m}$ (year) & $\sigma$ (years) & $\Delta t$ (years) & $\alpha$ \\
\hline

20    & 36.8      & 4.22$\pm$0.52 & 1996.50$\pm$0.14 & 0.62$\pm$0.14 & 0.00 & 0.21 \\
      & 22.2      & 4.65$\pm$0.14 & 1996.64$\pm$0.07 & 0.17$\pm$0.05 & 0.14$\pm$0.16 & \\
      & 14.5      & 4.22$\pm$0.0 & 1996.64$\pm$0.06 & 0.10$\pm$0.03 & 0.14$\pm$0.15 & \\
      & 8.0       & 4.46$\pm$0.36 & 1996.80$\pm$0.02 & 0.52$\pm$0.17 & 0.30$\pm$0.14 & \\
      & 4.8       & 4.37$\pm$0.22 & 1996.85$\pm$0.03 & 0.47$\pm$0.20 & 0.35$\pm$0.14 & \\ \hline 
21    & 36.8      & 4.41$\pm$0.36 & 1997.83$\pm$0.02 & 0.18$\pm$0.05 & 0.00 & $-$0.04\\
      & 22.2      & 3.92$\pm$0.40 & 1997.91$\pm$0.06 & 0.21$\pm$0.10 & 0.08$\pm$0.06 & \\
      & 14.5      & 2.99$\pm$0.56 & 1997.93$\pm$0.00 & 0.22$\pm$0.03 & 0.10$\pm$0.02 & \\
      & 8.0       & 3.20$\pm$0.00 & 1998.04$\pm$0.15 & 0.48$\pm$0.09 & 0.21$\pm$0.15 & \\
      & 4.8       & 2.94$\pm$0.04 & 1998.02$\pm$0.10 & 0.21$\pm$0.09 & 0.19$\pm$0.10 & \\ \hline       
22    & 22.2      & 3.45$\pm$0.46 & 1998.31$\pm$0.01 & 0.17$\pm$0.08 & 0.00 & 0.31 \\
      & 14.5      & 2.92$\pm$0.46 & 1998.33$\pm$0.02 & 0.17$\pm$0.02 & 0.02$\pm$0.02 & \\
      & 4.8       & 3.50$\pm$0.00 & 1998.34$\pm$0.09 & 0.14$\pm$0.00 & 0.03$\pm$0.09 & \\ \hline        
23    & 36.8      & 3.98$\pm$0.54 & 1998.43$\pm$0.14 & 0.63$\pm$0.07 & 0.00 & 0.25 \\
      & 22.2      & 3.92$\pm$0.02 & 1998.68$\pm$0.03 & 0.19$\pm$0.06 & 0.25$\pm$0.14 & \\
      & 14.5      & 2.92$\pm$0.46 & 1998.72$\pm$0.01 & 0.17$\pm$0.02 & 0.29$\pm$0.14 & \\
      & 8.0       & 2.34$\pm$0.12 & 1998.86$\pm$0.02 & 0.56$\pm$0.23 & 0.43$\pm$0.14 & \\
      & 4.8       & 2.31$\pm$0.10 & 1998.87$\pm$0.05 & 0.58$\pm$0.19 & 0.44$\pm$0.15 & \\ \hline      
24    & 36.8      & 4.43$\pm$0.46 & 2000.03$\pm$0.16 & 0.42$\pm$0.24 & 0.00 & $-$0.28\\
      & 22.2      & 4.21$\pm$0.18 & 2000.16$\pm$0.09 & 0.31$\pm$0.02 & 0.13$\pm$0.18 & \\
      & 14.5      & 3.71$\pm$0.32 & 2000.20$\pm$0.09 & 0.31$\pm$0.12 & 0.17$\pm$0.18 & \\
      & 8.0       & 3.24$\pm$0.10 & 2000.26$\pm$0.04 & 0.28$\pm$0.10 & 0.23$\pm$0.16 & \\
      & 4.8       & 2.54$\pm$0.20 & 2000.28$\pm$0.16 & 0.33$\pm$0.1 & 0.25$\pm$0.16 & \\ \hline      
25    & 36.8      & 1.50$\pm$0.14 & 2001.04$\pm$0.17 & 0.28$\pm$0.05 & 0.00 & 0.38 \\
      & 22.2      & 1.38$\pm$0.12 & 2001.05$\pm$0.12 & 0.30$\pm$0.06 & 0.01$\pm$0.21 & \\
      & 14.5      & 1.43$\pm$0.14 & 2001.06$\pm$0.09 & 0.21$\pm$0.06 & 0.02$\pm$0.19 & \\
      & 8.0       & 1.59$\pm$0.40 & 2001.00$\pm$0.12 & 0.24$\pm$0.04 & $-$0.04$\pm$0.21 & \\
      & 4.8       & 1.19$\pm$0.18 & 2001.02$\pm$0.08 & 0.26$\pm$0.06 & $-$0.02$\pm$0.19 & \\ \hline        
26    & 36.8      & 2.30$\pm$0.34 & 2002.14$\pm$0.09 & 0.51$\pm$0.13 & 0.00 & 0.48 \\
      & 22.2      & 2.07$\pm$0.48 & 2002.15$\pm$0.11 & 0.52$\pm$0.08 & 0.01$\pm$0.14 & \\
      & 14.5      & 1.55$\pm$0.26 & 2002.28$\pm$0.06 & 0.17$\pm$0.04 & 0.14$\pm$0.11 & \\
      & 4.8       & 1.09$\pm$0.38 & 2002.15$\pm$0.10 & 0.49$\pm$0.15 & 0.01$\pm$0.13 & \\ \hline       
27    & 36.8      & 2.56$\pm$0.26 & 2002.54$\pm$0.08 & 0.32$\pm$0.11 & 0.00 & 0.41 \\
      & 22.2      & 2.07$\pm$0.16 & 2002.63$\pm$0.04 & 0.10$\pm$0.03 & 0.09$\pm$0.09 & \\
      & 14.5      & 1.58$\pm$0.48 & 2002.73$\pm$0.05 & 0.28$\pm$0.00 & 0.19$\pm$0.09 & \\
      & 8.0       & 1.29$\pm$0.32 & 2002.66$\pm$0.08 & 0.40$\pm$0.05 & 0.12$\pm$0.08 & \\
      & 4.8       & 0.94$\pm$0.40 & 2002.62$\pm$0.05 & 0.15$\pm$0.07 & 0.08$\pm$0.09 & \\ \hline  
28    & 36.8      & 6.19$\pm$0.30 & 2005.12$\pm$0.16 & 0.45$\pm$0.23 & 0.00 & 0.09 \\
      & 14.5      & 4.53$\pm$0.40 & 2005.40$\pm$0.01 & 0.47$\pm$0.18 & 0.8$\pm$0.16 & \\
      & 8.0       & 3.47$\pm$0.30 & 2005.54$\pm$0.16 & 0.64$\pm$0.04 & 0.42$\pm$0.16 & \\
      & 4.8       & 2.4$\pm$0.46 & 2005.63$\pm$0.02 & 0.54$\pm$0.01 & 0.51$\pm$0.16 & \\ \hline       
29    & 36.8      & 4.83$\pm$0.06 & 2002.58$\pm$0.04 & 0.21$\pm$0.10 & 0.00 & 0.19 \\
      & 22.2      & 4.25$\pm$0.16 & 2005.82$\pm$0.09 & 0.24$\pm$0.08 & 0.24$\pm$0.10 & \\
      & 14.5      & 4.47$\pm$0.12 & 2005.99$\pm$0.08 & 0.35$\pm$0.07 & 0.41$\pm$0.09 & \\\hline
30    & 36.8      & 3.1$\pm$0.02 & 2007.05$\pm$0.04 & 0.24$\pm$0.11 & 0.00 & 0.31 \\
      & 22.2      & 2.81$\pm$0.56 & 2007.18$\pm$0.00 & 0.22$\pm$0.01 & 0.13$\pm$0.04 & \\   \hline   
31    & 36.8      & 3.93$\pm$0.48 & 2007.77$\pm$0.14 & 0.36$\pm$0.00 & 0.00 & 0.32 \\
      & 22.2      & 3.70$\pm$0.04 & 2007.84$\pm$0.04 & 0.17$\pm$0.08 & 0.07$\pm$0.14 & \\
      & 14.5      & 3.29$\pm$0.02 & 2007.75$\pm$0.14 & 0.33$\pm$0.12 & $-$0.02$\pm$0.14 & \\
      & 8.0       & 2.46$\pm$0.24 & 2007.73$\pm$0.01 & 0.42$\pm$0.21 & $-$0.04$\pm$0.14 & \\
      & 4.8       & 2.05$\pm$0.20 & 2007.93$\pm$0.03 & 0.24$\pm$0.07 & 0.16$\pm$0.14 & \\ \hline   
32    & 36.8      & 2.17$\pm$0.26 & 2008.58$\pm$0.04 & 0.36$\pm$0.07 & 0.00 & 0.12 \\
      & 22.2      & 1.92$\pm$0.50 & 2008.90$\pm$0.01 & 0.35$\pm$0.17 & 0.32$\pm$0.04 & \\
      & 14.5      & 1.32$\pm$0.04 & 2008.66$\pm$0.04 & 0.08$\pm$0.02 & 0.08$\pm$0.06 & \\
      & 8.0       & 1.48$\pm$0.08 & 2008.68$\pm$0.15 & 0.23$\pm$0.03 & 0.10$\pm$0.15 & \\ \hline 
33    & 36.8      & 3.84$\pm$0.52 & 2010.14$\pm$0.11 & 0.14$\pm$0.00 & 0.00 & 0.36 \\
      & 22.2      & 3.59$\pm$0.40 & 2010.13$\pm$0.07 & 0.14$\pm$0.04 & $-$0.01$\pm$0.13 & \\
      & 14.5      & 3.24$\pm$0.18 & 2010.24$\pm$0.02 & 0.87$\pm$0.28 & 0.10$\pm$0.11 & \\ 
      & 8.0       & 3.28$\pm$0.06 & 2010.15$\pm$0.11 & 0.21$\pm$0.04 & 0.01$\pm$0.16 & \\
      & 4.8       & 2.96$\pm$0.34 & 2010.23$\pm$0.02 & 0.17$\pm$0.08 & 0.09$\pm$0.11 & \\ \hline              
34    & 36.8      & 4.51$\pm$0.40 & 2010.87$\pm$0.02 & 0.22$\pm$0.09 & 0.00 & 0.39 \\
      & 22.2      & 3.74$\pm$0.14 & 2011.26$\pm$0.05 & 0.21$\pm$0.07 & 0.39$\pm$0.05 & \\ \hline 
35    & 36.8      & 4.36$\pm$0.22 & 2011.29$\pm$0.01 & 0.31$\pm$0.11 & 0.00 & 0.32\\
      & 22.2      & 3.31$\pm$0.50 & 2011.46$\pm$0.02 & 0.09$\pm$0.03 & 0.17$\pm$0.02 & \\
      & 14.5      & 3.12$\pm$0.22 & 2011.46$\pm$0.05 & 0.13$\pm$0.04 & 0.17$\pm$0.05 & \\\hline 
36    & 36.8      & 7.90$\pm$0.32 & 2011.98$\pm$0.04 & 0.36$\pm$0.15 & 0.00 & \\
      & 22.2      & 7.36$\pm$0.24 & 2012.16$\pm$0.05 & 0.21$\pm$0.03 & 0.18$\pm$0.10 & \\
      & 14.5      & 6.37$\pm$0.26 & 2012.27$\pm$0.01 & 0.17$\pm$0.08 & 0.29$\pm$0.04 & \\
      & 8.0       & 5.19$\pm$0.38 & 2011.99$\pm$0.06 & 0.46$\pm$0.18 & 0.01$\pm$0.07 & \\
      & 4.8       & 3.94$\pm$0.18 & 2012.15$\pm$0.15 & 0.36$\pm$0.09 & 0.17$\pm$0.15 & \\ \hline 
37    & 36.8      & 9.50$\pm$0.04 & 2013.19$\pm$0.14 & 0.65$\pm$0.03 & 0.00 & \\
      & 22.2      & 9.26$\pm$0.54 & 2013.19$\pm$0.13 & 0.62$\pm$0.09 & 0.00$\pm$0.09 & \\\hline 
38    & 36.8      & 6.77$\pm$0.10 & 2014.33$\pm$0.12 & 0.10$\pm$0.03 & 0.00 & \\
      & 22.2      & 6.48$\pm$0.16 & 2014.33$\pm$0.03 & 0.10$\pm$0.11 & 0.00$\pm$0.11 & \\\hline
\end{tabular}}
\label{BLLAC2}
\end{table}

\begin{figure*}
{\includegraphics[scale=0.33]{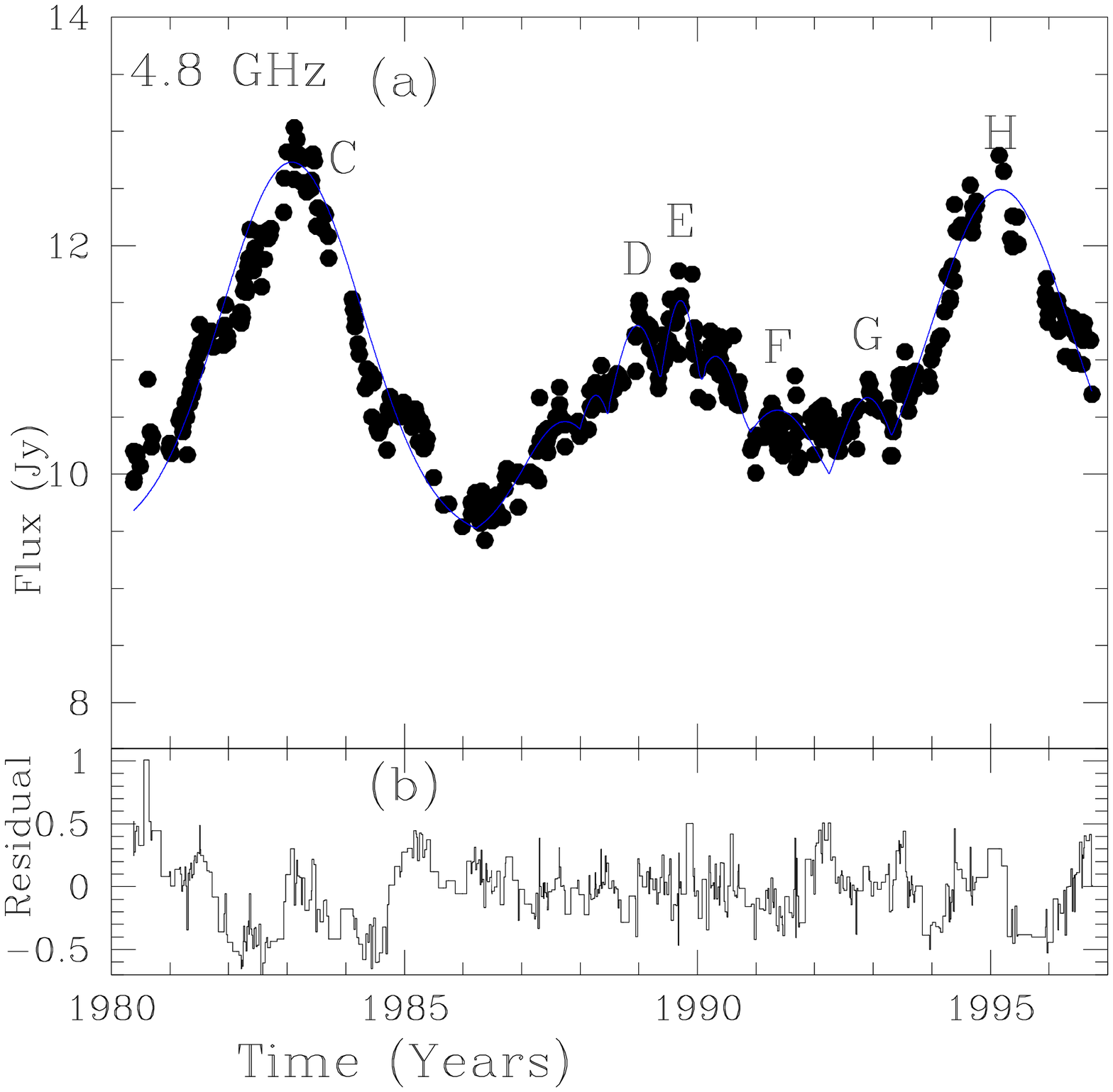}}
\hspace*{0.3in}{\includegraphics[scale=0.33]{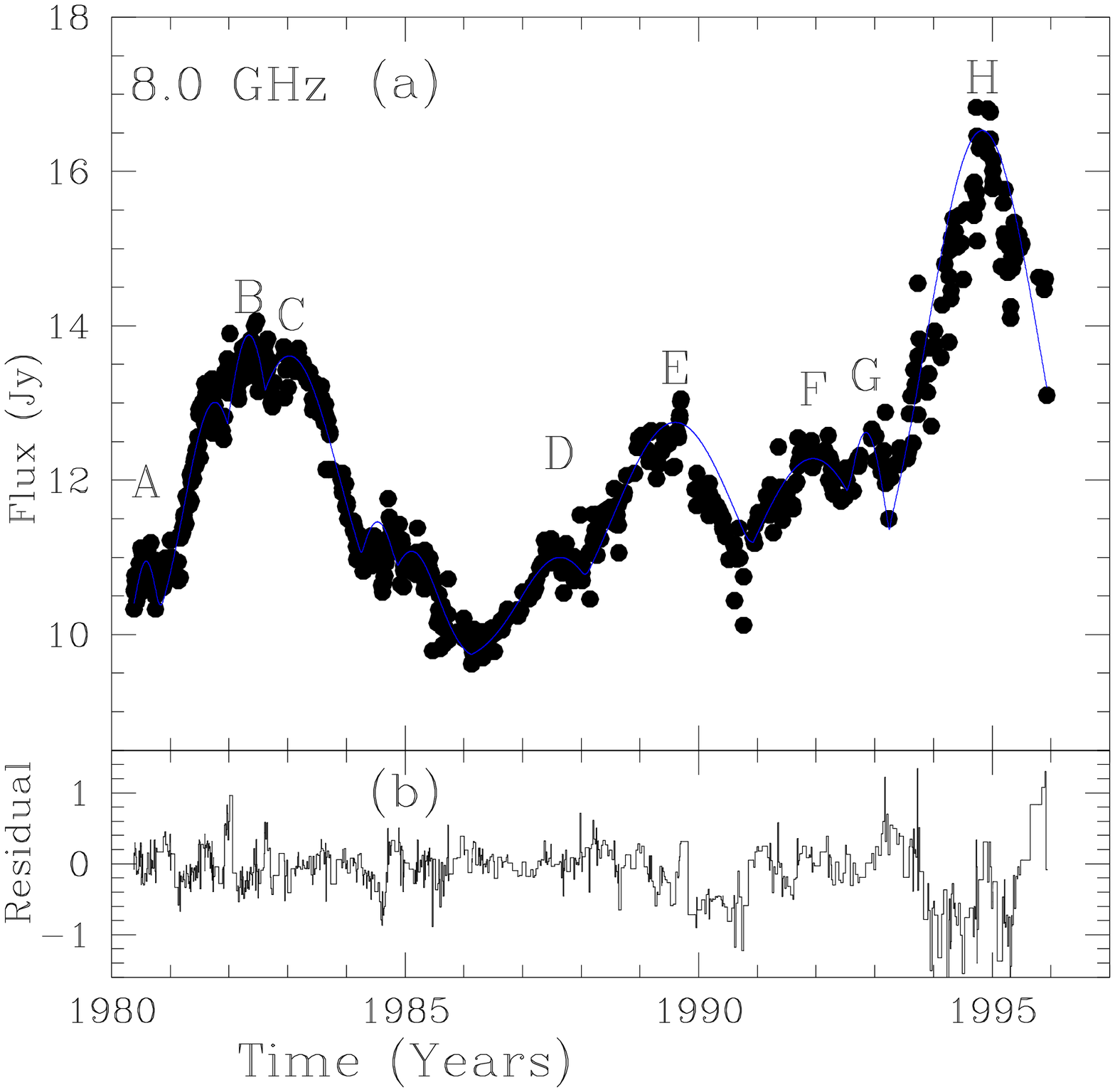}}
{\includegraphics[scale=0.33]{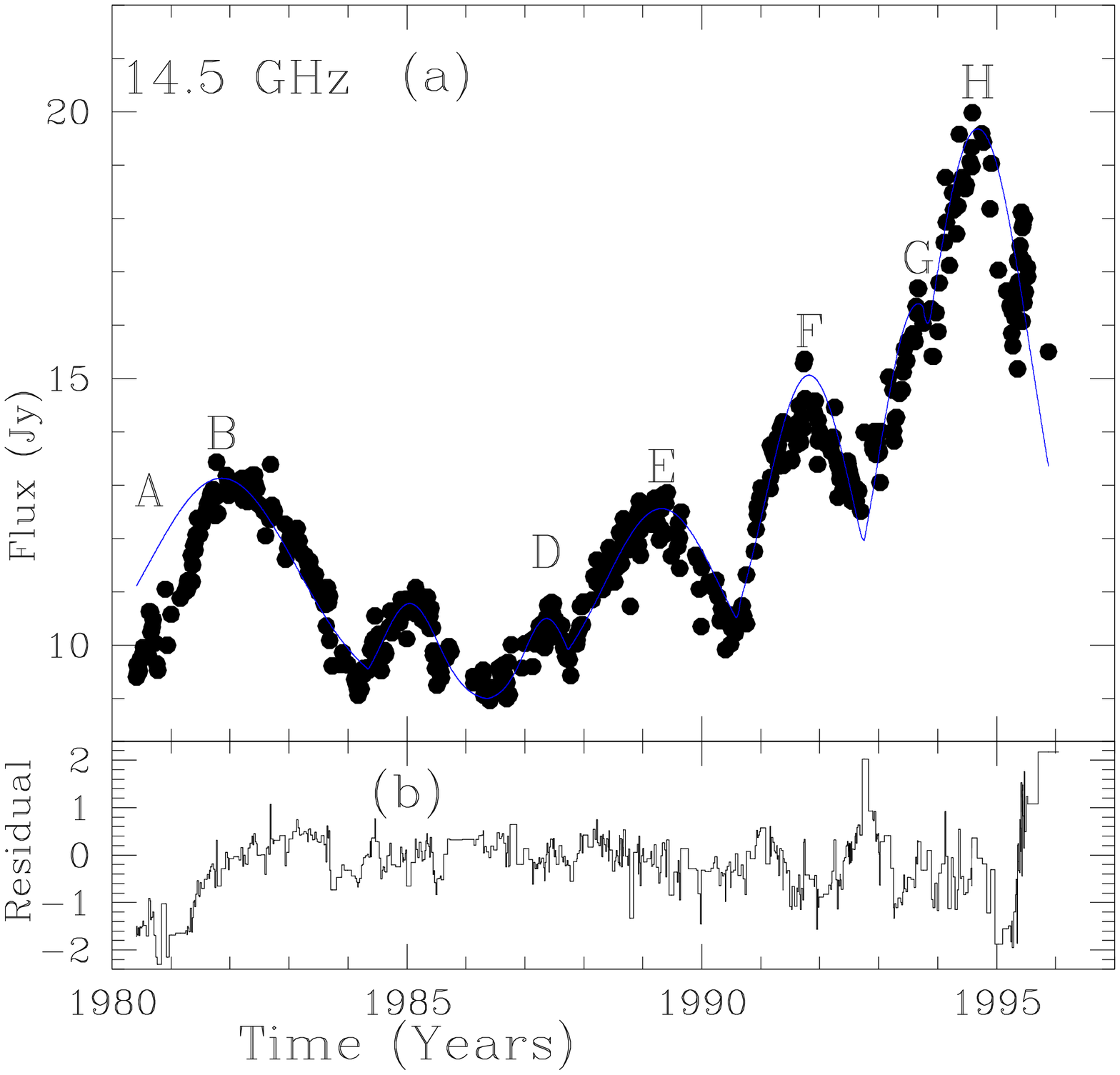}}
\hspace*{0.3in}{\includegraphics[scale=0.33]{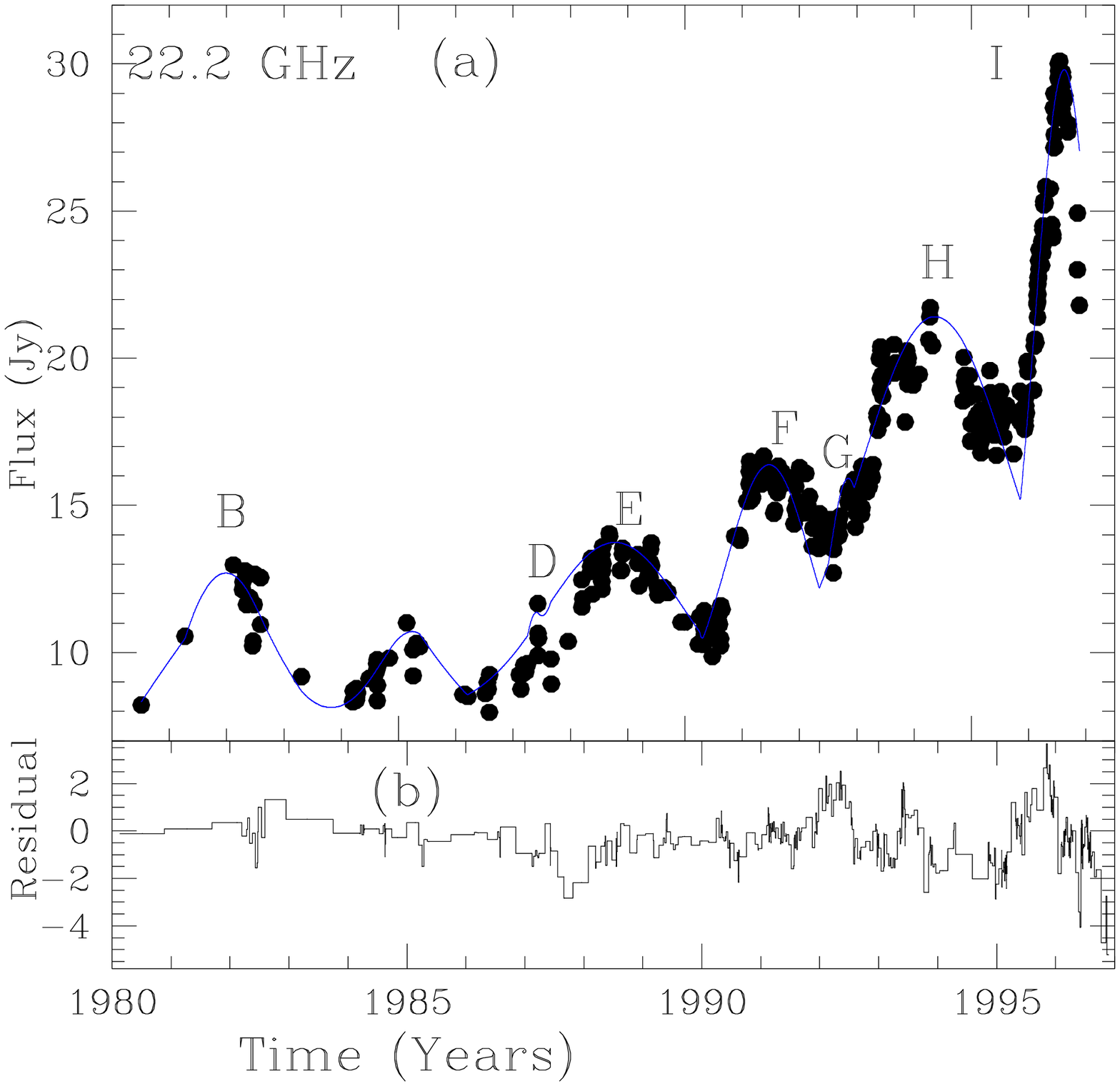}}
{\includegraphics[scale=0.33]{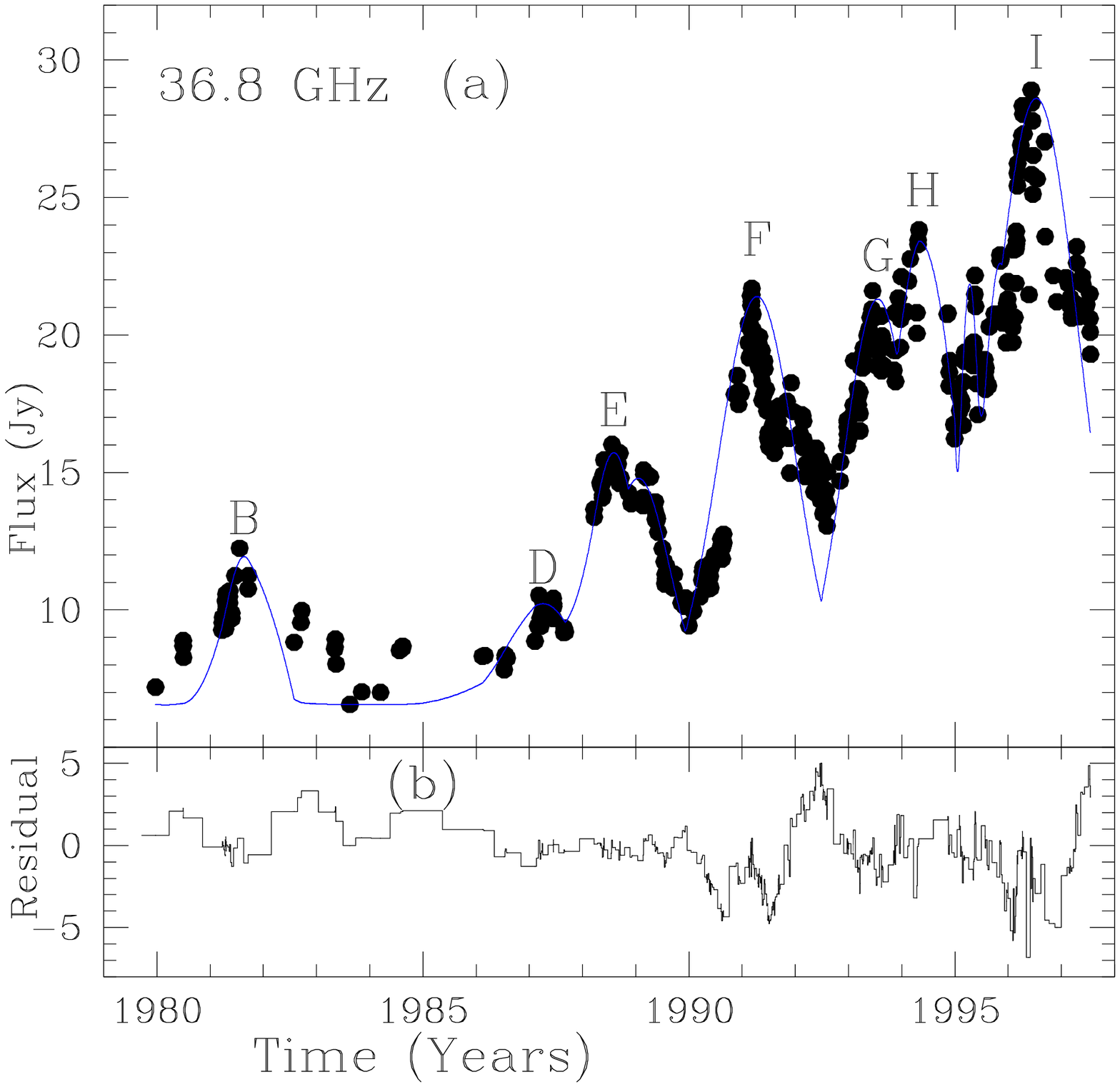}}
\caption{Segment 1 (beginning of observations to $\sim$ 1998.0) of 3C 279 at 4.8 GHz, 8.0 GHz, 14.5 GHz, 22.2 GHz and 36.8 GHz light curve
flares fit with a piecewise Gaussian function. The residual in the lower panels are calculated
as $[y(A_i,\overline{m}_i,\sigma_i)-x_i(t_i)]/{\rm Standard ~Deviation(x_i (t_i))}$.}
\end{figure*}

\begin{figure*}
{\includegraphics[scale=0.33]{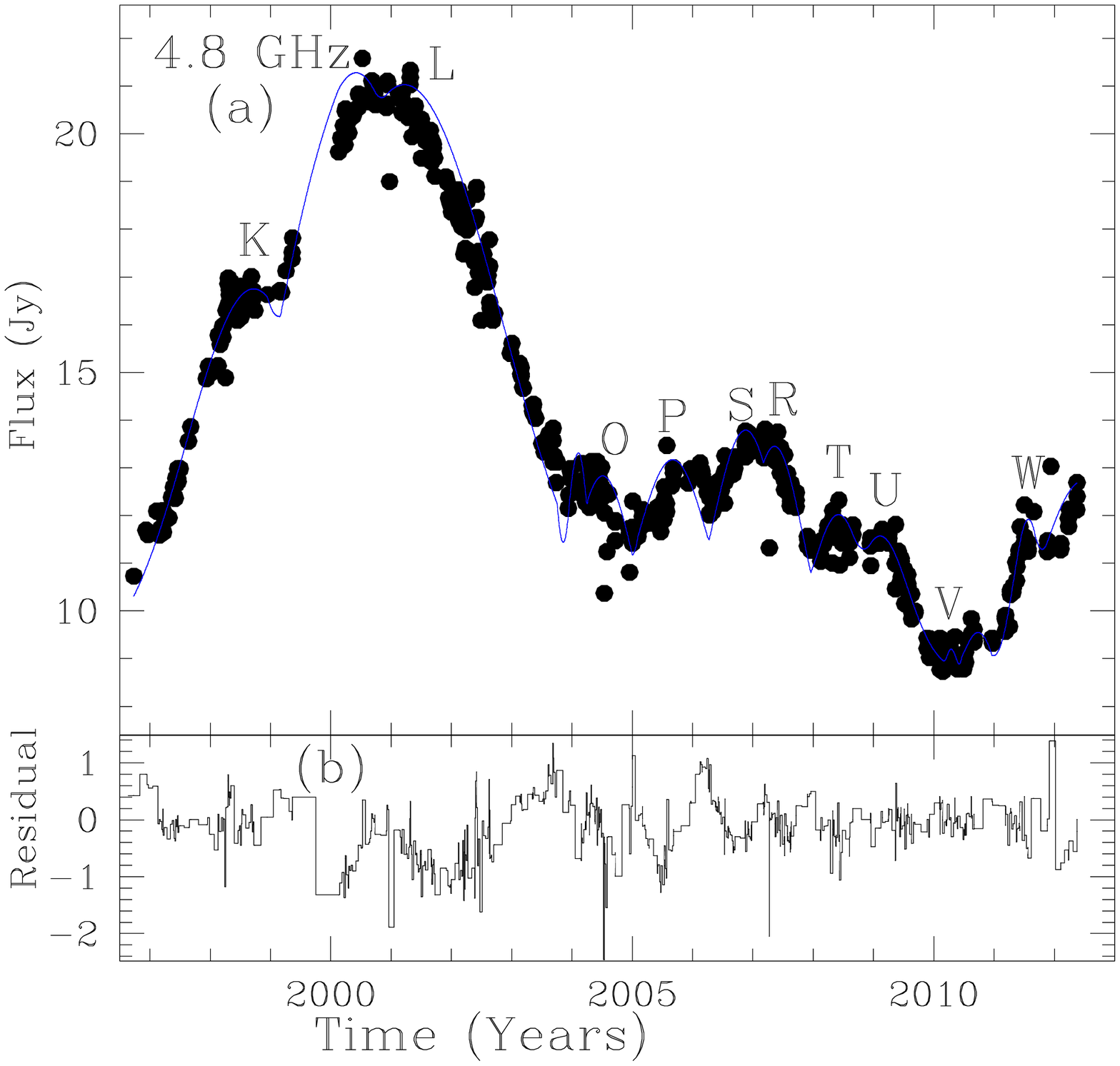}}
\hspace*{0.3in}{\includegraphics[scale=0.33]{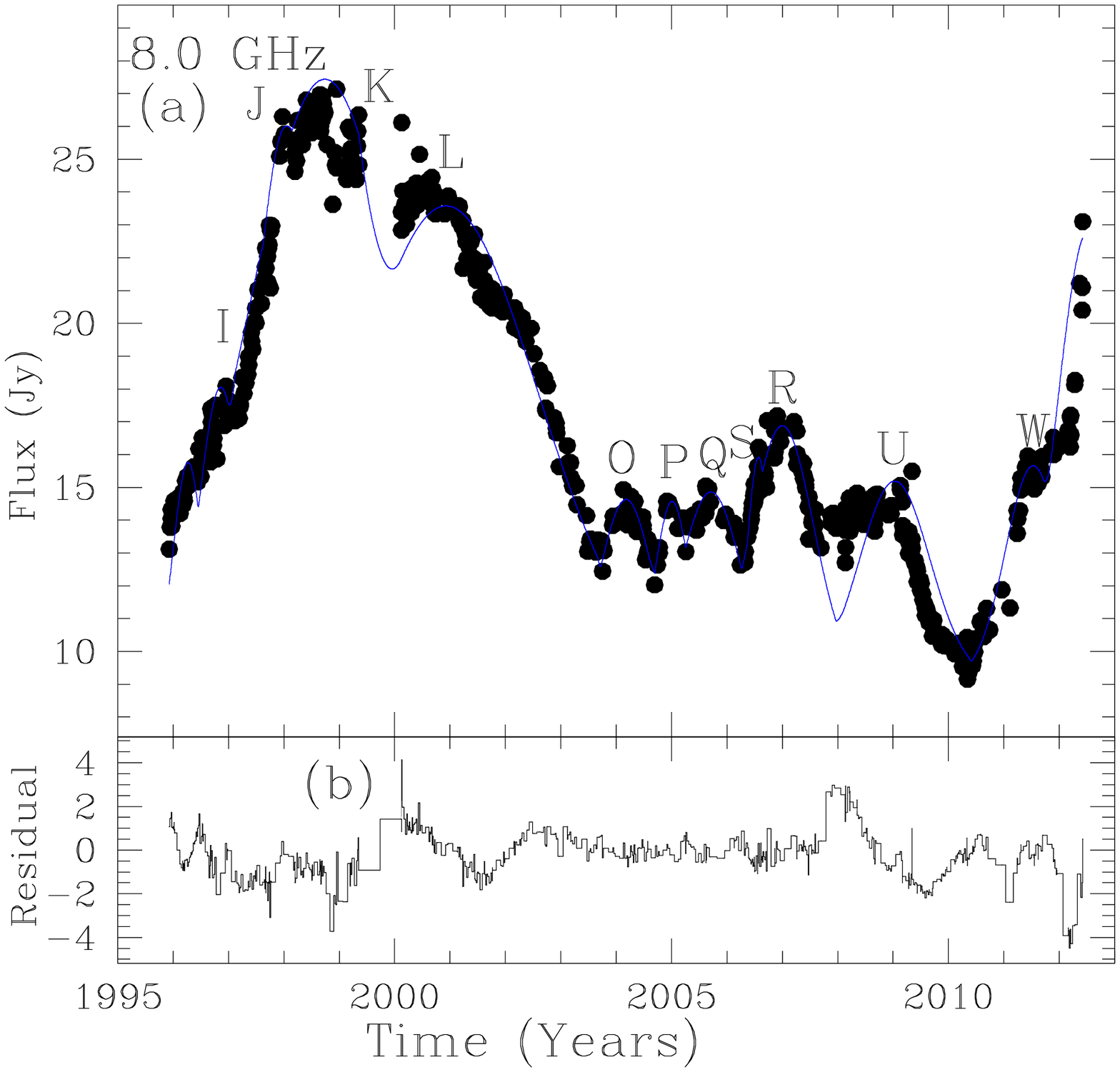}}
{\includegraphics[scale=0.33]{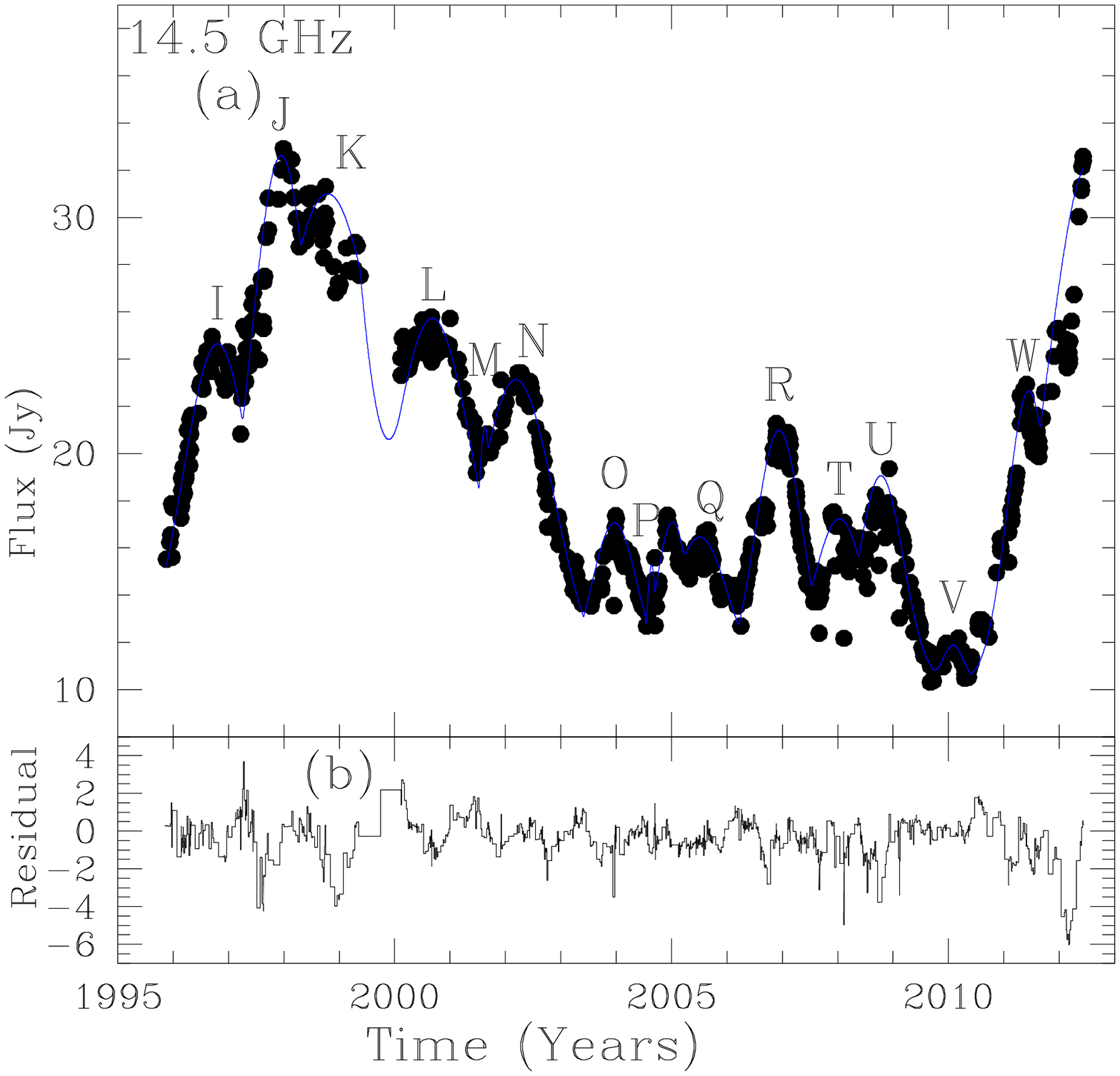}}
\hspace*{0.3in}{\includegraphics[scale=0.33]{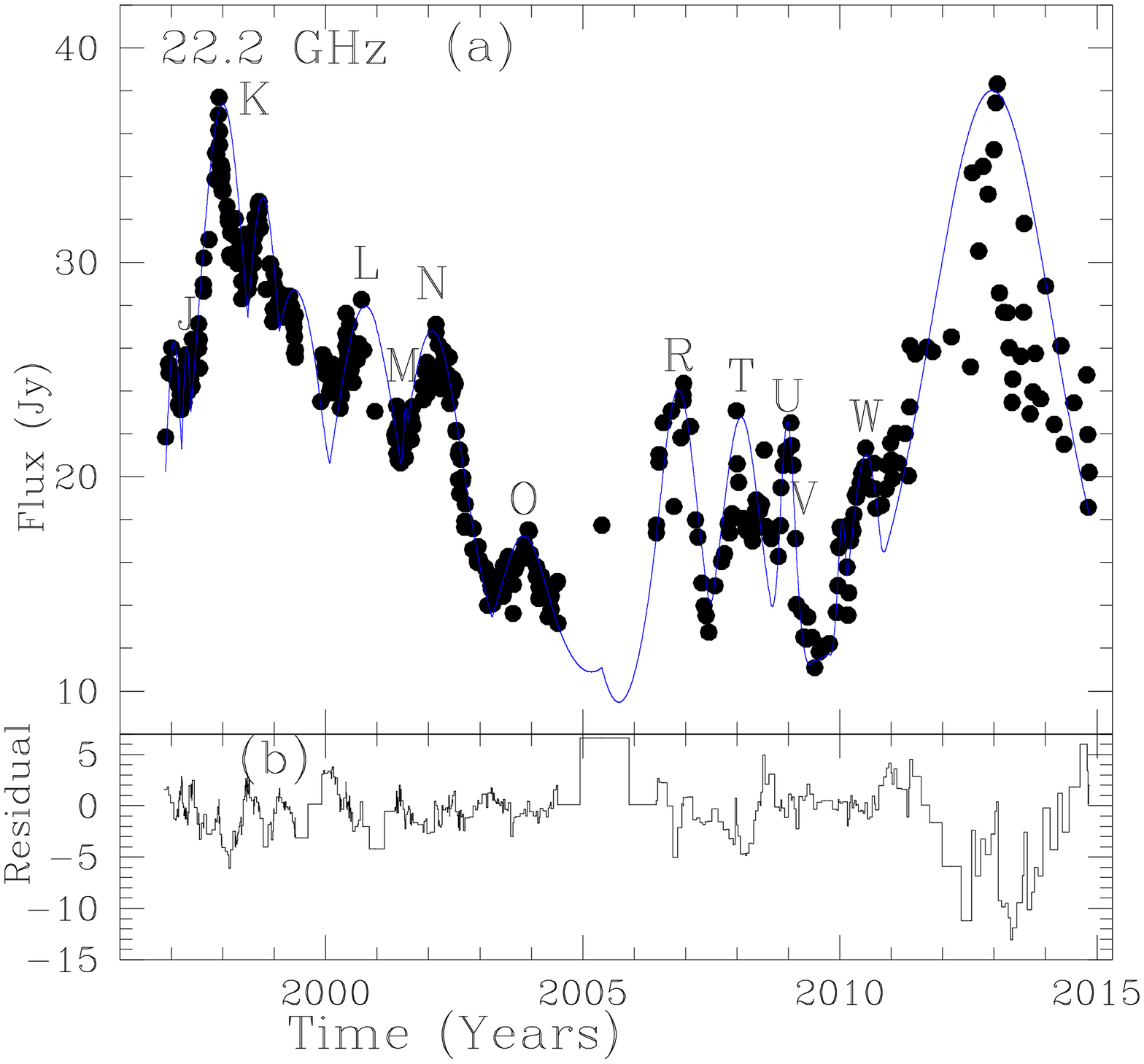}}
{\includegraphics[scale=0.33]{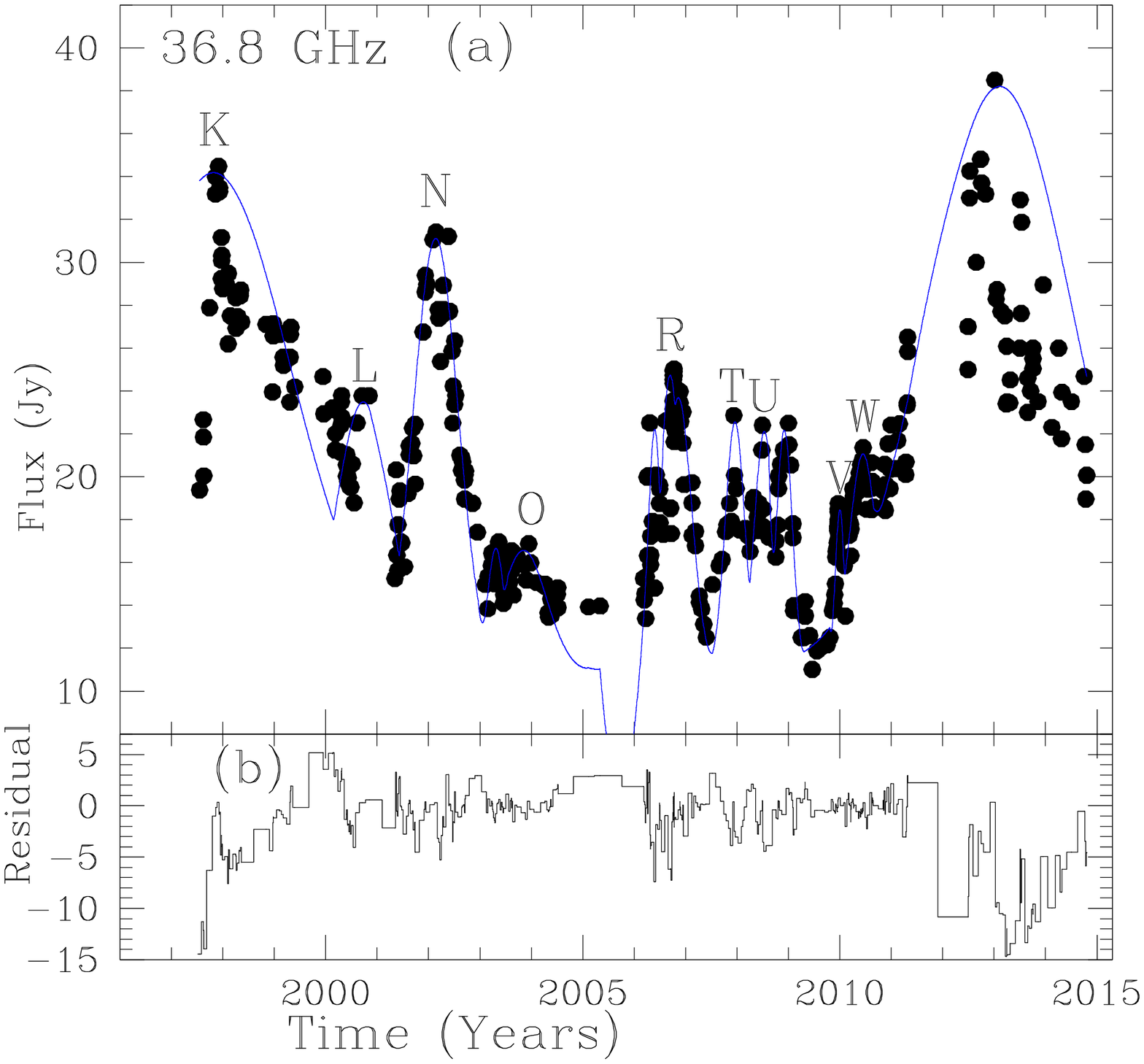}}
\caption{Segment 2 (1998.0 to end of observations) of 3C 279 at 4.8 GHz, 8.0 GHz, 14.5 GHz, 22.2 GHz and 36.8 GHz light curve flares fit 
with a piecewise Gaussian
function. The residual in the lower panels are calculated as $[y(A_i,\overline{m}_i,\sigma_i)-x_i(t_i)]/{\rm Standard ~Deviation(x_i (t_i))}$.}
\end{figure*}

\begin{figure*}
{\includegraphics[scale=0.33]{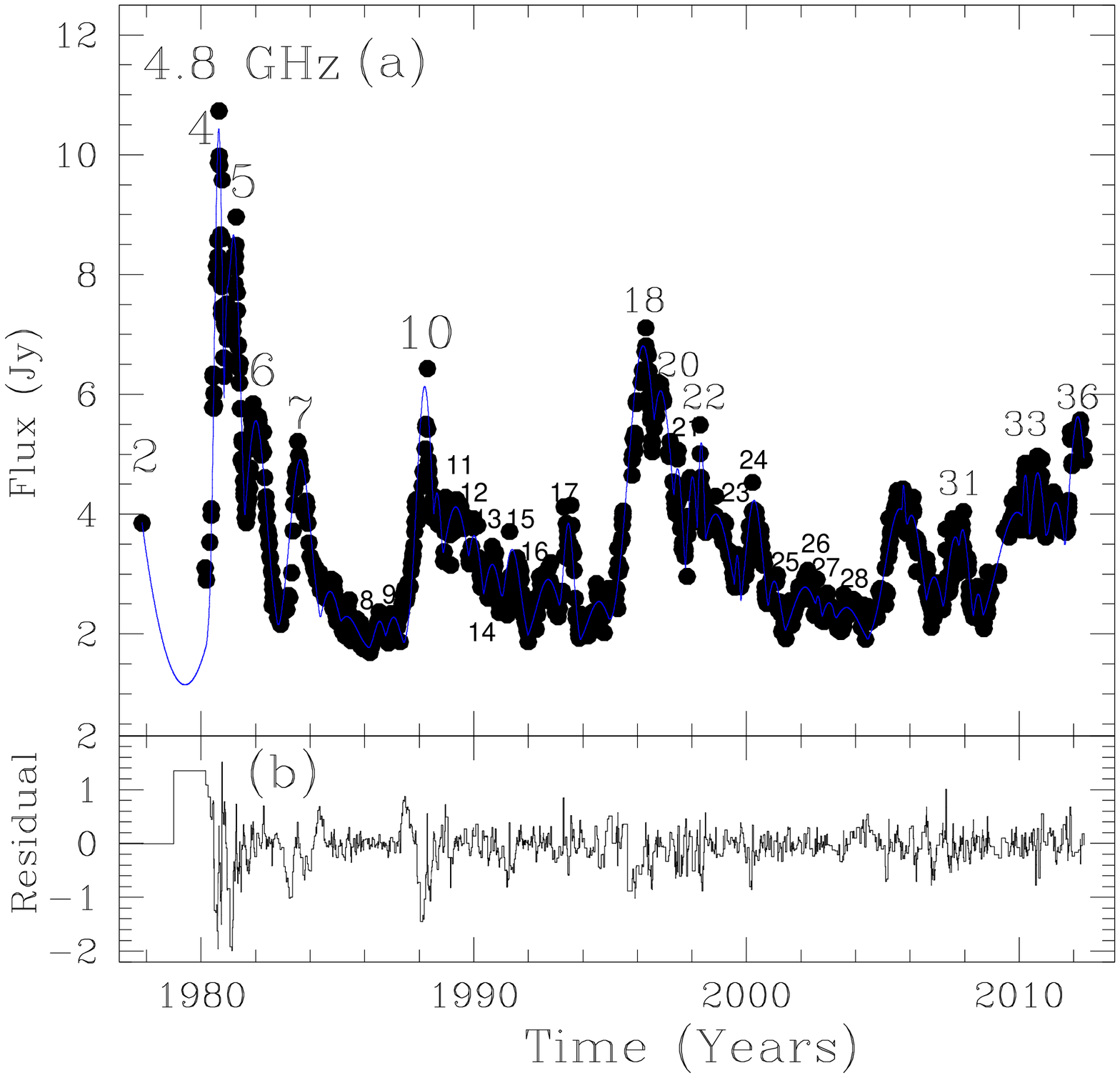}}
\hspace*{0.3in}{\includegraphics[scale=0.33]{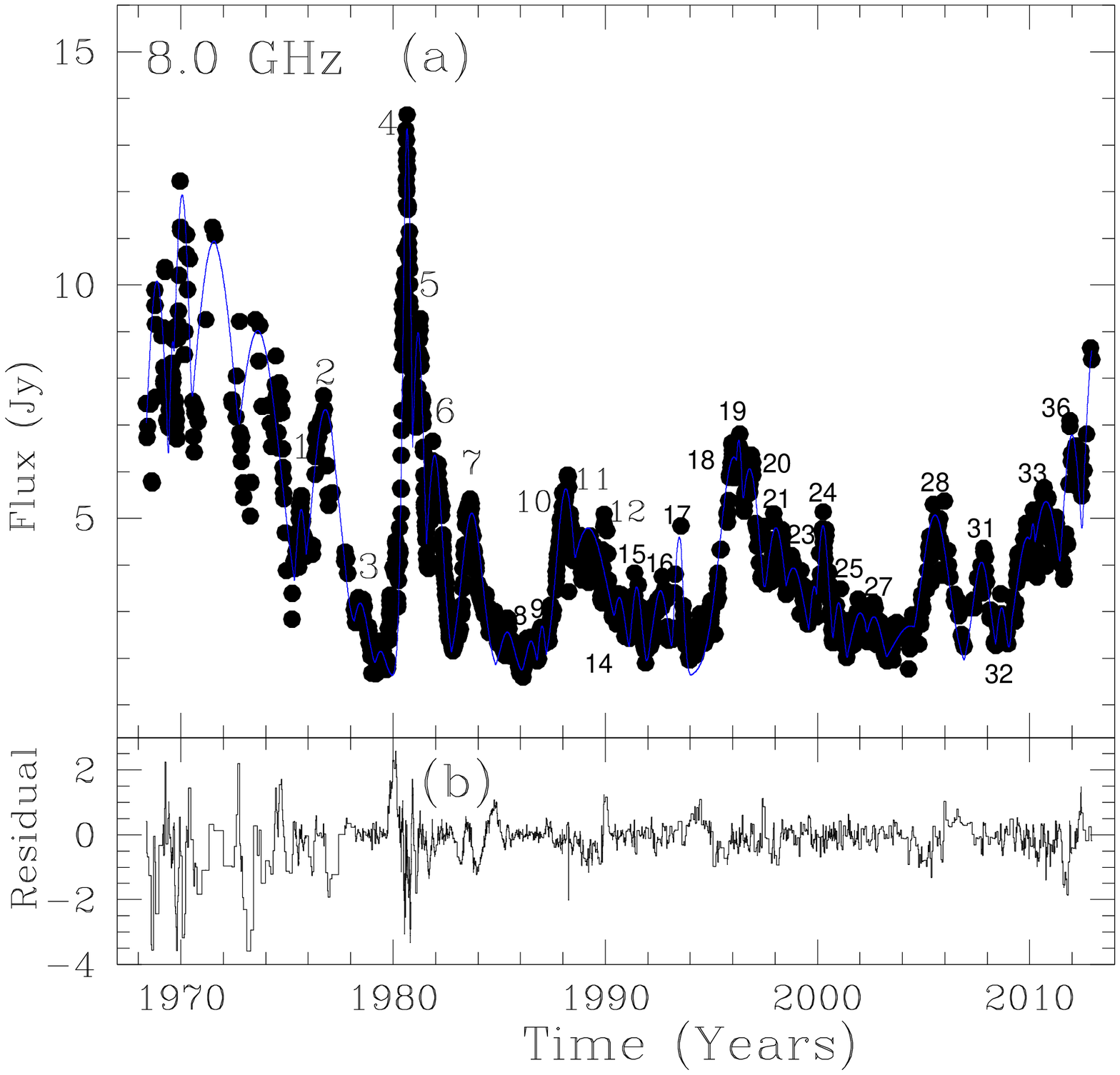}}
{\includegraphics[scale=0.33]{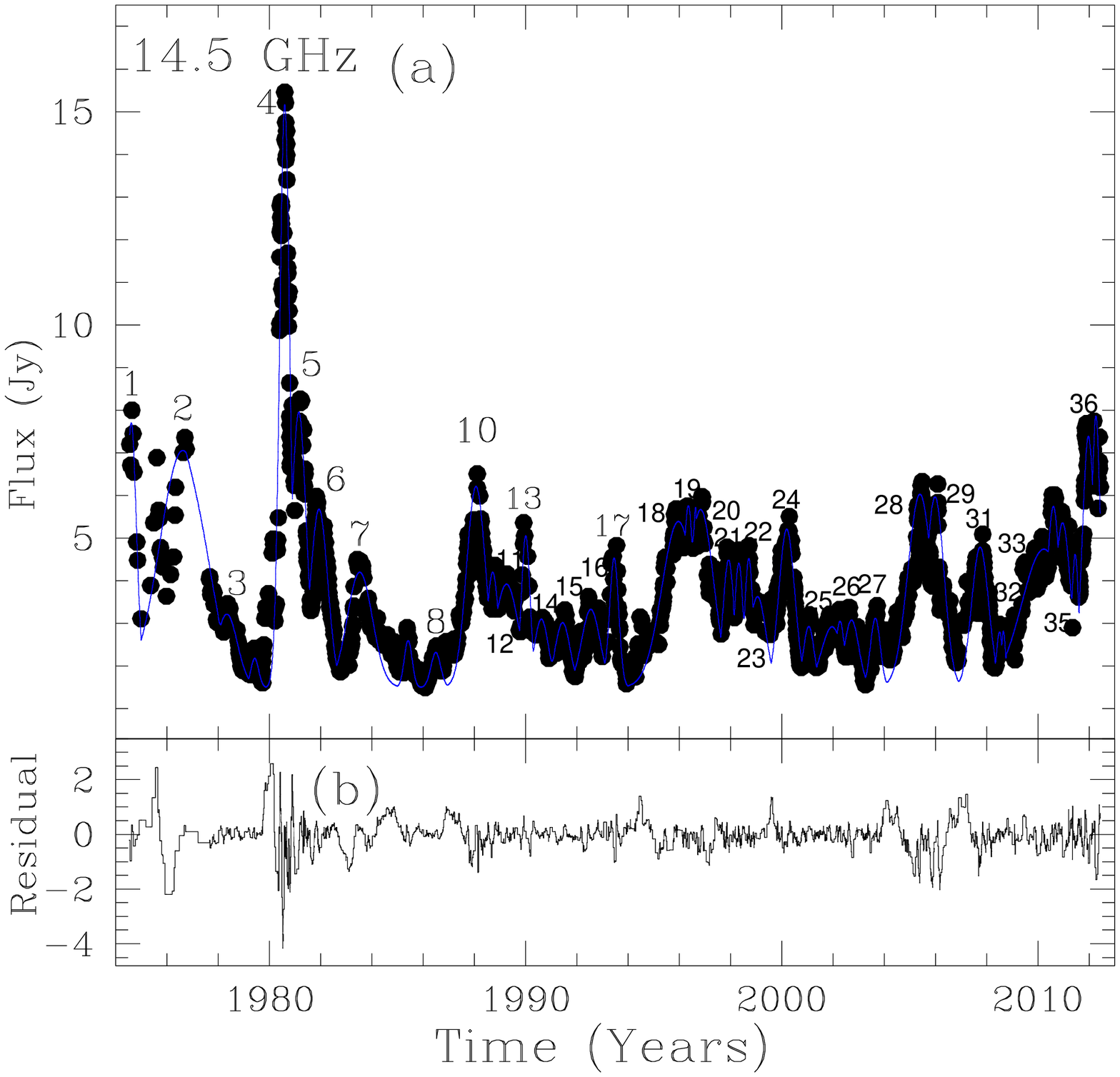}}
\hspace*{0.3in}{\includegraphics[scale=0.33]{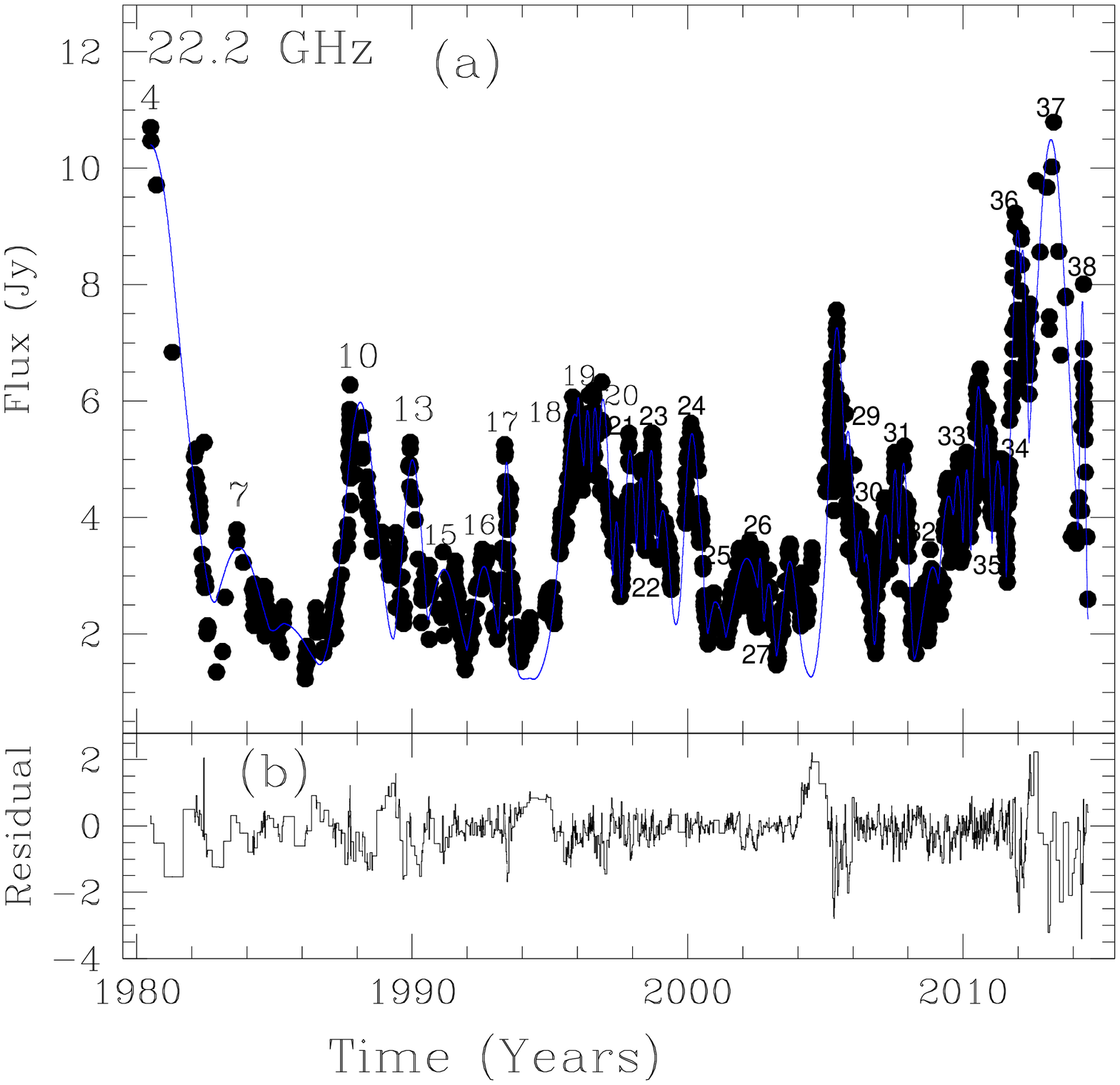}}
{\includegraphics[scale=0.33]{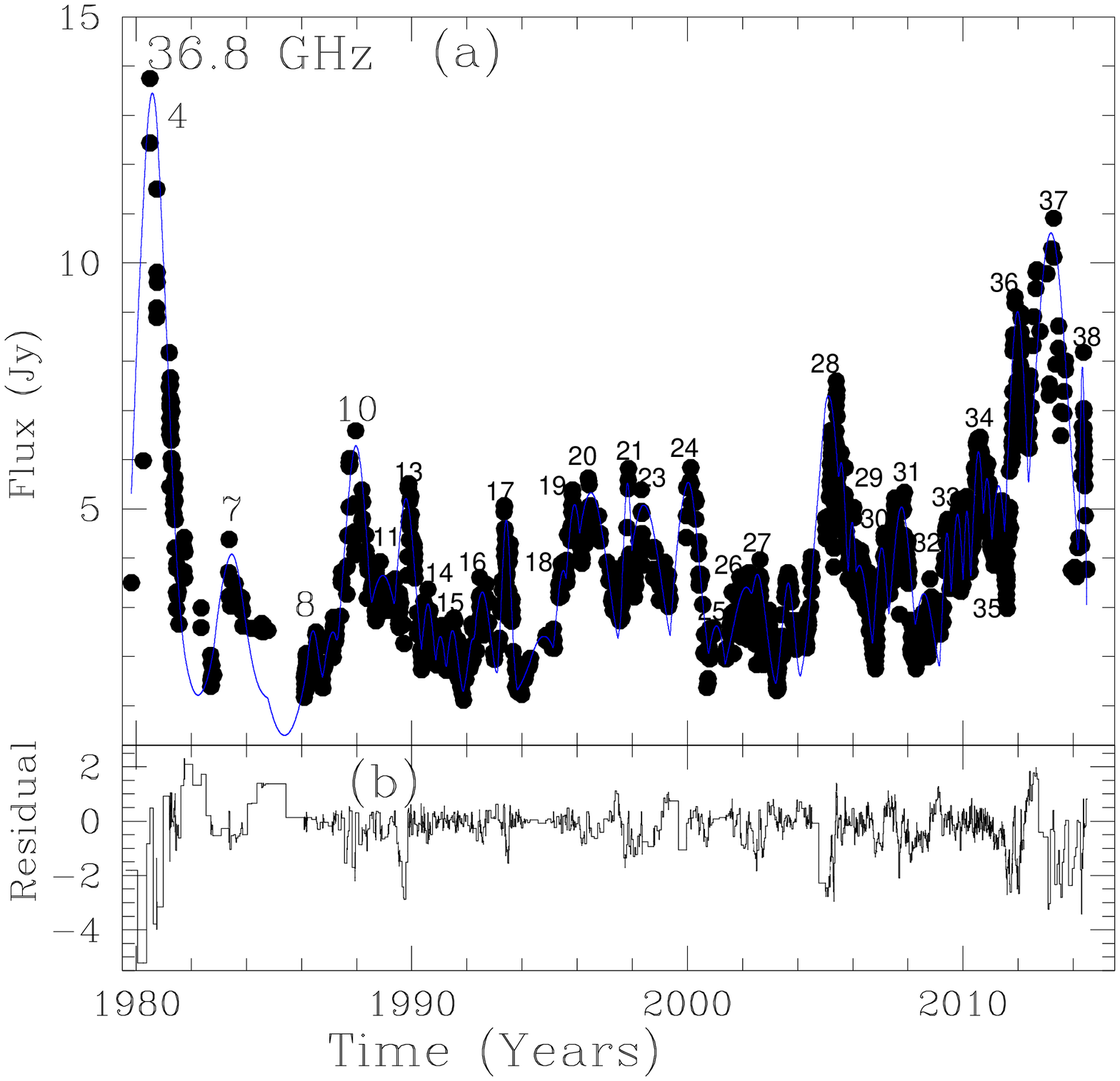}}
\caption{4.8 GHz, 8.0 GHz, 14.5 GHz, 22.2 GHz and 36.8 GHz light curve flares fit with a piecewise Gaussian
function for BL Lacertae. The residual in the lower panels are calculated as $[y(A_i,\overline{m}_i,\sigma_i)-x_i(t_i)]/{\rm Standard ~Deviation(x_i (t_i))}$.}

\end{figure*}

\section{Results and discussion}

\subsection{Core shift effect}

The highest frequency at which the flare occurred has been used as the reference frequency for measuring time lags ($\Delta t$) and are plotted
against observation frequency in Figure 8 for S5 0716+714, in Figs. 9 and 10 for 3C 279 and Figs. 11 and 12 for BL Lacertae. They generally indicate a
power law and were thus fitted with a function $\Delta t = a \nu^{-1/k_r}+b$ using a weighted non-linear least squares method to calculate $a$, $b$,
and $k_{r}$. The median of each parameter and the associated median deviation are 

\begin{enumerate}
\item For S5 0716+714\\
b = $-$0.03 $\pm$ 0.02 ~ yr\\
a = 1.01 $\pm$ 0.08 ~ yr~(GHz$)^{-1/k_r}$\\
k$_r$ = 1.03 $\pm$ 0.09 .\\

\item First segment of 3C 279: beginning of observation -- 1997.0 \\
b = $-$0.30 $\pm$ 0.04 ~ yr\\
a = 340.2 $\pm$ 48.3 ~ yr~(GHz)$^{-1/k_r}$\\
k$_r$ = 1.84 $\pm$ 0.13 ;\\

\item Second segment of 3C 279: 1998.0 -- end of observation\\
b = $-$0.09 $\pm$ 0.01 ~ yr\\
a = 1.86 $\pm$ 0.28 ~ yr~(GHz$)^{-1/k_r}$\\
k$_r$ = 1.02 $\pm$ 0.07 .\\

\item For BL Lacertae \\
b = $-$0.13 $\pm$ 0.05 ~ yr\\
a = 1.52 $\pm$ 0.16 ~ yr~(GHz)$^{-1/k_r}$\\
k$_r$ = 1.06 $\pm$ 0.22 ;\\

\end{enumerate}

From the time lag vs. frequency information (Figs. 8 - 12), the inferred $k_{r}$ are consistent with the equipartition between the
magnetic and particle energy densities. The weighted mean $k_r$ in all sources $\sim$ 1 except for Segment 1 of 3C 279 where k$_r \sim 1.8$
indicating either a real effect or an anomaly in the data. Large time lags ($>$ 0.5 yr) dominate its 4.8 GHz light curve indicating
that opacity effects may not be linear with frequency, resulting in differences compared to expectations from a single shock
propagating downstream causing the flare. We assumed $k = 1$ in this case to remain consistent with the results for further
parameter estimates. From the above fits, we have information on the amplitude and hence the peak flux for each flare at the observed frequencies.

As $r_{core}$ is dependent on $k_{r}$, the inferred values can help in understanding the jet geometry and also provides information on the distribution
of magnetic and electron number densities along the jet. 
As $S_\nu \propto \nu^{\alpha}$, the amplitude versus frequency data is fit with a function $A = N~ \nu^{\alpha}$ to determine the spectral index
$\alpha$ for each flare. We find $\alpha$in the range $-$16.45 and 2.41 for S5 0716+714. These are positive in 20 out of 24 cases, and negative for
the remaining implying that amplitude increase with frequency is the dominant trend in S5 0716+714 as given in Table 1. The
$\alpha$ for 3C 279 Segment 1 are in the range 0.02 to 1.08 and are given in the last column of Table 2 while for Segment 2, $\alpha$
are in the range 0.00 to 1.04 and are given in Column 7 of Table 3. The $\alpha$ for 3C 279 are positive in all cases with a trend of
increase in amplitude with the observing frequency. For BL Lacertae, we found $\alpha$ in the range of $-$16.67 and 1.20 as presented in Column 7 of
Table 4 and the amplitude was found to increase with observing frequency, with positive $\alpha$ for 22 out of 35 cases
while the opposite trend with negative $\alpha$ was found in the remaining cases.

Using the time lags derived from the Gaussian fit procedure, we calculated the core position offset for each flare as it is further used to
calculate the deprojected distance of the VLBI core from the base of the jet at a given frequency ($r_{core}$). Using above two quantities, along
with Equations 6 to 9, we then obtained magnetic field strength at 1 pc from the base of the jet and also weighted mean value of the magnetic field
in the core. All above parameters are given in Table 5 for S5 0716+714, Tables 6 and 7 for Segment 1 and 2 of 3C 279, respectively and in Table
8 for BL Lacertae. The columns of Tables 5$-$8 give the (1) flare nomenclature, (2) observing frequency ($\nu$), (3) core position offset
(mas), (4) core position offset in pcGHz, (5) distance of the emitting core from the base of the jet, (6) magnetic field strength at 1 pc distance
(in G), (7) equipartition magnetic field strength in mG at the emitting region.

Using the frequency dependent time delays obtainable from the core shift, associated jet parameters can be estimated independent of full VLBI
observations.
Proper motions of the jet components can be inferred from VLBI observations (e.g. Lister et al.\ 2013). Pushkarev et al. (2009) measured proper
motions for S5 0716+714 jet components in range 0.6 mas yr$^{-1}$ to 2.3 mas yr$^{-1}$ from which we adopted a median value of 1.4 mas yr$^{-1}$.
For 3C 279, they found proper motions in the range 0.005 mas yr$^{-1}$ to 0.651 mas yr$^{-1}$ between 1995 to 2010 from which we adopted a median
value of $\mu = 0.35$ mas yr$^{-1}$. Similarly, for BL Lacertae we adopted a median proper motion of 1.08 mas yr$^{-1}$.

In addition, we adopted $\theta = 5^{\circ}.2$, $\phi = 1^{\circ}.6$, $\delta = 10.9$, $z = 0.30~$ and $D_L = 1.60$ Mpc for
S5 0716+714; $\theta = 2^{\circ}.4$, $\phi = 0^{\circ}.4$, $\delta = 24.1$, $z = 0.536~$ and $D_L = 3.08$ Gpc for 3C 279;
and $\theta = 7^{\circ}.7$, $\phi = 1^{\circ}.9$, $\delta = 7.2$, $z = 0.069~$ and $D = 3.07$ Gpc for BL Lacertae (Jorstad et al.\ 2010;
Pushkarev et al.\ 2009; Algaba, Gabuzda, \& Smith 2012). The core position offset ($\Omega_{r\nu}$) for S5 0716+714 is $19.81 \pm 10.49$ pcGHz;
for Segment 1 of 3C 279 is $6.69 \pm 2.77$ pcGHz and for Segment 2 is $8.24 \pm 2.93$ pcGHz; for BL Lacertae,
\citet{1990ApJ...352...81M} measured an $\Omega_{r\nu}$ of 3.7 pcGHz based on 5.0 and 10.6 GHz observations while
\citet{2009MNRAS.400...26O} measured an $\Omega_{r\nu}$ of 3.4 $\pm$ 0.3 pcGHz. Our estimate of $\Omega_{r\nu} = 4.14 \pm 2.58$ pcGHz is
consistent with these studies.

It is expected for $r_{core}$ to decrease with increasing frequency as evident from Equation (\ref{rc}). For 4 out of 24 flares (G, J, O, R) of
S5 0716+714, above trend was inferred with removal of 14.5 GHz point from the flare R while the opposite trend was dominant in 3 out of
the remaining (M, Q, S) after removal of 4.8 GHz point from flare M and 8.0 GHz data point from flare S. But no conclusion could be drawn
for remaining 17 cases (A, B, C, D, E, F, H, I, K, L, N, P, T, U, V, W, X) as number of data points are at most just two. In case of 3C 279,
the expected trend was found in 8 out of 23 (D, E, F, G, L, O, U, V) flares after removing 1 of the data point in some of these such as:
22.2 GHz from flares G, L and U while 14.5 GHz from flare D and F whereas 4.8 GHz from flare E. The opposite trend was also found in
4 out of remaining 15 (B, K, R, W) with removal of 22.2 GHz point from K and W flares while 4.8 GHz from flare R. No conclusion could be drawn
for remaining 11 flares, namely A, C, H, I, J, M, N, P, Q, S, T, due to one or two data points. Similarly for BL Lacertae, $r_{core}$ was found
to decrease with increasing frequency for 6 out of 36 flares (10, 14, 16, 20, 25, 31) with removal of 22.2 GHz data point from flares
16, 20 and 31, while 14.5 GHz from flare 25. On the other hand,  $r_{core}$ increased with frequency in 9 of the remaining cases
(4, 7, 13, 15, 17, 23, 24, 27, 32) after ignoring 22.2 GHz point from flare R and 14.5 GHz data point from flares 4, 7, 17, 23, 24, and 27.
The remaining cases were not consistent with either of the above patterns and are thus inconclusive owing to lesser number of data points.
In the synchrotron opacity model, $r_{\rm core}$ is the opacity $\tau = 1$ surface where the core transitions from optically thick to
thin due to synchrotron self absorption. This transition is expected to occur at smaller $r_{\rm core}$ for increasing observation frequency,
closer to the true core distance in the context of a single shock propagating downstream causing the flaring core. Large time delays
($>$ 0.5 yr) dominating some lower frequency light curves (e.g. 4.8 GHz in 3C 279) result in non-uniformly distributed $r_{\rm core}$ 
estimates (uncertainty in the location of the $\tau = 1$ surface) thus indicating that opacity effects may not be linear with frequency, and
implying that the synchrotron opacity model may not be valid in all cases.

For S5 0716+714, 3C 279 (Segments 1 and 2) and BL Lacertae, the inferred $B$ are 0.22 $\pm$ 0.36 G, 0.0005 $\pm$ 0.0004 G, 0.35 $\pm$ 0.45 G and 0.02 $\pm$ 0.06 G respectively. 
We obtained a weighted mean core magnetic field strength $B_{core} = 15 \pm 5  mG$ for S5 0716+714, for 3C 279 $B_{core} = 0.014 \pm 0.03 mG$
(Segment 1), $B_{core} = 17 \pm 6 mG$ (Segment 2) and $B_{core} = 8 \pm 2 mG$ for BL Lacertae. For 3C 279, our Segment 2 results are consistent
with \citet{2005ApJ...619...73H} who obtained B = 12 $\pm$ 8 mG.

The 4.8 - 36.8 GHz light curves of S5 0716+714, 3C 279 (segments 1 and 2), and BL Lacertae are analyzed to infer the PSD shape and the results are
summarized in Tables 10 - 12 respectively while are plotted in Figures 14 - 17. No statistically significant QPO is detected in any case. For the
light curves of S5 0716+714, the power law PSD shape is the better model with a mean slope of $-1.3 \pm 0.2$. For the segment 1 light curves of 3C
279, the power law PSD shape is the better model with a mean slope of $-2.2 \pm 0.4$. For the segment 2 light curves of 3C 279, the power law PSD
model is the better shape for 3/5 light curves (4.8 GHz, 22.2 GHz and 36.8 GHz) with a mean slope of $-2.3 \pm 0.4$, indicating that the PSD slope
has not changed much between segments 1 and 2. For the remaining segment 2 light curves (8 GHz, 14.5 GHz), either both models could describe the
PSD shape or both may not suitably describe them which then necessitates the use of other PSD models such as the generalized bending power law
\citep{2004MNRAS.348..783M}, the damped random walk \citep{2009ApJ...698..895K, 2010ApJ...721.1014M, 2014ApJ...786..143S} or the continuous auto
regressive moving average process \citep{2014ApJ...788...33K}. For the light curves of BL Lacertae, the power law PSD shape is the better model with a
mean slope of $-1.8 \pm 0.3$.
The PSD from AGN light curves are typically power law shaped with higher power density at low frequencies due to long term,
higher amplitude variations. If smaller observation frequencies (e.g. 4.8 GHz) are associated with variability mechanisms
operational downstream in the jet, the slope is expected to steepen with decreasing observation frequency. We infer similar
slopes for a given blazar at all observation frequencies thus indicating no such regular trend. The flares can then arise
from multiple propagating shocks originating possibly in distinct events of injecting relativistic electrons into the jet with
similar energies. As the inferred $k_r \sim 1$ indicates equipartition, a strong magnetic field energy density can also participate
in the origin of flares through magnetic instabilities \cite[e.g.][]{2011MmSAI..82...95K} including reconnections or kink
instabilities \cite[e.g.][]{2003NewAR..47..513L,2016ASSL..427..473U} and is supported by simulations which suggest that
power law electron distributions can be produced with efficient acceleration \cite[e.g.][]{2015MNRAS.450..183S}.

\subsection{Tests of the MAD model}

We apply the formalism developed earlier to test the MAD scenario for S5 0716+714, 3C 279 and BL Lacertae. The $B~h$ in Gpc values are obtained from the
best fit slopes using the data of $B_{\rm core}(r_{\rm core})$ including all the flares for each source. 

The data in the log-log space is fit with the model $\log B_{\rm core} = m \log r_{\rm core} + \log c$ resulting in
\begin{enumerate}
\item BL Lacertae: ($m$, $c$) = (-0.094,  8.53 pc mG)
\item S5 0716+714: ($m$, $c$) = (-0.127,  31.44 pc mG)
\item 3C 279: ($m$, $c$) = (-0.194, 61.86 pc mG)
\end{enumerate}

The resulting $Bh$ at $r_{\rm core} = 1$ pc, $w (Bh)$ and $a$ are presented in Table 9.  
The eqn. (\ref{eqnaw}) predicts that the values of $0.296 < w < 1$ are disallowed since they yield spin values $a > 1$ (see Fig. \ref{aw}). 
We find $w$ outside the disallowed region $0.296 - 1$; the fact that $w$ values are of unity and reasonable values of spin are obtained is interesting and it implies that the magnetic arrest may be operating at some level in these systems. The higher values of spin correlate well with the luminosities which in 
addition is associated with the strong magnetic fields. Both these facts corroborate the MAD model 
with a black hole spin assisted relativistic jet (Blandford-Znajek mechanism) possibly operational in this class of objects.

\begin{table}
{\bf Table 5.}  Core position offsets, distance from jet base and magnetic field strengths inferred for the light curves of S5 0716+714.
\centering
\scalebox{.73}{
\begin{tabular}{|l|l|l|l|l|l|l|}\hline
Flare &  Frequency & $\Delta r$ & $\Omega_{r\nu }$ & $r_{core}$ & $B$ & $B_{core}$ \\
      & (GHz)      & (mas)     & (pc GHz$^{1/k_r}$)         & (pc)       & (G)       & (mG) \\ \hline  
A   & 4.8 & 658$\pm$ 238 & 13.3$\pm$ 5.02 & 32.1$\pm$ 12.91 & 0.22$\pm$ 0.31 & 7$\pm$ 10 \\
    & 8.0 & 168$\pm$ 252 & 6.21$\pm$ 9.36 & 9.14$\pm$ 13.87 & 0.13$\pm$ 0.22 & 14$\pm$ 32 \\ \hline
B  & 4.8 & 84$\pm$ 210 & 1.7$\pm$ 4.25 & 4.1$\pm$ 10.27 & 0.05$\pm$ 0.11 & 12$\pm$ 40 \\
 & 8.0 & 252$\pm$ 210 & 9.32$\pm$ 7.88 & 13.71$\pm$ 11.86 & 0.17$\pm$ 0.25 & 12$\pm$ 21 \\ \hline 
C  & 4.8 & 238$\pm$ 266 & 4.81$\pm$ 5.4 & 11.61$\pm$ 13.14 & 0.1$\pm$ 0.17 & 9$\pm$ 18 \\ \hline
D  & 4.8 & 154$\pm$ 140 & 3.11$\pm$ 2.85 & 7.51$\pm$ 6.96 & 0.08$\pm$ 0.11 & 10$\pm$ 18 \\
   & 14.5 & 42$\pm$ 126 & 3.59$\pm$ 10.79 & 2.97$\pm$ 8.95 & 0.08$\pm$ 0.22 & 28$\pm$ 112 \\\hline   
E  & 14.5 & 98 $\pm$ 28 & 8.37 $\pm$ 2.83 & 6.93 $\pm$ 2.85 & 0.16 $\pm$ 0.22 & 23 $\pm$ 33 \\\hline 
F  & 4.8 & 994 $\pm$ 84 & 20.09$\pm$ 2.77 & 48.49$\pm$ 9.48 & 0.3$\pm$ 0.41 & 6 $\pm$ 9 \\
   & 8.0 & 756$\pm$ 84 & 27.95$\pm$ 5.07 & 41.13$\pm$ 10.61 & 0.38$\pm$ 0.53 & 9$\pm$ 13 \\ \hline      
G  & 4.8 & 560$\pm$ 322 & 11.32$\pm$ 6.62 & 27.32$\pm$ 16.43 & 0.2$\pm$ 0.28 & 7$\pm$ 11 \\
   & 8.0 & 140$\pm$ 322 & 5.18$\pm$ 11.93 & 7.62$\pm$ 17.61 & 0.11$\pm$ 0.24 & 14$\pm$ 46 \\   
   & 14.5 & 98 $\pm$ 196 & 8.37 $\pm$ 16.81 & 6.93 $\pm$ 14.0 & 0.16 $\pm$ 0.32 & 23 $\pm$ 65 \\\hline 
H  & 14.5 & 546 $\pm$ 294 & 46.65 $\pm$ 26.48 & 38.59 $\pm$ 23.71 & 0.56$\pm$ 0.8 & 15 $\pm$ 23 \\
   & 22.2 & 322 $\pm$ 378. & 62.4$\pm$ 74.33 & 34.46$\pm$ 42.11 & 0.7$\pm$ 1.14 & 20 $\pm$ 41 \\ \hline      
I  & 4.8 & 112 $\pm$ 294 & 2.26$\pm$ 5.95 & 5.46$\pm$ 14.37 & 0.06$\pm$ 0.14 & 11 $\pm$ 39 \\   
   & 14.5 & 448 $\pm$ 266 & 38.28$\pm$ 23.75 & 31.66$\pm$ 21.01 & 0.49$\pm$ 0.7 & 15 $\pm$ 24 \\ \hline  
J  & 4.8 & 280 $\pm$ 224 & 5.66$\pm$ 4.57 & 13.66$\pm$ 11.19 & 0.12$\pm$ 0.17 & 9 $\pm$ 15 \\   
   & 8.0 & 140 $\pm$ 322  & 5.18$\pm$ 11.93 & 7.62$\pm$ 17.61 & 0.11$\pm$ 0.24 & 14 $\pm$ 46  \\    
   & 14.5 & 70 $\pm$ 294  & 5.98$\pm$ 25.14 & 4.95$\pm$ 20.83 & 0.12$\pm$ 0.42 & 25 $\pm$ 134  \\\hline   
K   & 8.0 & 238 $\pm$ 266  & 8.8$\pm$ 9.92 & 12.95$\pm$ 14.78 & 0.16$\pm$ 0.26 & 13 $\pm$ 25  \\\hline
L  & 8.0 & 182 $\pm$ 98  & 6.73$\pm$ 3.75 & 9.9$\pm$ 5.81 & 0.13$\pm$ 0.19 & 13 $\pm$ 21  \\
   & 14.5 & 28 $\pm$ 84  & 2.39$\pm$ 7.19 & 1.98$\pm$ 5.97 & 0.06$\pm$ 0.16 & 31 $\pm$ 125 \\
   & 22.2 & 28 $\pm$ 140  & 5.43$\pm$ 27.15 & 3 $\pm$ 15.02 & 0.11$\pm$ 0.45 & 38 $\pm$ 242  \\\hline 
M  & 4.8 & 980 $\pm$ 70  & 19.81$\pm$ 2.59 & 47.81$\pm$ 9.09 & 0.3$\pm$ 0.41 & 6 $\pm$ 9 \\
   & 8.0 & 70 $\pm$ 56 & 2.59 $\pm$ 2.1 & 3.81 $\pm$ 3.17 & 0.07$\pm$ 0.1 & 17 $\pm$ 29  \\   
   & 14.5 & 70 $\pm$ 84  & 5.98$\pm$ 7.26 & 4.95$\pm$ 6.11 & 0.12$\pm$ 0.2 & 25 $\pm$ 51  \\
   & 22.2 & 98 $\pm$ 224  & 18.99$\pm$ 43.58 & 10.49$\pm$ 24.23 & 0.29$\pm$ 0.63 & 28 $\pm$ 87  \\\hline   
N   & 8.0 & 154 $\pm$ 224 & 5.69$\pm$ 8.32 & 8.38$\pm$ 12.34 & 0.12$\pm$ 0.21 & 14 $\pm$ 32  \\\hline
O  & 8.0 & 546 $\pm$ 70 & 20.19$\pm$ 3.88 & 29.71$\pm$ 7.89 & 0.3$\pm$ 0.41 & 10 $\pm$ 14 \\
   & 14.5 & 392 $\pm$ 196 & 33.49$\pm$ 17.8 & 27.7$\pm$ 16.1 & 0.44$\pm$ 0.62 & 16 $\pm$ 24 \\
   & 22.2 & 168 $\pm$ 168 & 32.55$\pm$ 33.21 & 17.98$\pm$ 18.98 & 0.43$\pm$ 0.67 & 24 $\pm$ 45 \\\hline
P  & 22.2 & 350 $\pm$ 84  & 67.82$\pm$ 21.28 & 37.46$\pm$ 15.56 & 0.74$\pm$ 1.03 & 20 $\pm$ 29  \\\hline
Q   & 4.8 & 378 $\pm$ 210  & 7.64$\pm$ 4.33 & 18.44$\pm$ 10.75 & 0.15$\pm$ 0.21 & 8 $\pm$ 12  \\
   & 8.0  & 364 $\pm$ 196 & 13.46$\pm$ 7.5 & 19.8$\pm$ 11.62 & 0.22$\pm$ 0.32 & 11 $\pm$ 17 \\
   & 14.5 & 308 $\pm$ 70 & 26.32$\pm$ 7.62 & 21.77$\pm$ 8.13 & 0.37$\pm$ 0.51 & 17 $\pm$ 24 \\
   & 22.2 & 224 $\pm$ 56 & 43.41$\pm$ 13.96 & 23.97$\pm$ 10.1 & 0.53$\pm$ 0.74 & 22 $\pm$ 32 \\\hline
R  & 4.8 & 154 $\pm$ 168 & 3.11$\pm$ 3.41 & 7.51$\pm$ 8.3 & 0.08$\pm$ 0.12 & 10 $\pm$ 19 \\
   & 8.0 & 126 $\pm$ 112  & 4.66$\pm$ 4.19 & 6.86$\pm$ 6.3 & 0.1$\pm$ 0.15 & 15 $\pm$ 26  \\
   & 14.5 & 518 $\pm$ 112 & 44.26$\pm$ 12.44 & 36.61$\pm$ 13.43 & 0.54$\pm$ 0.75 & 15 $\pm$ 21 \\
   & 22.2 & 42 $\pm$ 112  & 8.14$\pm$ 21.77 & 4.49$\pm$ 12.08 & 0.15$\pm$ 0.37 & 34 $\pm$ 123  \\ \hline 
S  & 4.8 & 84 $\pm$ 140 & 1.7$\pm$ 2.84 & 4.1$\pm$ 6.87 & 0.05$\pm$ 0.09 & 12 $\pm$ 29 \\
   & 8.0 & 252 $\pm$ 280  & 9.32 $\pm$ 10.44 & 13.71$\pm$ 15.57 & 0.17$\pm$ 0.27 & 12 $\pm$ 24 \\
   & 14.5 & 84 $\pm$ 140  & 7.18$\pm$ 12.03 & 5.94$\pm$ 10.05 & 0.14$\pm$ 0.26 & 24 $\pm$ 59 \\
   & 22.2 & 112 $\pm$ 84  & 21.7$\pm$ 16.86 & 11.99$\pm$ 9.87 & 0.32$\pm$ 0.47 & 27 $\pm$ 45 \\\hline
T  & 4.8 & 28 $\pm$ 154  & 0.57$\pm$ 3.11 & 1.37$\pm$ 7.52 & 0.02$\pm$ 0.09 & 15 $\pm$ 108 \\
   & 8.0 & 252 $\pm$ 140  & 9.32$\pm$ 5.35 & 13.71$\pm$ 8.26 & 0.17$\pm$ 0.24 & 12 $\pm$ 19  \\
   & 14.5 & 98 $\pm$ 70 & 8.37$\pm$ 6.17 & 6.93$\pm$ 5.36 & 0.16$\pm$ 0.23 & 23 $\pm$ 38 \\\hline 
U  & 4.8 & 196 $\pm$ 238 & 3.96$\pm$ 4.83 & 9.56$\pm$ 11.73 & 0.09$\pm$ 0.15 & 9 $\pm$ 19  \\
   & 14.5 & 28 $\pm$ 14  & 2.39$\pm$ 1.27 & 1.98$\pm$ 1.15 & 0.06$\pm$ 0.09 & 31 $\pm$ 48 \\
   & 22.2 & 98$\pm$ 28 & 18.99$\pm$ 6.65 & 10.49$\pm$ 4.65 & 0.29$\pm$ 0.4 & 28$\pm$ 40 \\\hline
V  & 14.5 & 168 $\pm$ 196 & 14.35$\pm$ 16.94 & 11.87$\pm$ 14.29 & 0.23$\pm$ 0.38 & 20 $\pm$ 40 \\\hline
W  & 22.2 & 28 $\pm$ 238 & 5.43$\pm$ 46.13 & 3.0$\pm$ 25.49 & 0.11$\pm$ 0.74 & 38 $\pm$ 405 \\\hline
X  & 22.2 & 28 $\pm$ 196 & 5.43$\pm$ 38  & 3 $\pm$ 21 & 0.11$\pm$ 0.61 & 38 $\pm$ 335 \\\hline
\end{tabular}}
\label{tab4s55a}
\end{table}

\begin{table}
{\bf Table 6.} Core position offsets, distance from jet base and magnetic field strengths inferred for segment 1 light curves for 3C 279.
\centering
\scalebox{.67}{
\begin{tabular}{|l|l|l|l|l|l|l|l|}\hline
Flare &  Frequency & $\Delta r$ & $\Omega_{r\nu }$ & $r_{core}$ & $B$ & $B_{core}$ \\
      & (GHz)      & (mas)     & (pc GHz$^{1/k_r}$)         & (pc)       & (G)       & (mG) \\ \hline  
      
A     & 8.0 & 14$\pm$ 45.5 & 0.37$\pm$ 1.22 & 2.89$\pm$ 9.4 & 0.00006$\pm$ 0.0002 & 0.021$\pm$ 0.086 \\\hline
B     & 8.0 & 63 $\pm$ 42 & 1.69$\pm$ 1.12 & 13.01$\pm$ 8.74 & 0.00019$\pm$ 0.0002 & 0.015$\pm$ 0.015 \\
      & 14.5 & 77 $\pm$ 24.5 & 4.04$\pm$ 1.3 & 22.57$\pm$ 7.6 & 0.00037$\pm$ 0.0002 & 0.016$\pm$ 0.012 \\
      & 22.2 & 133$\pm$ 66.5 & 14.24$\pm$ 7.17 & 63.48$\pm$ 32.77 & 0.00098$\pm$ 0.0007 & 0.015$\pm 0.014$ \\\hline
C     & 4.8 & 175$\pm$ 21& 2.99$\pm$ 0.36 & 30.43$\pm$ 4.11 & 0.0003$\pm$ 0.0002 & 0.01$\pm$ 0.006 \\\hline
D     & 4.8 & 602 $\pm$ 49  & 10.27$\pm$ 0.86 & 104.67$\pm$ 10.73 & 0.00076$\pm$ 0.0005 & 0.007$\pm$ 0.005 \\
      & 8.0 & 133 $\pm$ 28 & 3.56$\pm$ 0.76 & 27.47$\pm$ 6.23 & 0.00034$\pm$ 0.0002 & 0.012$\pm$ 0.008 \\
      & 14.5 & 38.5$\pm$ 35  & 2.02$\pm$ 1.84 & 11.28$\pm$ 10.33 & 0.00022$\pm$ 0.0002 & 0.019$\pm$ 0.025 \\
      & 22.2 & 56 $\pm$ 38.5 & 6.0$\pm$ 4.14 & 26.73$\pm$ 18.69 & 0.0005$\pm$ 0.0004 & 0.019$\pm$ 0.02 \\\hline
E     & 4.8 & 392 $\pm$ 31.5 & 6.69$\pm$ 0.56 & 68.16$\pm$ 6.93 & 0.00055$\pm$ 0.0003 & 0.008$\pm$ 0.005 \\
      & 8.0 & 353.5$\pm$ 42. & 9.46$\pm$ 1.17 & 73.0$\pm$ 10.65 & 0.00071$\pm$ 0.0004 & 0.01$\pm$ 0.006 \\
      & 14.5 & 255.5$\pm$ 56. & 13.4$\pm$ 3. & 74.89$\pm$ 18.38 & 0.00093$\pm$ 0.0006 & 0.012$\pm$ 0.008 \\
      & 22.2 & 66.5$\pm$ 42. & 7.12$\pm$ 4.52 & 31.74$\pm$ 20.46 & 0.00058$\pm$ 0.0004 & 0.018$\pm$ 0.018 \\\hline 
F     & 4.8 & 287 $\pm$ 105 & 4.9$\pm$ 1.79 & 49.9$\pm$ 18.52 & 0.00043$\pm$ 0.0003 & 0.009$\pm$0.007 \\
      & 8.0 & 234.5$\pm$ 73.5 & 6.27$\pm$ 1.98 & 48.43$\pm$ 15.72 & 0.00052$\pm$ 0.0003 & 0.011$\pm$ 0.008 \\
      & 14.5 & 189 $\pm$ 73.5 & 9.91$\pm$ 3.88 & 55.39$\pm$ 22.4 & 0.00074$\pm$ 0.0005 & 0.013$\pm$ 0.011 \\
      & 22.2 & 66.5$\pm$ 70. & 7.12$\pm$ 7.51 & 31.74$\pm$ 33.66 & 0.00058$\pm$ 0.0006 & 0.018$\pm$ 0.027 \\ \hline 
G  & 4.8 & 273 $\pm$ 42 & 4.66$\pm$ 0.72 & 47.47$\pm$ 7.88 & 0.00042$\pm$ 0.0003 & 0.009$\pm$ 0.006 \\
  & 8.0 & 154 $\pm$ 66.5 & 4.12$\pm$ 1.78 & 31.8$\pm$ 13.99 & 0.00038$\pm$ 0.0003 & 0.012$\pm$ 0.01 \\
 & 14.5 & 101.5$\pm$ 63. & 5.32$\pm$ 3.31 & 29.75$\pm$ 18.76 & 0.00046$\pm$ 0.0004 & 0.015$\pm$ 0.015 \\
 & 22.2 & 7 $\pm$ 35 & 0.75$\pm$ 3.75 & 3.34$\pm$ 16.71 & 0.0001$\pm$ 0.0004 & 0.031$\pm$ 0.194 \\\hline 
H  & 22.2 & 31.5$\pm$ 73.5 & 3.37$\pm$ 7.87 & 15.04$\pm$ 35.14 & 0.00032$\pm$ 0.0006 & 0.022$\pm$ 0.065 \\\hline 
\end{tabular}}
\label{tab279a}
\end{table}

\begin{table}
{\bf Table 7.} Core position offsets, distance from jet base and magnetic field strengths inferred for segment 2 light curves of 3C 279.
\centering
\scalebox{.67}{
\begin{tabular}{|l|l|l|l|l|l|l|l|}\hline
Flare &  Frequency & $\Delta r$ & $\Omega_{r\nu }$ & $r_{core}$  & $B$ & $B_{core}$ \\
      & (GHz)      & (mas)     & (pc GHz$^{1/k_r}$)         & (pc)       & (G)       & (mG) \\ \hline
      
I   & 8.0 & 21$\pm$ 63. & 1.01$\pm$ 3.03 & 3.15$\pm$ 9.47 & 0.07$\pm$ 0.2 & 23 $\pm$ 88 \\\hline
J   & 4.8 & 259 $\pm$ 31.5 & 6.78$\pm$ 1. & 34.89$\pm$ 6.34 & 0.3$\pm$ 0.3 & 9 $\pm$ 9 \\
   & 14.5 & 227.5$\pm$ 24.5 & 25.35$\pm$ 4.44 & 44.28$\pm$ 11.13 & 0.8$\pm$ 0.8 & 18$\pm$ 20 \\\hline  
K  & 4.8 & 315 $\pm$ 17.5 & 8.24$\pm$ 0.83 & 42.44$\pm$ 6.2 & 0.34$\pm$ 0.4 & 8 $\pm$9 \\
   & 8.0 & 322 $\pm$ 49  & 15.47$\pm$ 2.91 & 48.33$\pm$ 11.33 & 0.55$\pm$ 0.6 & 11$\pm$ 12 \\
   & 14.5 & 339.5$\pm$ 31.5 & 37.83$\pm$ 6.29 & 66.08$\pm$ 16.21 & 1.08$\pm$ 1.1 & 16 $\pm$ 18 \\
   & 22.2 & 63$\pm$59.5 & 15.96$\pm$15.27 & 18.54$\pm$18.16 & 0.56$\pm$0.7 & 30$\pm$49 \\\hline
L  & 4.8 & 171.5$\pm$ 63. & 4.49$\pm$ 1.69 & 23.11$\pm$ 9.04 & 0.22$\pm$ 0.2 & 9$\pm$ 11 \\
   & 8.0 & 70$\pm$ 42& 3.36$\pm$ 2.05 & 10.51$\pm$ 6.58 & 0.18$\pm$ 0.2 & 17$\pm$ 22\\
   & 14.5 & 17.5$\pm$ 38.5 & 1.95$\pm$ 4.3 & 3.41$\pm$ 7.53 & 0.12$\pm$ 0.2 & 34$\pm$ 101 \\
   & 22.2 & 14 $\pm$ 178.5 & 3.55$\pm$ 45.21 & 4.12$\pm$ 52.54 & 0.18$\pm$ 1.8 & 44 $\pm$ 710 \\\hline 
M  & 14.5 & 24.5$\pm$ 21. & 2.73$\pm$ 2.37 & 4.77$\pm$ 4.23 & 0.15$\pm$ 0.2 & 32$\pm$ 48\\\hline
N  & 14.5 & 17.5$\pm$ 21. & 1.95$\pm$ 2.36 & 3.41$\pm$ 4.16 & 0.12$\pm$ 0.2 & 34$\pm$ 63 \\
   & 22.2 & 24.5$\pm$ 59.5 & 6.21$\pm$ 15.1 & 7.21$\pm$ 17.61 & 0.28$\pm$ 0.6 & 39$\pm$ 124 \\\hline 
O  & 4.8 & 231 $\pm$ 42  & 6.04$\pm$ 1.21 & 31.12$\pm$ 7.05 & 0.27$\pm$ 0.3 & 9 $\pm$ 9 \\
   & 8.0 & 115.5$\pm$ 56. & 5.55$\pm$ 2.76 & 17.34$\pm$ 8.96 & 0.26$\pm$ 0.3 & 15 $\pm$ 18  \\
   & 14.5 & 42 $\pm$ 21 & 4.68$\pm$ 2.43 & 8.17$\pm$ 4.49 & 0.23$\pm$ 0.3 & 28 $\pm$ 34\\
   & 22.2 & 3.5$\pm$ 42. & 0.89$\pm$ 10.64 & 1.03$\pm$ 12.36 & 0.06$\pm$ 0.6 & 63$\pm$ 946 \\\hline 
P   & 4.8 & 360.5$\pm$ 52.5 & 9.43$\pm$ 1.59 & 48.57$\pm$ 9.65 & 0.38$\pm$ 0.4 & 8$\pm$ 8 \\
   & 8.0 & 129.5$\pm$ 42 & 6.22$\pm$ 2.13 & 19.44$\pm$ 7.19 & 0.28$\pm$ 0.3 & 14$\pm$ 16 \\ \hline
Q  & 8.0 & 70$\pm$ 66.5 & 3.36$\pm$ 3.22 & 10.51$\pm$ 10.16 & 0.18$\pm$ 0.2 & 17$\pm$ 27 \\ \hline
R  & 4.8 & 231$\pm$ 28 & 6.04$\pm$ 0.89 & 31.12$\pm$ 5.65 & 0.27$\pm$ 0.3 & 9$\pm$ 9 \\
   & 8.0 & 105$\pm$ 49 & 5.04$\pm$ 2.42 & 15.76$\pm$ 7.87 & 0.24$\pm$ 0.3 & 15$\pm$ 18 \\
   & 14.5 & 84 $\pm$ 35  & 9.36$\pm$ 4.11 & 16.35$\pm$ 7.76 & 0.38$\pm$ 0.4 & 23 $\pm$ 28  \\
   & 22.2 & 56$\pm$ 52.5 & 14.18$\pm$ 13.48 & 16.48$\pm$ 16.03 & 0.52$\pm$ 0.7 & 31$\pm$ 50 \\ \hline 
S  & 4.8 & 108.5$\pm$ 59.5 & 2.84$\pm$ 1.57 & 14.62$\pm$ 8.26 & 0.15$\pm$ 0.2 & 11$\pm$ 13 \\ \hline
T   & 4.8 & 161$\pm$ 38.5 & 4.21$\pm$ 1.07 & 21.69$\pm$ 5.96 & 0.21$\pm$ 0.2 & 10$\pm$ 11 \\
   & 14.5 & 24.5$\pm$ 52.5 & 2.73$\pm$ 5.86 & 4.77$\pm$ 10.28 & 0.15$\pm$ 0.3 & 32$\pm$ 91 \\
   & 22.2 & 42$\pm$ 35 & 10.64$\pm$ 9.02 & 12.36$\pm$ 10.79 & 0.42$\pm$ 0.5 & 34$\pm$ 51 \\\hline
U  & 4.8 & 203$\pm$ 10.5 & 5.31$\pm$ 0.52 & 27.35$\pm$ 3.96 & 0.25$\pm$ 0.3 & 9$\pm$ 10 \\
   & 8.0 & 175$\pm$ 17.5 & 8.41$\pm$ 1.25 & 26.27$\pm$ 5.37 & 0.35$\pm$ 0.4 & 13$\pm$ 14 \\
   & 14.5 & 87.5$\pm$ 38.5 & 9.75$\pm$ 4.5 & 17.03$\pm$ 8.43 & 0.39$\pm$ 0.4 & 23$\pm$ 28 \\
   & 22.2 & 157.5$\pm$ 10.5 & 39.89$\pm$ 6.74 & 46.35$\pm$ 12.44 & 1.12$\pm$ 1.2 & 24$\pm$ 26 \\\hline 
V   & 4.8 & 101.5$\pm$ 28. & 2.66$\pm$ 0.77 & 13.68$\pm$ 4.2 & 0.15$\pm$ 0.2 & 11$\pm$ 12 \\
   & 14.5 & 28 $\pm$ 59.5 & 3.12$\pm$ 6.64 & 5.45$\pm$ 11.65 & 0.17$\pm$ 0.3 & 31 $\pm$ 87  \\
   & 22.2 & 17.5$\pm$ 31.5 & 4.43$\pm$ 8.01 & 5.15$\pm$ 9.37 & 0.22$\pm$ 0.4 & 42$\pm$ 105 \\\hline
W  & 4.8 & 392$\pm$ 63 & 10.26$\pm$ 1.86 & 52.81$\pm$ 11.09 & 0.41$\pm$ 0.4 & 8$\pm$ 8 \\
   & 8. & 378$\pm$ 70. & 18.16$\pm$ 3.91 & 56.74$\pm$ 14.59 & 0.62$\pm$ 0.7 & 11$\pm$ 12 \\
   & 14.5 & 350$\pm$ 45.5 & 39.0$\pm$ 7.39 & 68.12$\pm$ 17.82 & 1.1$\pm$ 1.2 & 16$\pm$ 18 \\
   & 22. & 21$\pm$ 31.5 & 5.32$\pm$ 8.02 & 6.18$\pm$ 9.41 & 0.25$\pm$ 0.4 & 40$\pm$ 87 \\
\hline
\end{tabular}}
\label{tab279b}
\end{table}

\begin{table}
{\bf Table 8.} Core position offsets, distance from jet base and magnetic field strengths inferred for the light curves of BL Lacertae.
\centering
\scalebox{.67}{
\begin{tabular}{|l|l|l|l|l|l|l|l|}\hline
Flare &  Frequency & $\Delta r$ & $\Omega_{r\nu }$ & $r_{core}$  & $B_{F}$ & $B_{core}$ \\
      & (GHz)      & (mas)     & (pc GHz$^{1/k_r}$)         & (pc)       & (G)       & (mG) \\ \hline    
      
1    & 8.0 & 64.8$\pm$ 183.6 & 0.78$\pm$ 2.21 & 0.82$\pm$ 2.37 & 0.007$\pm$ 0.023 & 8$\pm$ 37 \\\hline
2  & 8.0 & 205.2$\pm$ 86.4 & 2.46$\pm$ 1.29 & 2.6$\pm$ 1.73 & 0.015$\pm$ 0.044 & 6$\pm$ 17 \\ \hline
3   & 4.8 & 334.8$\pm$ 118.8 & 2.22$\pm$ 0.95 & 3.8$\pm$ 2. & 0.014$\pm$ 0.041 & 4$\pm$ 11 \\
   & 8.0 & 118.8$\pm$ 172.8 & 1.42$\pm$ 2.12 & 1.51$\pm$ 2.33 & 0.01$\pm$ 0.032 & 7$\pm$ 23 \\\hline
4   & 4.8 & 64.8$\pm$ 108.0 & 0.43$\pm$ 0.72 & 0.73$\pm$ 1.26 & 0.004$\pm$ 0.014 & 6$\pm$ 21 \\
   & 8.0 & 75.6$\pm$ 43.2 & 0.91$\pm$ 0.59 & 0.96$\pm$ 0.74 & 0.007$\pm$ 0.022 & 8$\pm$ 23 \\
   & 14.5 & 10.8$\pm$ 10.8 & 0.3$\pm$ 0.32 & 0.18$\pm$ 0.21 & 0.003$\pm$ 0.01 & 18$\pm$ 59 \\
   & 22.2 & 97.2$\pm$ 43.2 & 5.99$\pm$ 3.76 & 2.45$\pm$ 2.14 & 0.029$\pm$ 0.085 & 12$\pm$ 36 \\\hline 
5  & 4.8 & 43.2$\pm$ 205.2 & 0.29$\pm$ 1.36 & 0.49$\pm$ 2.33 & 0.003$\pm$ 0.014 & 7$\pm$ 43 \\
   & 8.0 & 32.4$\pm$ 205.2 & 0.39$\pm$ 2.46 & 0.41$\pm$ 2.61 & 0.004$\pm$ 0.022 & 10$\pm$ 81 \\\hline
6   & 4.8 & 75.6$\pm$ 248.4 & 0.5$\pm$ 1.65 & 0.86$\pm$ 2.84 & 0.005$\pm$ 0.018 & 6$\pm$ 28 \\
   & 8.0 & 21.6$\pm$ 172.8 & 0.26$\pm$ 2.07 & 0.27$\pm$ 2.2 & 0.003$\pm$ 0.019 & 11$\pm$ 111 \\\hline 
7   & 4.8 & 183.6$\pm$ 259.2 & 1.22$\pm$ 1.74 & 2.08$\pm$ 3.05 & 0.009$\pm$ 0.028 & 4$\pm$ 15 \\
   & 8.0 & 259.2$\pm$ 226.8 & 3.11$\pm$ 2.89 & 3.29$\pm$ 3.34 & 0.018$\pm$ 0.053 & 5$\pm$ 17 \\
   & 14.5 & 64.8$\pm$ 172.8 & 1.77$\pm$ 4.78 & 1.07$\pm$ 2.95 & 0.012$\pm$ 0.042 & 11$\pm$ 49 \\
   & 22.2 & 259.2$\pm$ 172.8 & 15.97$\pm$ 12.78 & 6.54$\pm$ 6.56 & 0.058$\pm$ 0.173 & 9$\pm$ 28 \\\hline 
8   & 4.8 & 129.6$\pm$ 75.6 & 0.86$\pm$ 0.54 & 1.47$\pm$ 1.03 & 0.007$\pm$ 0.021 & 5$\pm$ 15 \\
   & 8.0 & 75.6$\pm$ 118.8 & 0.91$\pm$ 1.45 & 0.96$\pm$ 1.59 & 0.007$\pm$ 0.023 & 8$\pm$ 27 \\
   & 14.5 & 75.6$\pm$ 86.4 & 2.07$\pm$ 2.5 & 1.25$\pm$ 1.65 & 0.013$\pm$ 0.041 & 11$\pm$ 35 \\\hline
9  & 4.8 & 43.2$\pm$ 75.6 & 0.29$\pm$ 0.51 & 0.49$\pm$ 0.88 & 0.003$\pm$ 0.01 & 7$\pm $24 \\\hline
10 & 4.8 & 237.6$\pm$ 32.4 & 1.57$\pm$ 0.43 & 2.69$\pm$ 1.11 & 0.011$\pm$ 0.032 & 4$\pm$ 12 \\
   & 8.0 & 172.8$\pm$ 172.8 & 2.07$\pm$ 2.17 & 2.19$\pm$ 2.47 & 0.013$\pm$ 0.04 & 6$\pm$ 20 \\
   & 14.5 & 97.2$\pm$ 118.8 & 2.66$\pm$ 3.42 & 1.61$\pm$ 2.23 & 0.016$\pm$ 0.049 & 10$\pm$ 33 \\
   & 22.2 & 151.2$\pm$ 118.8 & 9.31$\pm$ 8.4 & 3.81$\pm$ 4.14 & 0.039$\pm$ 0.118 & 10$\pm$ 33 \\\hline 
11  & 4.8 & 324.0$\pm$ 183.6 & 2.15$\pm$ 1.32 & 3.67$\pm$ 2.53 & 0.014$\pm$ 0.04 & 4$\pm$ 11 \\
   & 8.0 & 162.0$\pm$ 140.4 & 1.94$\pm$ 1.79 & 2.06$\pm$ 2.07 & 0.013$\pm$ 0.038 & 6$\pm$ 20 \\
   & 14.5 & 248.4$\pm$ 140.4 & 6.8$\pm$ 4.68 & 4.12$\pm$ 3.56 & 0.031$\pm$ 0.093 & 8$\pm$ 24 \\\hline
12   & 4.8 & 86.4$\pm$ 140.4 & 0.57$\pm$ 0.94 & 0.98$\pm$ 1.64 & 0.005$\pm$ 0.017 & 5$\pm$ 19 \\
    & 8.0 & 75.6$\pm$ 205.2 & 0.91$\pm$ 2.48 & 0.96$\pm$ 2.65 & 0.007$\pm$ 0.026 & 8$\pm$ 34 \\\hline
13   & 4.8 & 237.6$\pm$ 194.4 & 1.57$\pm$ 1.34 & 2.69$\pm$ 2.44 & 0.011$\pm$ 0.033 & 4$\pm$ 13 \\
   & 14.5 & 237.6$\pm$ 194.4 & 6.5$\pm$ 5.9 & 3.94$\pm$ 4.13 & 0.03$\pm$ 0.091 & 8$\pm$ 24 \\
   & 22.2 & 237.6$\pm$ 194.4 & 14.63$\pm$ 13.62 & 5.99$\pm$ 6.66 & 0.054$\pm$ 0.164 & 9$\pm$ 29 \\\hline
14   & 4.8 & 97.2$\pm$ 10.8 & 0.64$\pm$ 0.17 & 1.1$\pm$ 0.45 & 0.006$\pm$ 0.017 & 5$\pm$ 15 \\
   & 8. & 75.6$\pm$ 10.8 & 0.91$\pm$ 0.31 & 0.96$\pm$ 0.51 & 0.007$\pm$ 0.022 & 8$\pm$ 23 \\
   & 14.5 & 32.4$\pm$ 118.8 & 0.89$\pm$ 3.27 & 0.54$\pm$ 2. & 0.007$\pm$ 0.028 & 13$\pm$ 73 \\\hline 
15   & 4.8 & 399.6$\pm$ 172.8 & 2.65$\pm$ 1.31 & 4.53$\pm$ 2.64 & 0.016$\pm$ 0.047 & 3$\pm$ 11 \\
   & 8.0 & 475.2$\pm$ 32.4 & 5.7$\pm$1.83 & 6.03$\pm$ 3.13 & 0.028$\pm$ 0.081 & 5$\pm$ 14 \\
   & 14.5 & 442.8$\pm$ 21.6 & 12.12$\pm$ 4.81 & 7.34$\pm$ 4.82 & 0.047$\pm$ 0.14 & 6$\pm$ 20 \\
   & 22. & 129.6$\pm$ 216. & 7.98$\pm$ 13.77 & 3.27$\pm$ 5.98 & 0.035$\pm$ 0.112 & 11$\pm$ 39 \\\hline
16   & 4.8 & 183.6$\pm$ 64.8 & 1.22$\pm$ 0.52 & 2.08$\pm$ 1.09 & 0.009$\pm$ 0.027 & 4$\pm$ 13 \\
   & 8.0 & 43.2$\pm$ 151.2 & 0.52$\pm$ 1.82 & 0.55$\pm$ 1.94 & 0.005$\pm$ 0.019 & 9$\pm$ 47 \\
   & 14.5 & 10.8$\pm$ 129.6 & 0.3$\pm$ 3.55 & 0.18$\pm$ 2.15 & 0.003$\pm$ 0.03 & 18$\pm$ 275 \\
   & 22.2 & 54$\pm$ 183.6 & 3.33$\pm$ 11.4 & 1.36$\pm$ 4.74 & 0.019$\pm$ 0.072 & 14$\pm$ 71 \\ \hline
17   & 4.8 & 766.8$\pm$ 43.2 & 5.08$\pm$ 1.24 & 8.69$\pm$ 3.42 & 0.025$\pm$ 0.074 & 3$\pm$ 9 \\
   & 8.0 & 550.8$\pm$ 75.6 & 6.6$\pm$ 2.26 & 6.99$\pm$ 3.72 & 0.031$\pm$ 0.09 & 4$\pm$ 13 \\
   & 14.5 & 507.6$\pm$ 162. & 13.89$\pm$ 7.04 & 8.41$\pm$ 6.13 & 0.052$\pm$ 0.155 & 6$\pm$ 19 \\
   & 22.2 & 442.8$\pm$ 86.4 & 27.27$\pm$ 13.21 & 11.17$\pm$ 8.66 & 0.085$\pm$ 0.252 & 8$\pm$ 23 \\ \hline 
19   & 8.0 & 410.4$\pm$ 183.6 & 4.92$\pm$ 2.69 & 5.21$\pm$ 3.55 & 0.025$\pm$ 0.073 & 5$\pm$ 14 \\
   & 14.5 & 486$\pm$ 118.8 & 13.3$\pm$6.16 & 8.06$\pm$ 5.64 & 0.051$\pm$ 0.15 & 6$\pm$ 19 \\
   & 22.2 & 129.6$\pm$ 97.2 & 7.98$\pm$ 6.96 & 3.27$\pm$ 3.47 & 0.035$\pm$ 0.105 & 11$\pm$ 34 \\\hline
20   & 4.8 & 378.0$\pm$ 151.2 & 2.51$\pm$ 1.17 & 4.29$\pm$ 2.39 & 0.015$\pm$ 0.045 & 4$\pm$ 11 \\
   & 8.0 & 324.0$\pm$ 151.2 & 3.88$\pm$ 2.18 & 4.11$\pm$ 2.86 & 0.021$\pm$ 0.062 & 5$\pm$ 15 \\
   & 14.5 & 151.2$\pm$ 162.0 & 4.14$\pm$ 4.72 & 2.51$\pm$ 3.15 & 0.022$\pm$ 0.067 & 9$\pm$ 29 \\
   & 22.2 & 151.2$\pm$ 172.8 & 9.31$\pm$ 11.42 & 3.81$\pm$ 5.22 & 0.039$\pm$ 0.12 & 10$\pm$ 35 \\\hline  
21   & 4.8 & 205.2$\pm$ 108.0 & 1.36$\pm$ 0.79 & 2.33$\pm$ 1.52 & 0.01$\pm$ 0.029 & 4$\pm$ 13 \\
   & 8.0 & 226.8$\pm$ 162.0 & 2.72$\pm$ 2.12 & 2.88$\pm$ 2.53 & 0.016$\pm$ 0.048 & 6$\pm$ 17 \\
   & 14.5 & 108.0$\pm$ 21.6 & 2.96$\pm$ 1.3 & 1.79$\pm$ 1.23 & 0.017$\pm$ 0.051 & 10$\pm$ 29 \\
   & 22.2 & 86.4$\pm$ 64.8 & 5.32$\pm$ 4.64 & 2.18$\pm$ 2.31 & 0.026$\pm$ 0.079 & 12$\pm$ 38 \\\hline   
22   & 4.8 & 32.4$\pm$ 97.2 & 0.21$\pm$ 0.65 & 0.37$\pm$ 1.11 & 0.003$\pm$ 0.009 & 7$\pm$ 33 \\
   & 14.5 & 21.6$\pm$ 21.6 & 0.59$\pm$ 0.64 & 0.36$\pm$ 0.43 & 0.005$\pm$ 0.016 & 15$\pm$ 49 \\\hline   
23   & 4.8 & 475.2$\pm$ 162.0 & 3.15$\pm$ 1.31 & 5.39$\pm$ 2.79 & 0.018$\pm$ 0.053 & 3$\pm$ 10 \\
   & 8.0 & 464.4$\pm$ 151.2 & 5.57$\pm$ 2.52 & 5.89$\pm$ 3.59 & 0.027$\pm$ 0.08 & 5$\pm$ 14 \\
   & 14.5 & 313.2$\pm$ 151.2 & 8.57$\pm$ 5.34 & 5.19$\pm$ 4.23 & 0.037$\pm$ 0.11 & 7$\pm$ 22 \\
   & 22.2 & 270.0$\pm$ 151.2 & 16.63$\pm$ 11.88 & 6.81$\pm$ 6.38 & 0.06$\pm$ 0.178 & 9$\pm$ 27 \\\hline 
24   & 4.8 & 270.0$\pm$ 172.8 & 1.79$\pm$ 1.22 & 3.06$\pm$ 2.29 & 0.012$\pm$ 0.036 & 4$\pm$ 12 \\
   & 8.0 & 21.6$\pm$ 205.2 & 0.26$\pm$ 2.46 & 0.27$\pm$ 2.61 & 0.003$\pm$ 0.022 & 11$\pm$ 131 \\
   & 14.5 & 21.6$\pm$ 205.2 & 0.59$\pm$ 5.62 & 0.36$\pm$ 3.41 & 0.005$\pm$ 0.04 & 15$\pm$ 181 \\
   & 22.2 & 10.8$\pm$ 226.8 & 0.67$\pm$ 13.97 & 0.27$\pm$ 5.72 & 0.006$\pm$ 0.09 & 21$\pm$ 560 \\\hline
   
\end{tabular}}
\label{tab4bl6}
\end{table}

\begin{table}
{\bf Table 8.} for BL Lacertae continued.
\centering
\scalebox{.67}{
\begin{tabular}{|l|l|l|l|l|l|l|l|}\hline
Flare &  Frequency & $\Delta r$ & $\Omega_{r\nu }$ & $r_{core}$  & $B_{F}$ & $B_{core}$ \\
      & (GHz)      & (mas)     & (pc GHz$^{1/k_r}$)         & (pc)       & (G)       & (mG) \\ \hline
   
25   & 4.8 & 10.8$\pm$ 140.4 & 0.07$\pm$ 0.93 & 0.12$\pm$ 1.59 & 0.001$\pm$ 0.012 & 10$\pm$ 156 \\
   & 14.5 & 151.2$\pm$ 118.8 & 4.14$\pm$ 3.64 & 2.51$\pm$ 2.56 & 0.022$\pm$ 0.066 & 9$\pm$ 28 \\
    & 22.2 & 10.8$\pm$ 151.2 & 0.67$\pm$ 9.32 & 0.27$\pm$3.82 & 0.006$\pm$ 0.062 & 21$\pm$ 376\\\hline 
26   & 4.8 & 86.4$\pm$ 97.2 & 0.57$\pm$ 0.66 & 0.98$\pm$ 1.17 & 0.005$\pm$ 0.016 & 5$\pm$ 18 \\
   & 8.0 & 129.6$\pm$ 86.4 & 1.55$\pm$ 1.14 & 1.64$\pm$ 1.39 & 0.011$\pm$ 0.032 & 7$\pm$ 20 \\
   & 14.5 & 205.2$\pm$ 97.2 & 5.62$\pm$ 3.46 & 3.4$\pm$ 2.75 & 0.027$\pm$ 0.081 & 8$\pm$ 25 \\
   & 22.2 & 97.2$\pm$ 97.2 & 5.99$\pm$ 6.55 & 2.45$\pm$ 3.07 & 0.029$\pm$ 0.087 & 12$\pm$ 38 \\\hline
27   & 4.8 & 550.8$\pm$172.8 & 3.65$\pm$ 1.44 & 6.25$\pm$ 3.12 & 0.02$\pm$ 0.059 & 3$\pm$ 10 \\
   & 8.0 & 453.6$\pm$ 172.8 & 5.44$\pm$ 2.68 & 5.76$\pm$ 3.69 & 0.027$\pm$ 0.079 & 5$\pm$ 14 \\
   & 14.5 & 864.0$\pm$ 172.8 & 23.64$\pm$ 10.44 & 14.32$\pm$ 9.81 & 0.077$\pm$ 0.227 & 5$\pm$ 16 \\\hline 
28   & 14.5 & 442.8$\pm$ 97.2 & 12.12$\pm$ 5.46 & 7.34$\pm$ 5.07 & 0.047$\pm$ 0.14 & 6$\pm$ 20 \\
   & 22.2 & 259.2$\pm$ 108.0 & 15.97$\pm$ 9.72 & 6.54$\pm$ 5.61 & 0.058$\pm$ 0.172 & 9$\pm$ 27 \\\hline 
29   & 22.2 & 140.4$\pm$ 43.2 & 8.65$\pm$ 4.67 & 3.54$\pm$ 2.87 & 0.037$\pm$ 0.11 & 11$\pm$ 32 \\ \hline
30  & 4.8 & 172.8$\pm$ 151.2 & 1.15$\pm$ 1.04 & 1.96$\pm$ 1.88 & 0.009$\pm$ 0.026 & 4$\pm$ 14 \\
   & 8.0 & 43.2$\pm$ 151.2 & 0.52$\pm$ 1.82 & 0.55$\pm$ 1.94 & 0.005$\pm$ 0.019 & 9$\pm$ 47 \\
   & 14.5 & 21.6$\pm$151.2 & 0.59$\pm$ 4.14 & 0.36$\pm$ 2.52 & 0.005$\pm$ 0.031 & 15$\pm$ 137 \\
   & 22.2 & 75.6$\pm$ 151.2 & 4.66$\pm$ 9.54 & 1.91$\pm$ 4.07 & 0.024$\pm $0.078 & 12$\pm$ 49 \\\hline
31   & 8.0 & 108.0$\pm$ 162.0 & 1.29$\pm$ 1.98 & 1.37$\pm$ 2.17 & 0.009$\pm$ 0.03 & 7$\pm$ 24 \\
   & 14.5 & 86.4$\pm$ 64.8 & 2.36$\pm$ 2.0 & 1.43$\pm$ 1.43 & 0.015$\pm$ 0.044 & 10$\pm$ 32 \\
   & 22.2 & 345.6$\pm$ 43.2 & 21.29$\pm$ 9.81 & 8.72$\pm$ 6.64 & 0.071$\pm$ 0.21 & 8$\pm$ 25 \\\hline 
32   & 4.8 & 97.2$\pm$ 118.8 & 0.64$\pm$ 0.8 & 1.1$\pm$ 1.41 & 0.006$\pm$ 0.018 & 5$\pm$ 17 \\
   & 8. & 10.8$\pm$ 172.8 & 0.13$\pm$ 2.07 & 0.14$\pm$ 2.19 & 0.002$\pm$ 0.021 & 13$\pm$ 261 \\
   & 14.5 & 108.0$\pm$ 118.8 & 2.96$\pm$ 3.45 & 1.79$\pm$ 2.29 & 0.017$\pm$ 0.052 & 10$\pm$ 32 \\
   & 22.2 & 10.8$\pm$ 140.4 & 0.67$\pm$ 8.65 & 0.27$\pm$ 3.55 & 0.006$\pm$ 0.058 & 21$\pm$ 350 \\\hline 
33   & 22.2 & 421.2$\pm$ 54.0 & 25.94$\pm$ 11.98 & 10.62$\pm$ 8.09 & 0.082$\pm$ 0.243 & 8$\pm$ 24 \\\hline
34   & 14.5 & 183.6$\pm$ 54.0 & 5.02$\pm$ 2.47 & 3.04$\pm$ 2.19 & 0.025$\pm$ 0.074 & 8$\pm$ 25 \\
   & 22.2 & 183.6$\pm$ 21.6 & 11.31$\pm$ 5.19 & 4.63$\pm$ 3.52 & 0.045$\pm$ 0.133 & 10$\pm$ 30 \\\hline
35   & 4.8 & 183.6$\pm$ 162.0 & 1.22$\pm$ 1.11 & 2.08$\pm$ 2.01 & 0.009$\pm$ 0.027 & 4$\pm$ 14 \\
   & 8.0 & 10.8$\pm$ 75.6 & 0.13$\pm$ 0.91 & 0.14$\pm$ 0.96 & 0.002$\pm$ 0.01 & 13$\pm$ 120 \\
   & 14.5 & 313.2$\pm$ 43.2 & 8.57$\pm$ 3.57 & 5.19$\pm$ 3.48 & 0.037$\pm$ 0.109 & 7$\pm$ 22 \\
   & 22.2 & 194.4$\pm$ 108. & 11.97$\pm$ 8.51 & 4.9$\pm$ 4.58 & 0.047$\pm$ 0.14 & 10$\pm$ 30 \\
  
\hline
\end{tabular}}
\label{tab4bl62}
\end{table}

\begin{table}
{\bf Table 9.} The table shows the bolometric luminosity in units of $10^{46}$ ergs, $Bh$, $w$ and spin value $a$ derived for the three sources.
\centering
\scalebox{1}{
\begin{tabular}{|l|l|c|c|c|c|}
\hline
 \# & Source & $L_{46}$ & $B h$ (Gpc) & $w(a)$ & $a$  \\
\hline
 1 & BL Lac & 0.1 & 8.5 & 0.006 & 0.16 \\
 2 & S5 0716+714 & 1.59 & 31.4 & 0.007 & 0.17 \\
 3 & 3C 279 & 199.5 & 62.8 & 0.23 & 0.9 \\
\hline
\end{tabular}
}
\end{table}
\label{spin}

\begin{figure}
\epsfig{figure= 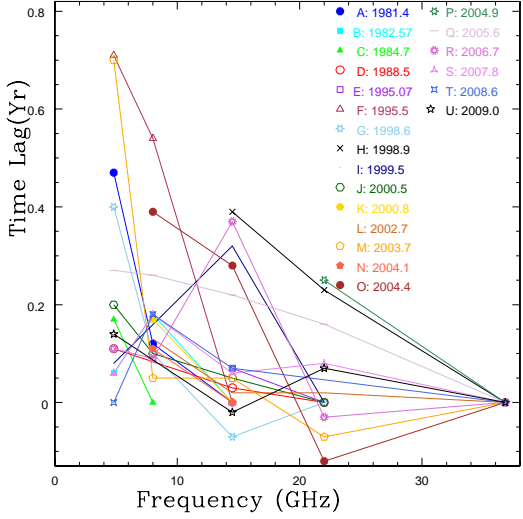,height=2.8in,width=2.8in,angle=0}
\caption{Time lags $\Delta t$ as a function of $\nu$ for the light curves of S5 0716+714.
A fit to $\Delta t = a \nu^{-1/k_r}+b$ yields $b = -0.03 \pm 0.02$ yr, $a = 1.01 \pm 0.08$ yr (GHz)$^{-1/k_r}$ and $k_r = 1.03 \pm 0.09$.
The $k_r$ value indicates consistency of the equipartition between magnetic field energy density and the particle kinetic energy.}
\label{tnu1}
\end{figure}

\begin{figure}
\epsfig{figure= 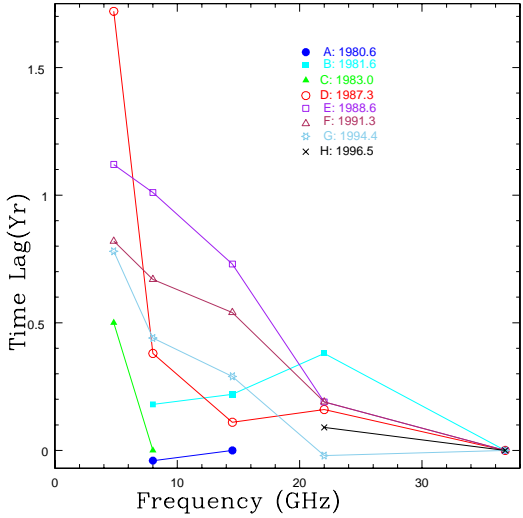,height=2.8in,width=2.8in,angle=0}
\caption{Time lags $\Delta t$ as a function of $\nu$ for segment 1 light curves of 3C 279.
A fit to $\Delta t = a \nu^{-1/k_r}+b$ yields $b = -0.30 \pm 0.04$ yr, $a = 340.2 \pm 48.3$ yr (GHz)$^{-1/k_r}$ and $k_r = 1.84 \pm 0.13$.}
\label{tnu2}
\end{figure}

\begin{figure}
\epsfig{figure= 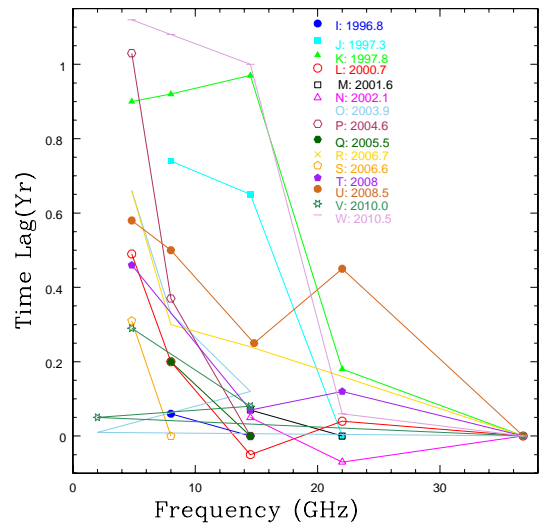,height=2.8in,width=2.8in,angle=0}
\caption{Fig \ref{tnu2} continued for segment 2 light curves of 3C 279 with $b = -0.09 \pm 0.01$ yr,
$a = 1.86 \pm 0.28$ yr (GHz)$^{-1/k_r}$ and $k_r = 1.02 \pm 0.07$.
The $k_r$ value indicates consistency of the equipartition between magnetic field energy density and the particle kinetic energy.}
\label{tnu3}
\end{figure}

\begin{figure}
\epsfig{figure= 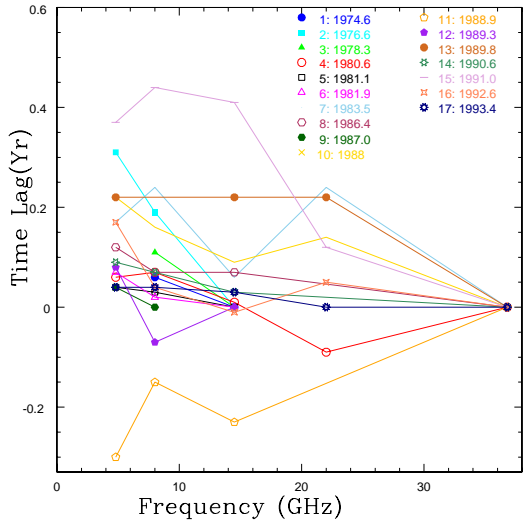,height=2.8in,width=2.8in,angle=0}
\caption{Time lags $\Delta t$ as a function of $\nu$ for light curves of BL Lacertae.
A fit to $\Delta t = a \nu^{-1/k_r}+b$ yields $b = -0.13 \pm 0.05$ yr, $a = 1.52 \pm 0.16$ yr (GHz)$^{-1/k_r}$ and $k_r = 1.06 \pm 0.22$.
The $k_r$ value indicates consistency of the equipartition between magnetic field energy density and the particle kinetic energy.}
\label{tnu4}
\end{figure}

\begin{figure}
\epsfig{figure= 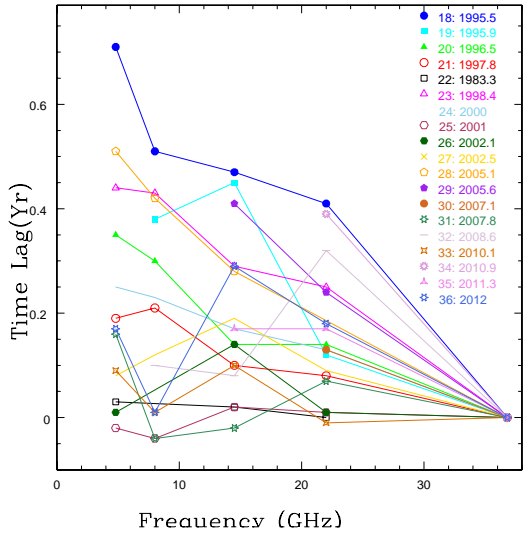,height=2.8in,width=2.8in,angle=0}
\caption{Fig \ref{tnu4} continued for rest of the flares.}
\label{tnu5}
\end{figure}

\begin{figure}
\centerline{\includegraphics[scale=0.5]{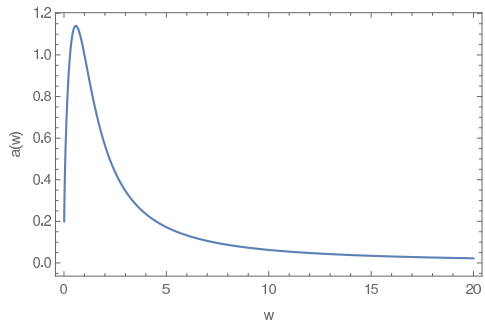}}
\caption{The spin parameter $a$ is plotted a function of the parameter $w$. The values of $0.296<w<1$ are disallowed since they yield spin values $a>1$.}
\label{aw}
\end{figure}

\begin{figure}
\centerline{\includegraphics[scale=0.155]{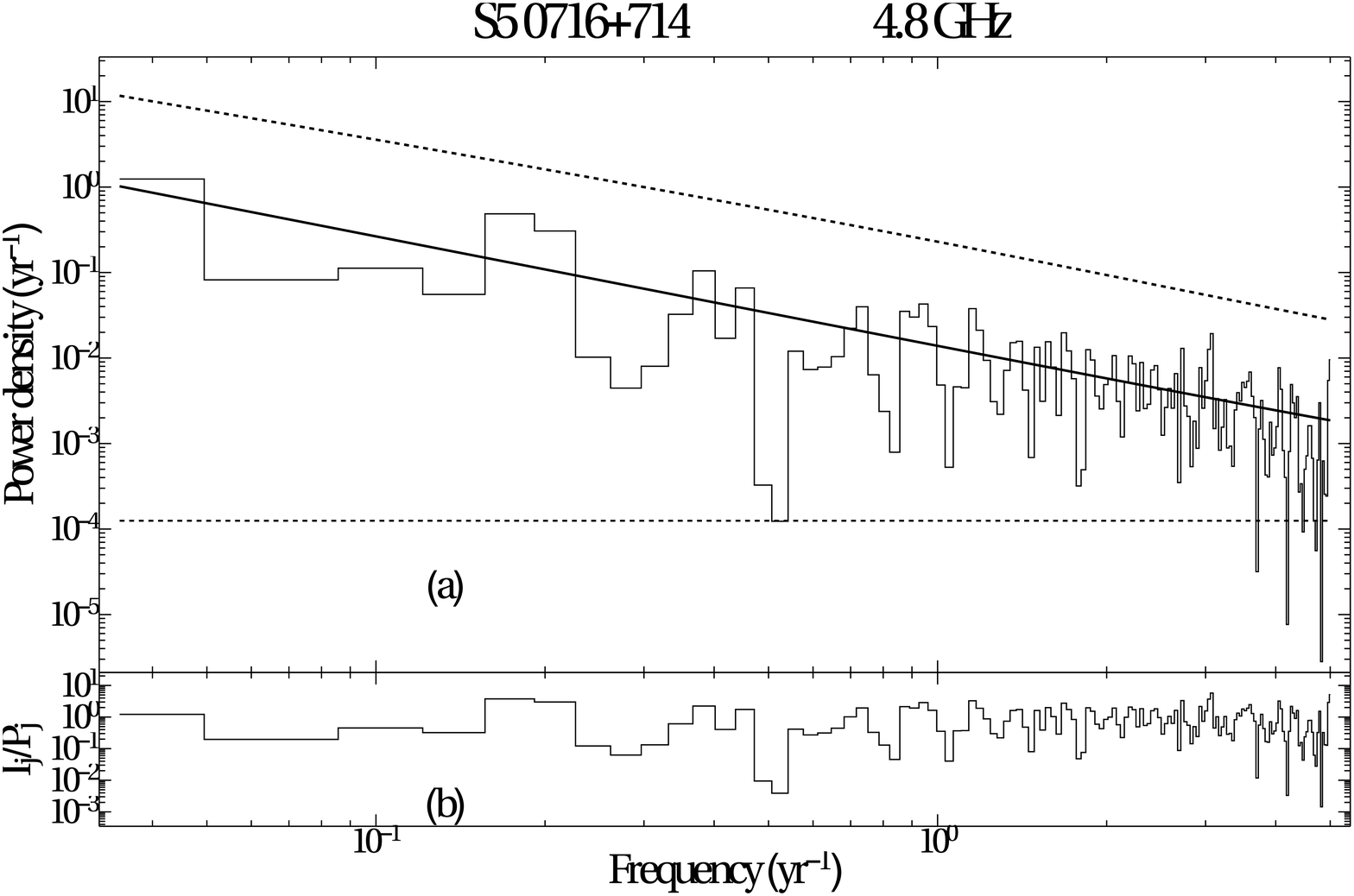}}
\centerline{\includegraphics[scale=0.155]{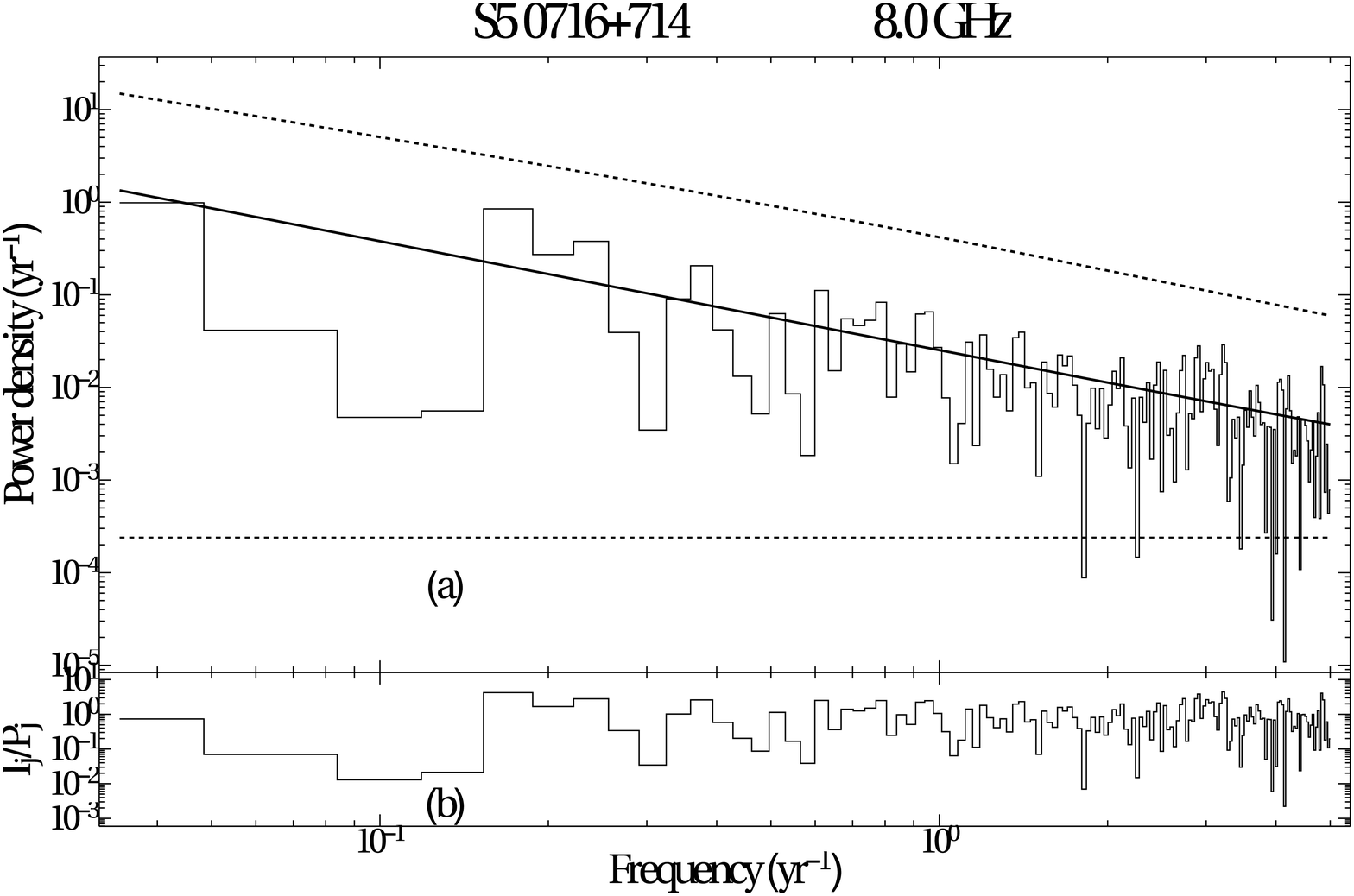}}
\centerline{\includegraphics[scale=0.155]{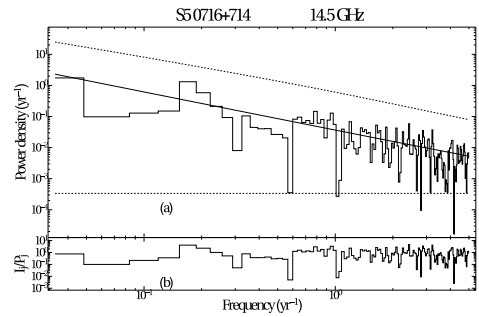}}
\centerline{\includegraphics[scale=0.155]{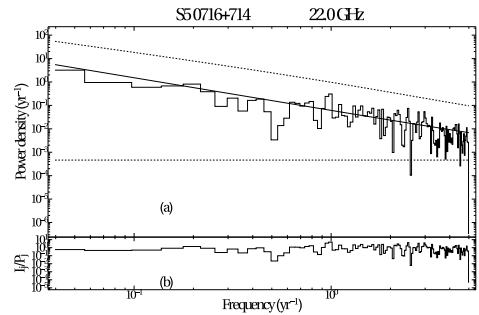}}
\centerline{\includegraphics[scale=0.155]{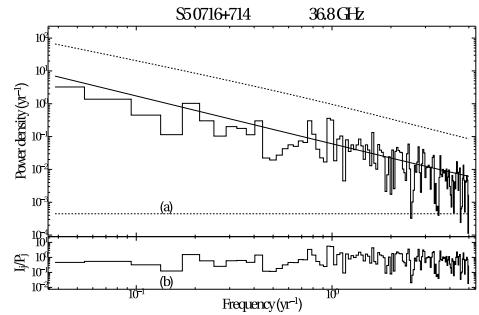}}
\caption{Periodogram analysis of S5 0716+714.
The best fit model is the solid curve, the dashed curve above it is the 99 \% significance contour which can identify statistically significant
quasi-periodic components, the dot-dashed horizontal line is the white noise level and the plot below each periodogram panel shows the fit
residuals.}
\label{psda}
\end{figure}

 \begin{table}
 {\bf Table 10.} Results from the parametric PSD models fit to the periodogram of the 4.8 - 36.8 GHz light curves of S5 0716+714.
Columns 1 -- 7 give the observation frequency and segment, the model (PL: power law + constant noise, BPL: bending power law + constant noise),
the best-fit parameters $\log(A)$, slope $\alpha$ and the bend frequency $f_b$ with their 95\% errors derived from $\Delta S$,
the AIC and the likelihood of a particular model. The best fit PSD is highlighted.
\centering
\scalebox{.73}{
\begin{tabular}{lllllll}
\hline
Observation & PSD       & \multicolumn{3}{c}{PSD Fit parameters}    & AIC & Model \\
Frequency   & model     & \multicolumn{3}{l}{} &     & likelihood\\
            &           & log(A) & $\alpha$ & log(f$_b$)            &     & \\ \hline
4.8 GHz & {\bf PL} & -1.86 $\pm$ 0.11 & -1.3 $\pm$ 0.2 & & 1169.59 & 1.00 \\
 & BPL & -1.20 $\pm$ 0.18 & -1.4 $\pm$ 0.2 & -1.46 $\pm$ 0.08 & 1169.59 & 0.36 \\ \hline
8.0 GHz & {\bf PL} & -1.60 $\pm$ 0.11 & -1.2 $\pm$ 0.2 & & 1000.61 & 1.00 \\
 & BPL & -1.06 $\pm$ 0.18 & -1.3 $\pm$ 0.2 & -1.47 $\pm$ 0.08 & 1002.70 & 0.35 \\ \hline
14.5 GHz & {\bf PL} & -1.44 $\pm$ 0.11 & -1.2 $\pm$ 0.2 & & 901.40 & 1.00 \\
 & BPL & -0.84 $\pm$ 0.18 & -1.3 $\pm$ 0.2 & -1.47 $\pm$ 0.08 & 903.65 & 0.32 \\ \hline
22.2 GHz & {\bf PL} & -1.22 $\pm$ 0.12 & -1.4 $\pm$ 0.3 & & 676.81 & 1.00 \\
 & BPL & -0.43 $\pm$ 0.18 & -1.5 $\pm$ 0.2 & -1.40 $\pm$ 0.08 & 679.83 & 0.22 \\ \hline
36.8 GHz & {\bf PL} & -1.22 $\pm$ 0.12 & -1.5 $\pm$ 0.3 & & 711.67 & 1.00 \\
 & BPL & -0.36 $\pm$ 0.18 & -1.6 $\pm$ 0.2 & -1.41 $\pm$ 0.08 & 715.06 & 0.18 \\ \hline
\end{tabular}}
\label{bllactab}
\end{table}

\begin{figure}
\centerline{\includegraphics[scale=0.155]{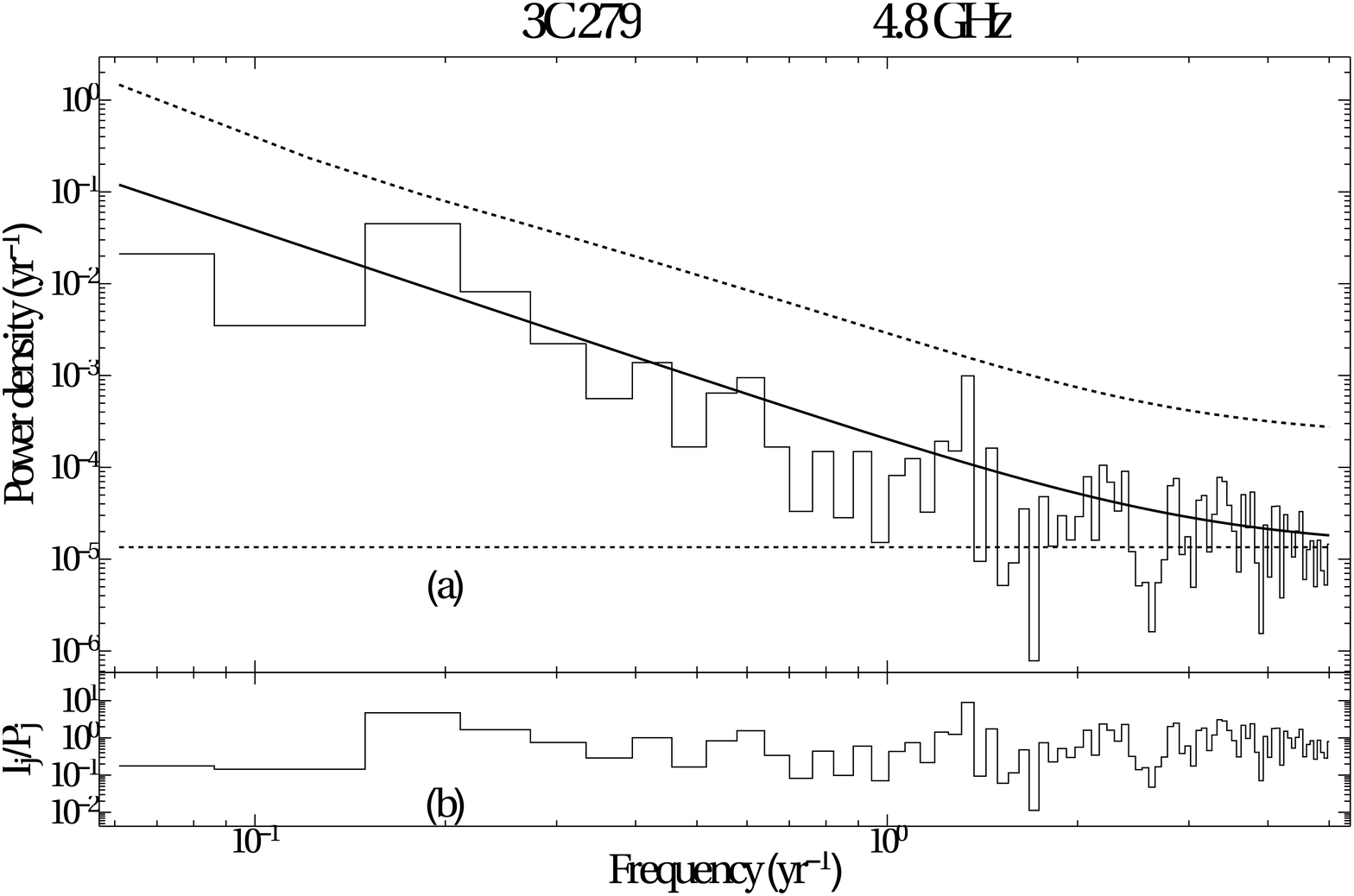}}
\centerline{\includegraphics[scale=0.155]{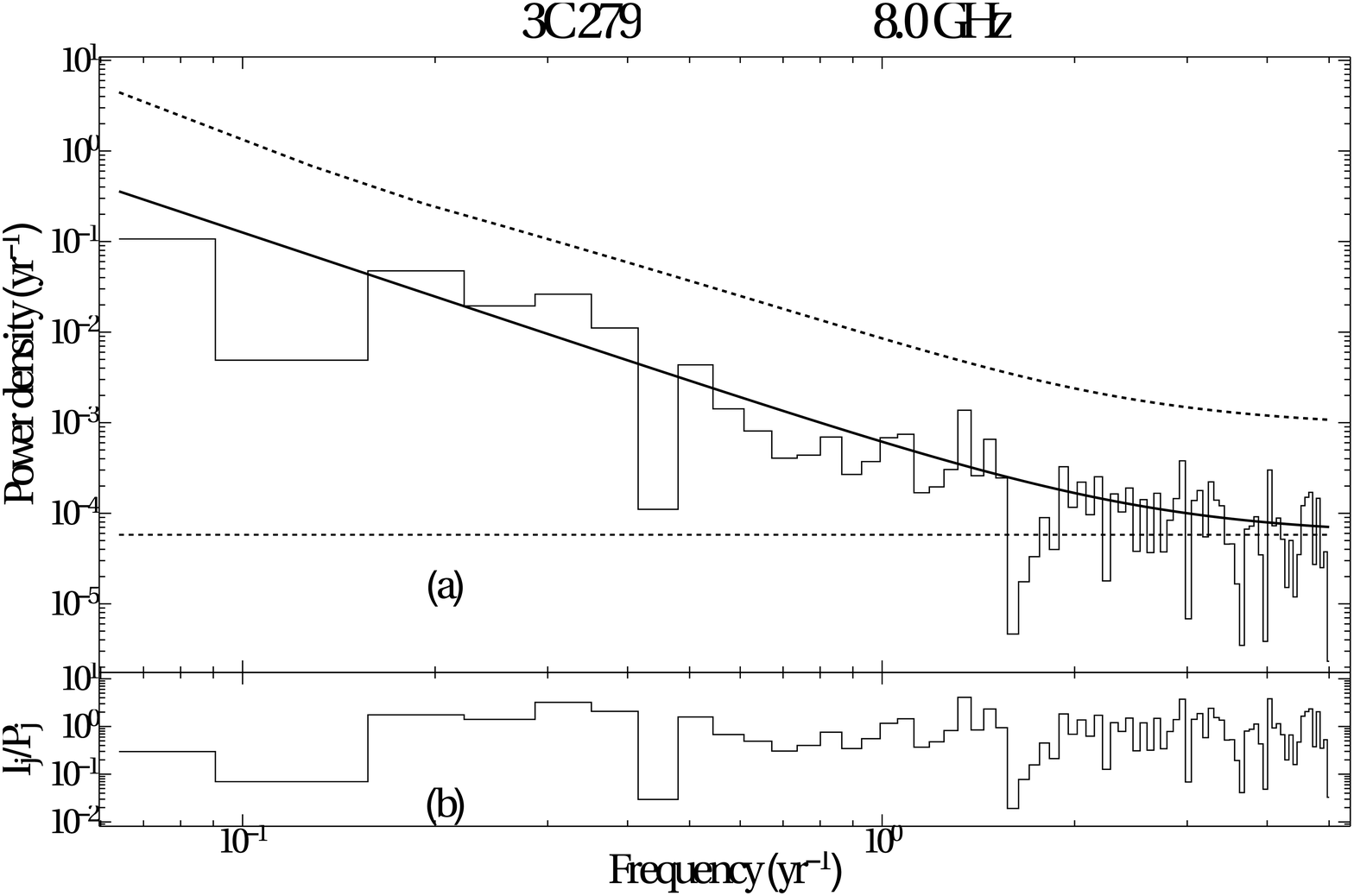}}
\centerline{\includegraphics[scale=0.155]{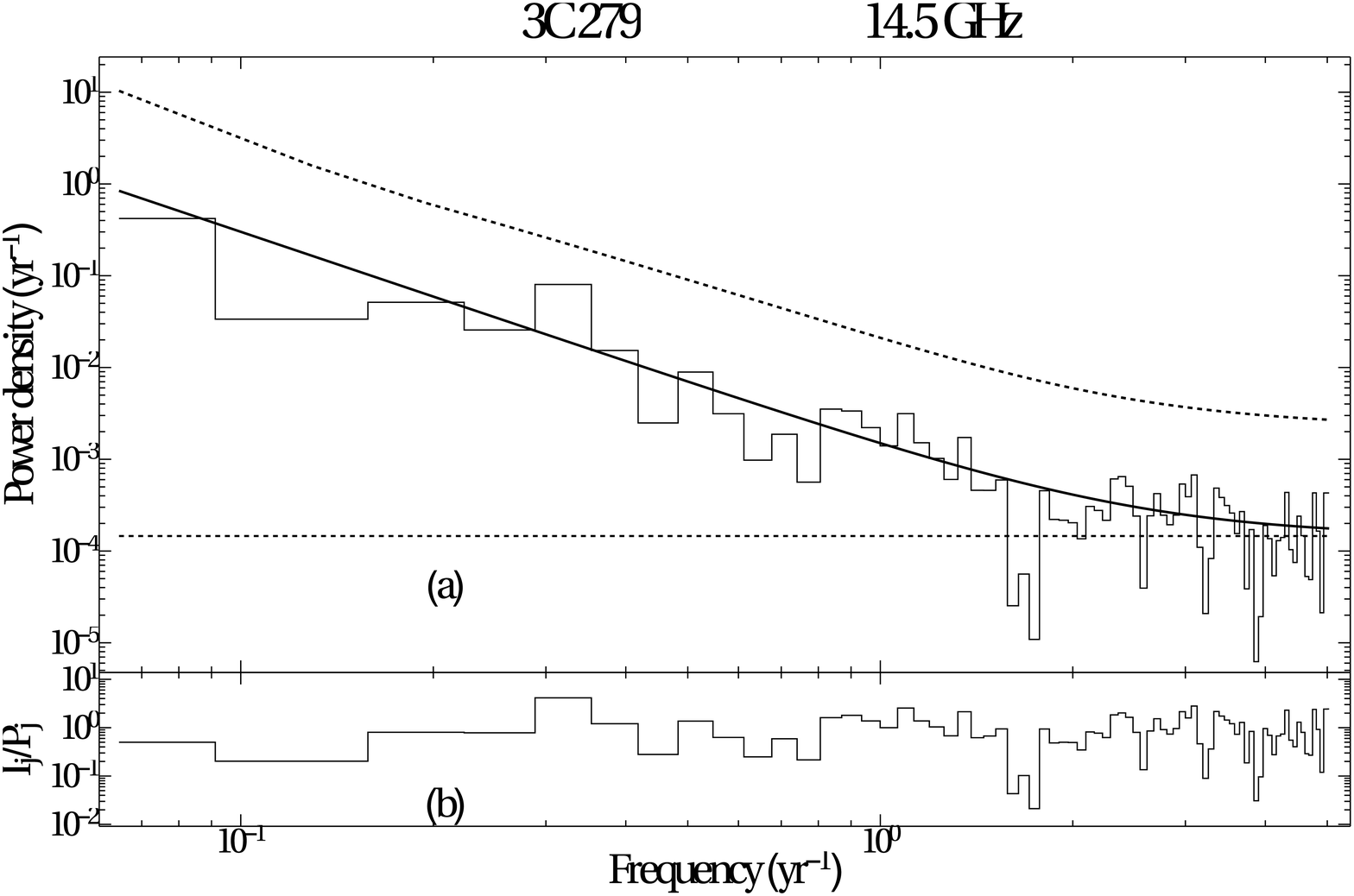}}
\centerline{\includegraphics[scale=0.155]{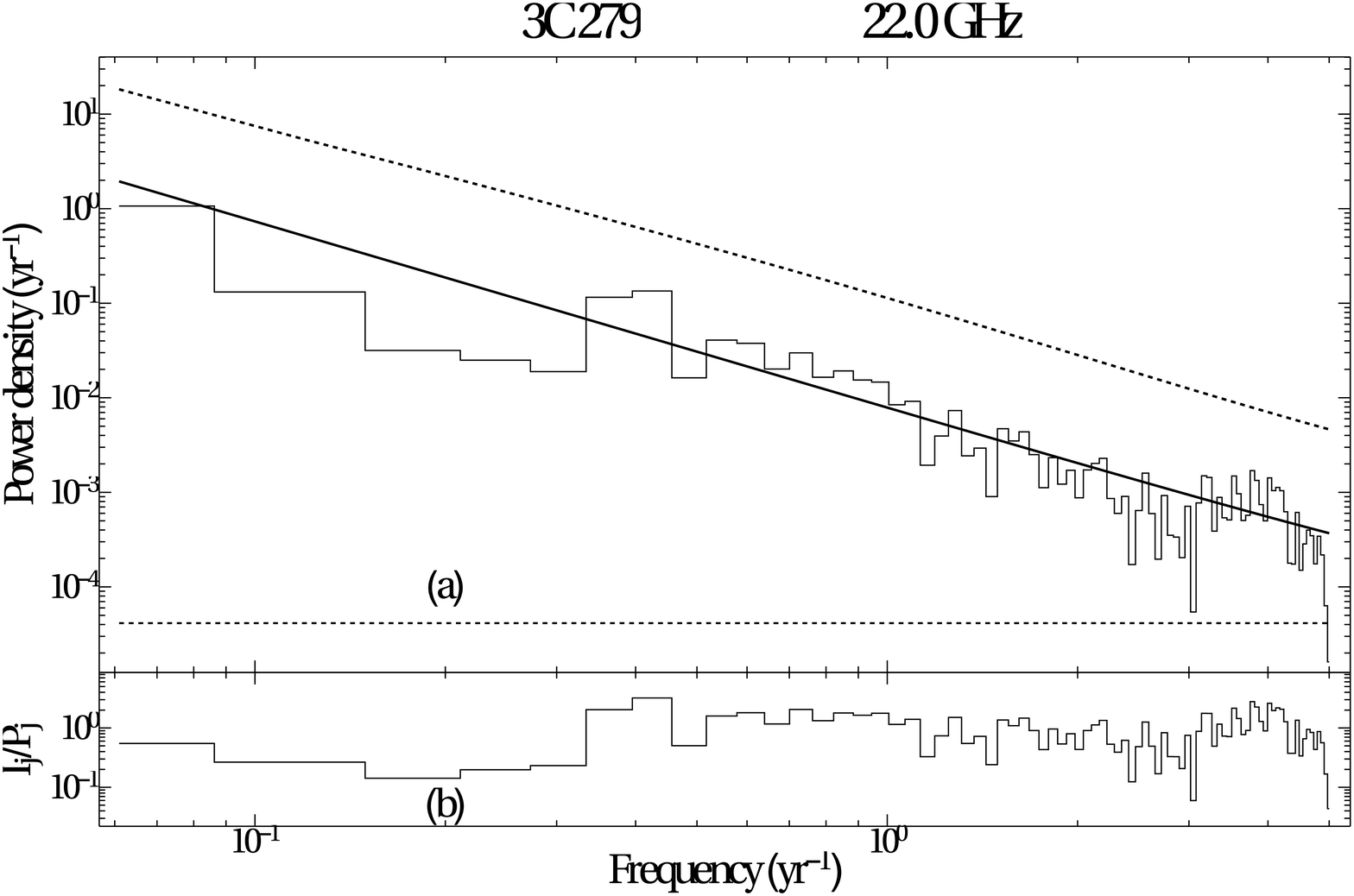}}
\centerline{\includegraphics[scale=0.155]{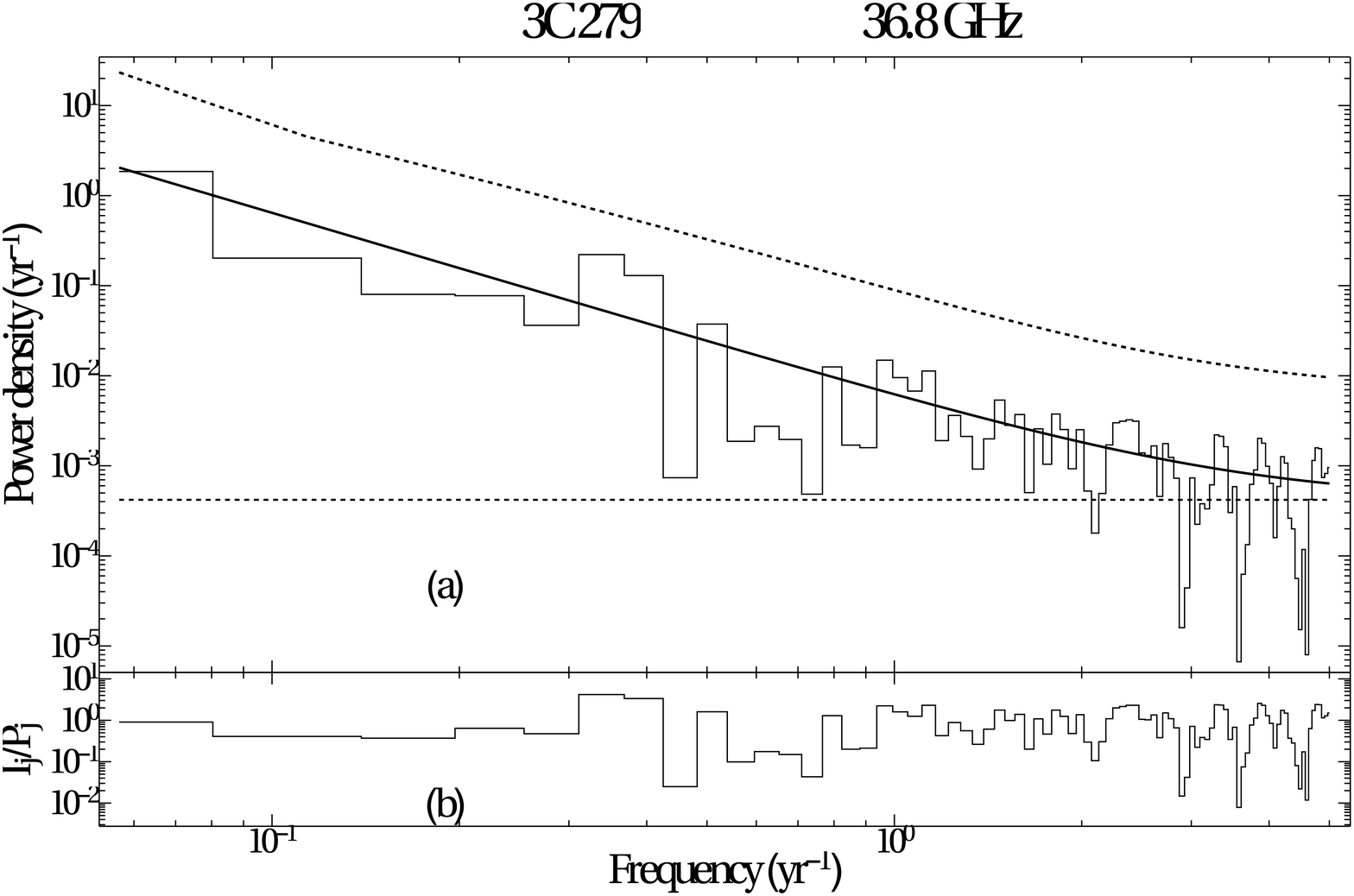}}
\caption{Periodogram analysis of segment 1 of 3C 279 (beginning of observations to $\sim$ 2007.0).
Plot details same as \ref{psda}.}
\label{seg1psda}
\end{figure}

\begin{figure}
\centerline{\includegraphics[scale=0.155]{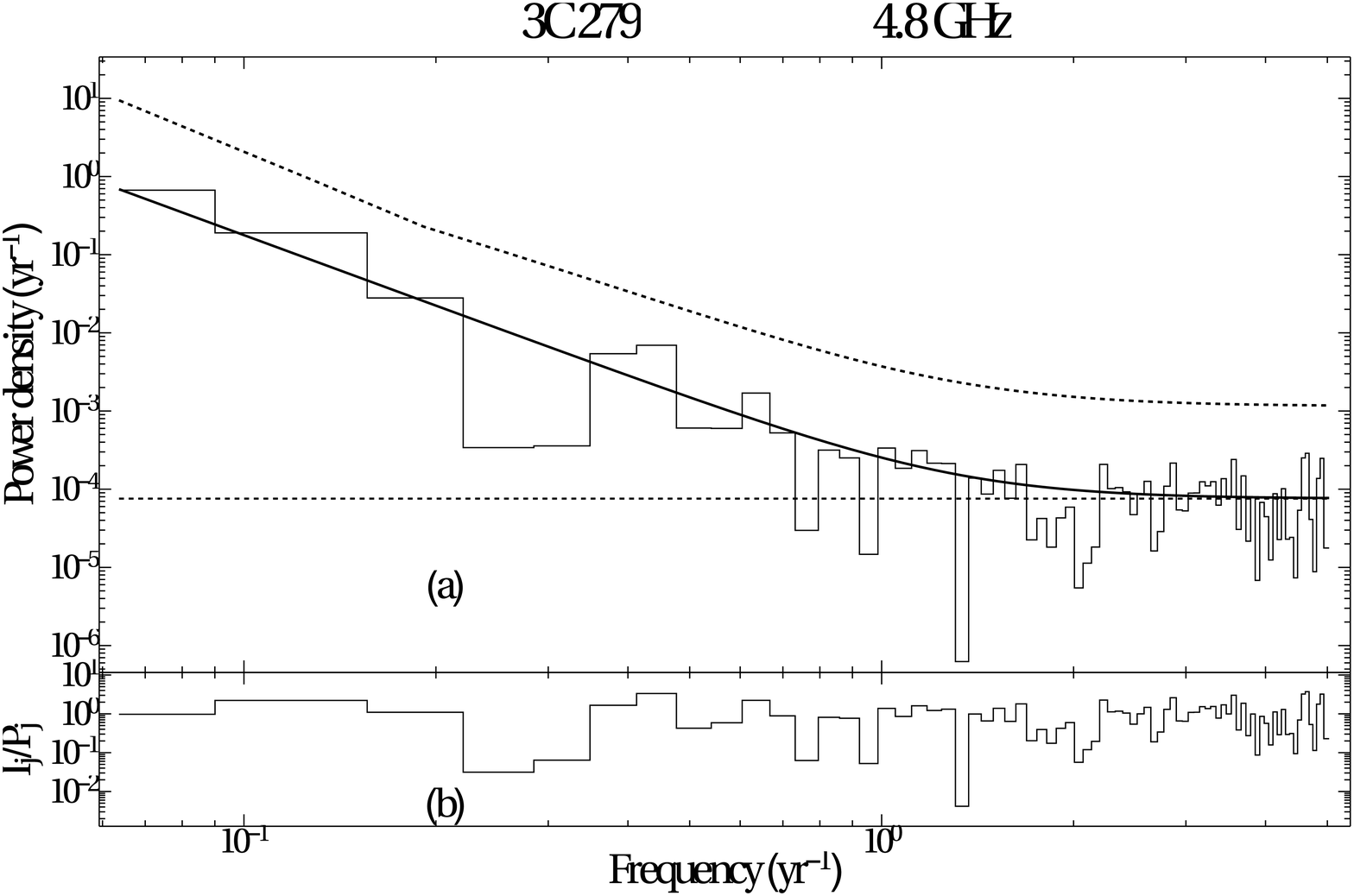}}
\centerline{\includegraphics[scale=0.155]{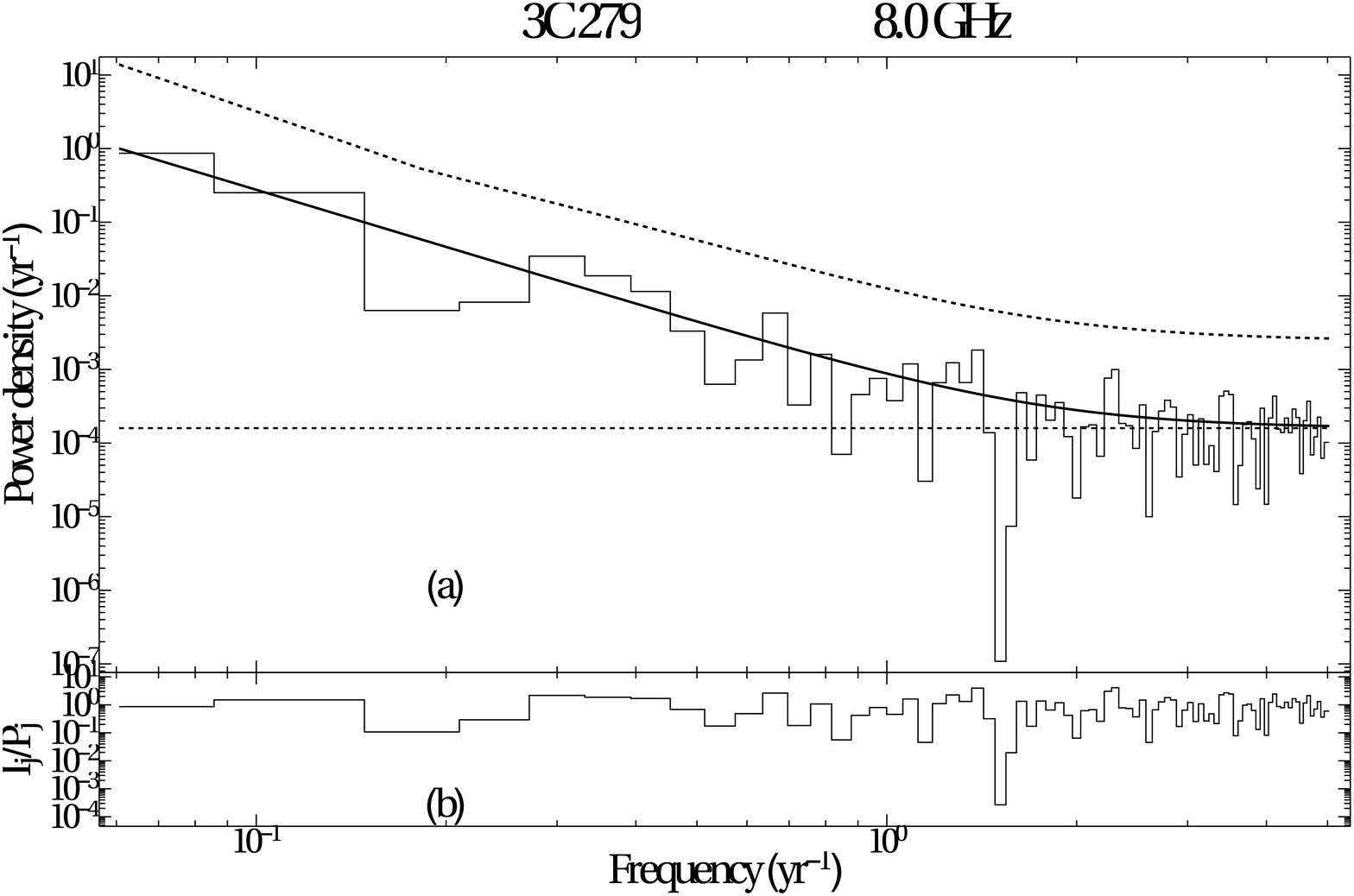}}
\centerline{\includegraphics[scale=0.155]{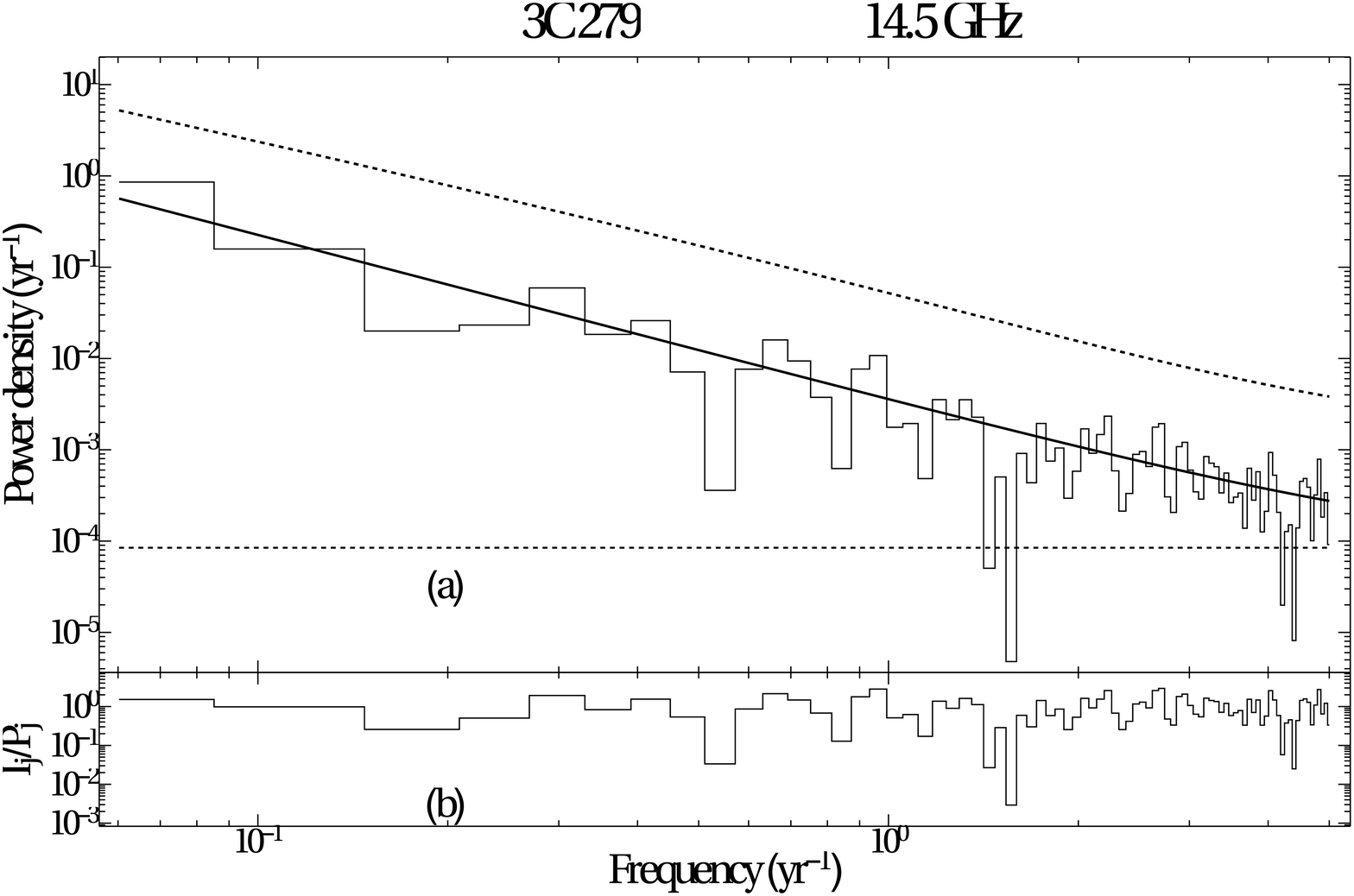}}
\centerline{\includegraphics[scale=0.155]{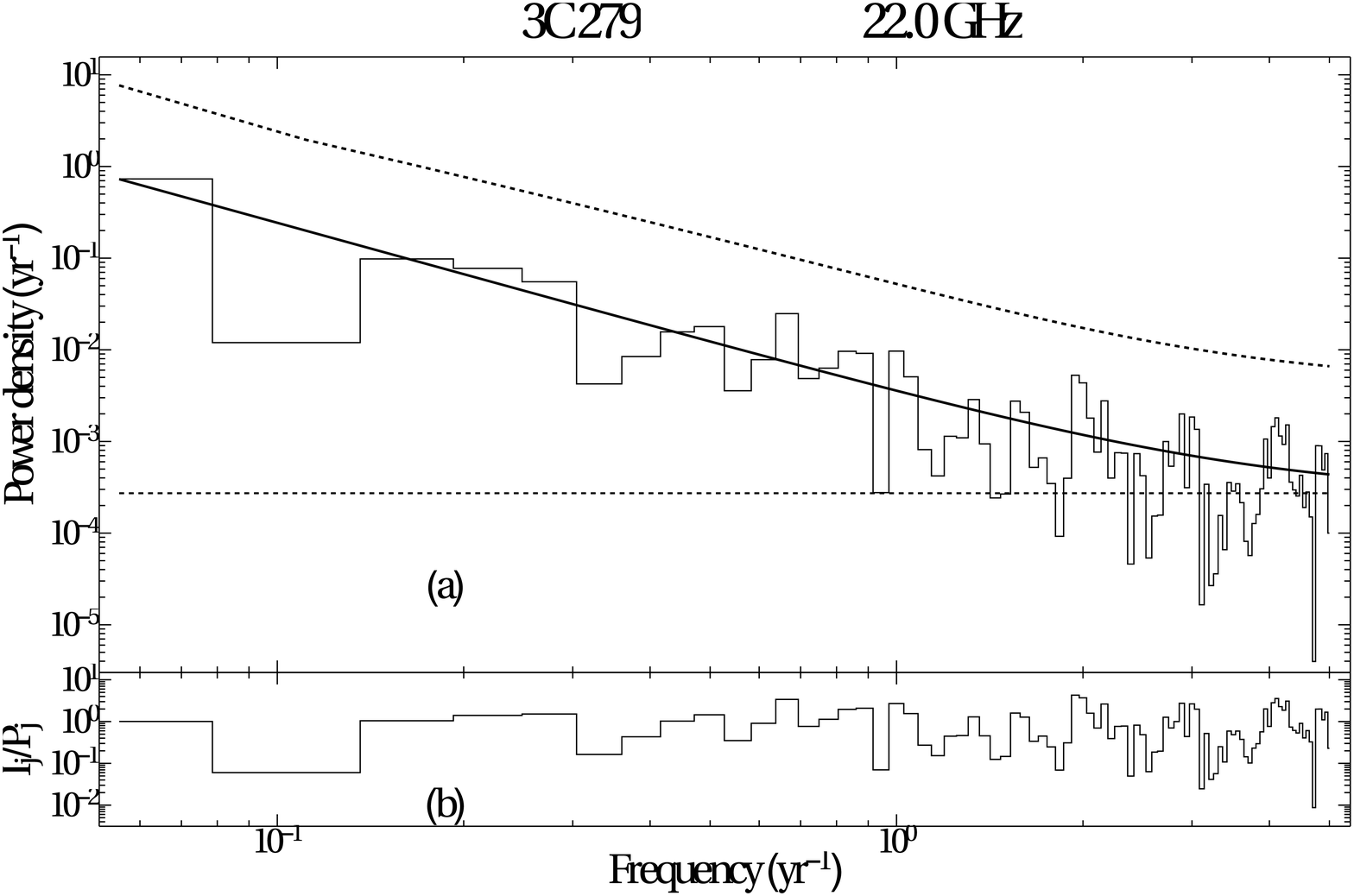}}
\centerline{\includegraphics[scale=0.155]{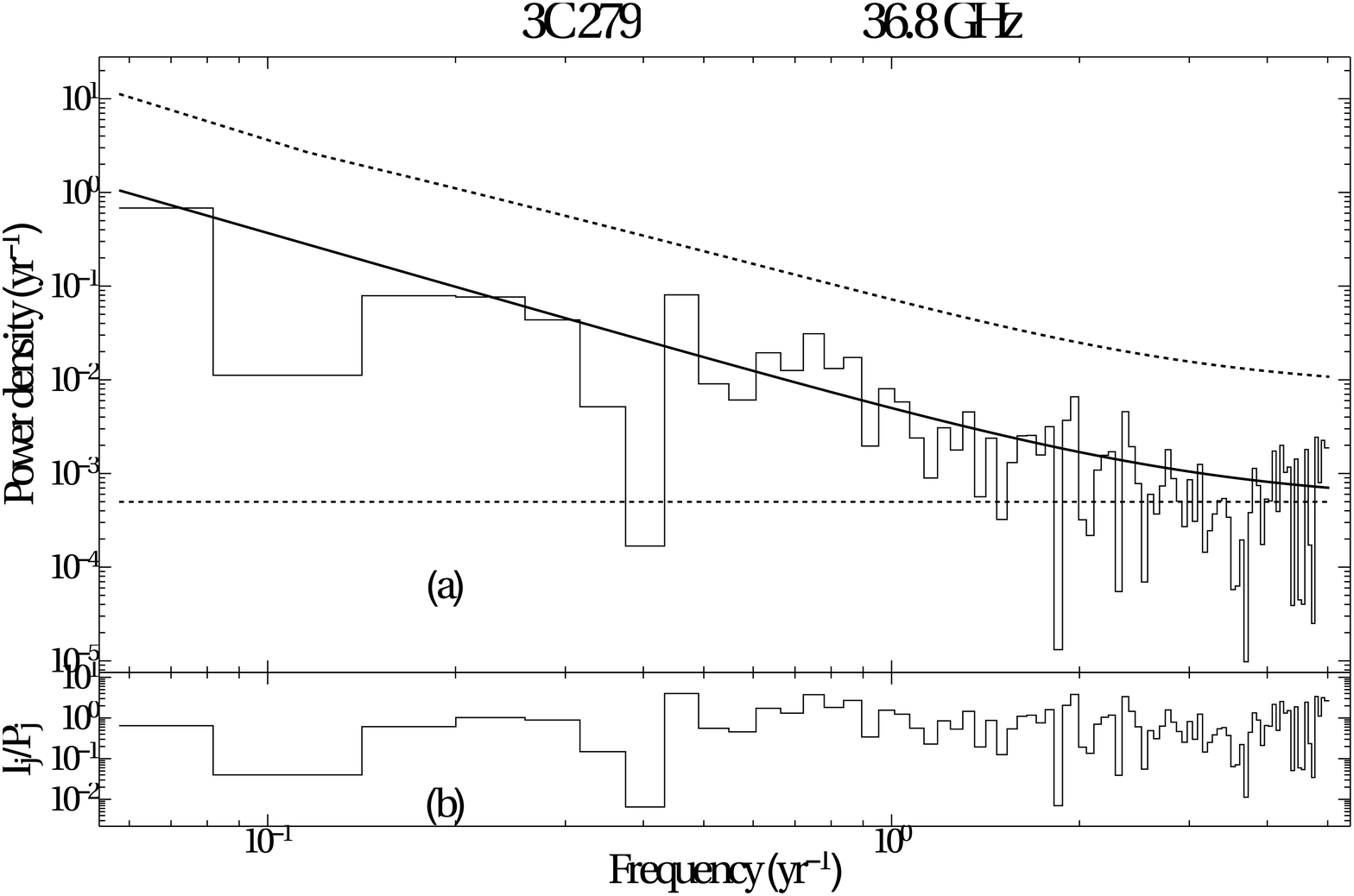}}
\caption{Fig.\ \ref{seg1psda} continued for segment 2 of 3C 279 (2007.0 to end of observations).}
\label{seg2psda}
\end{figure}

\begin{figure}
\centerline{\includegraphics[scale=0.155]{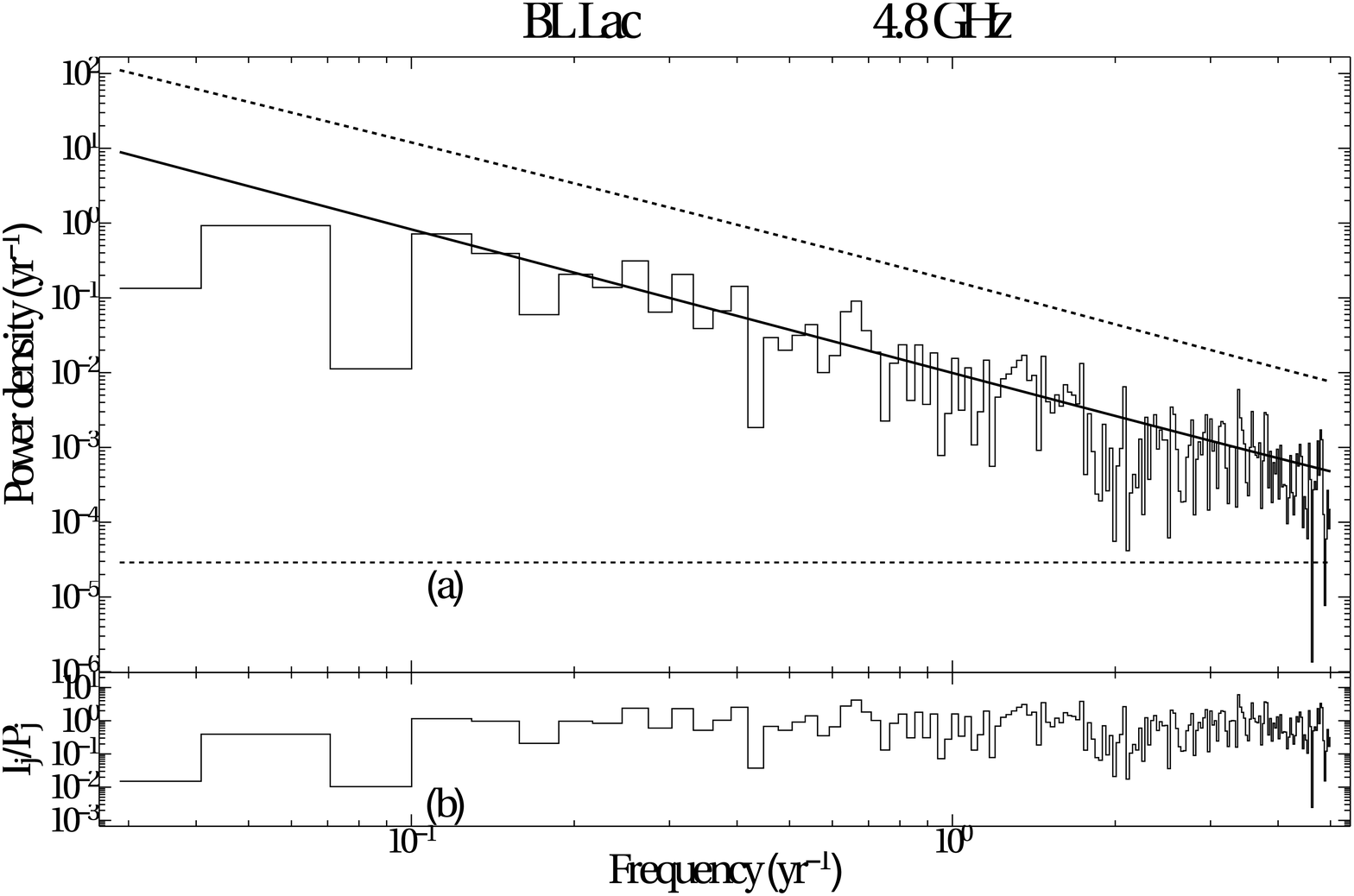}}
\centerline{\includegraphics[scale=0.155]{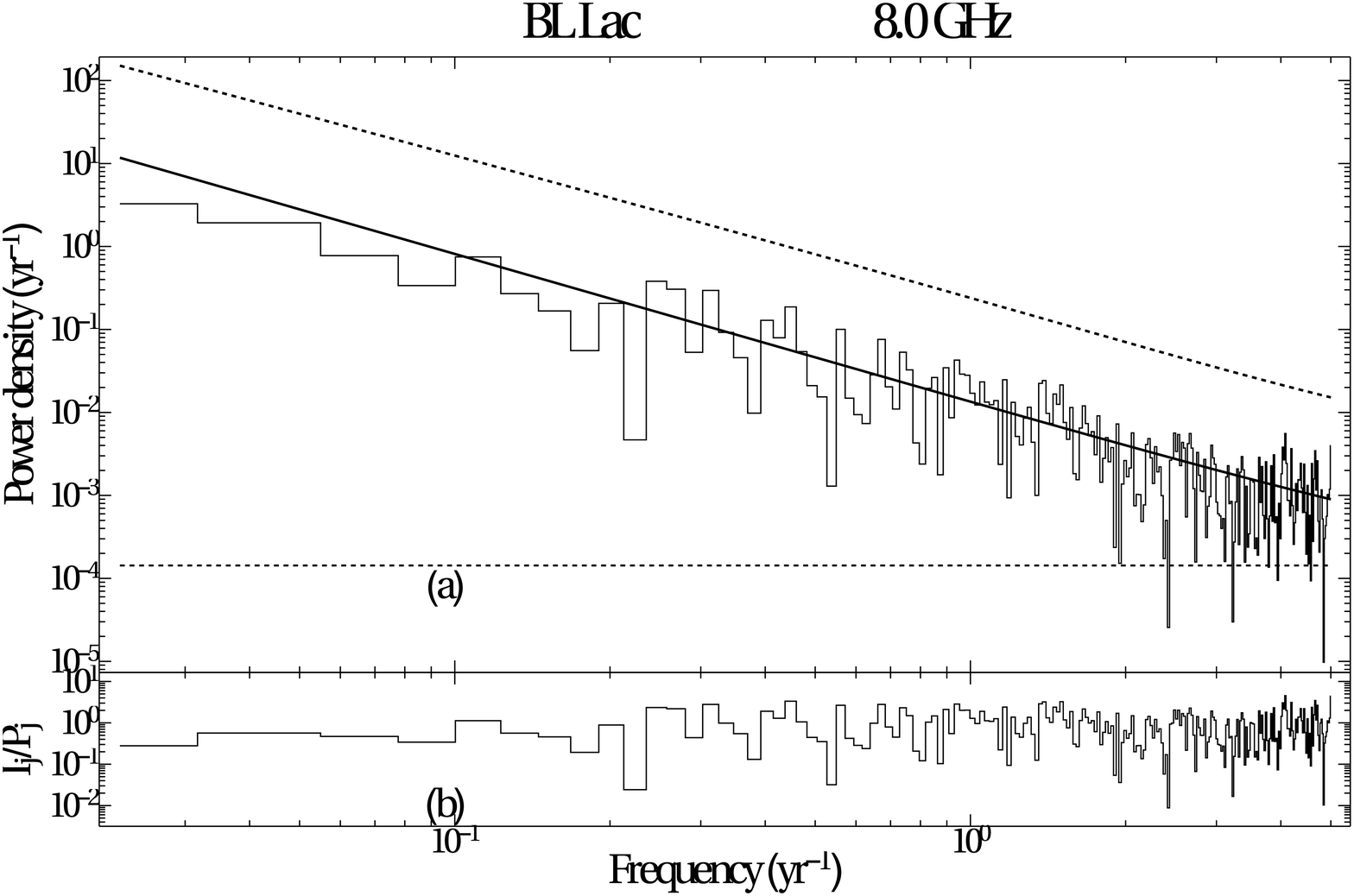}}
\centerline{\includegraphics[scale=0.155]{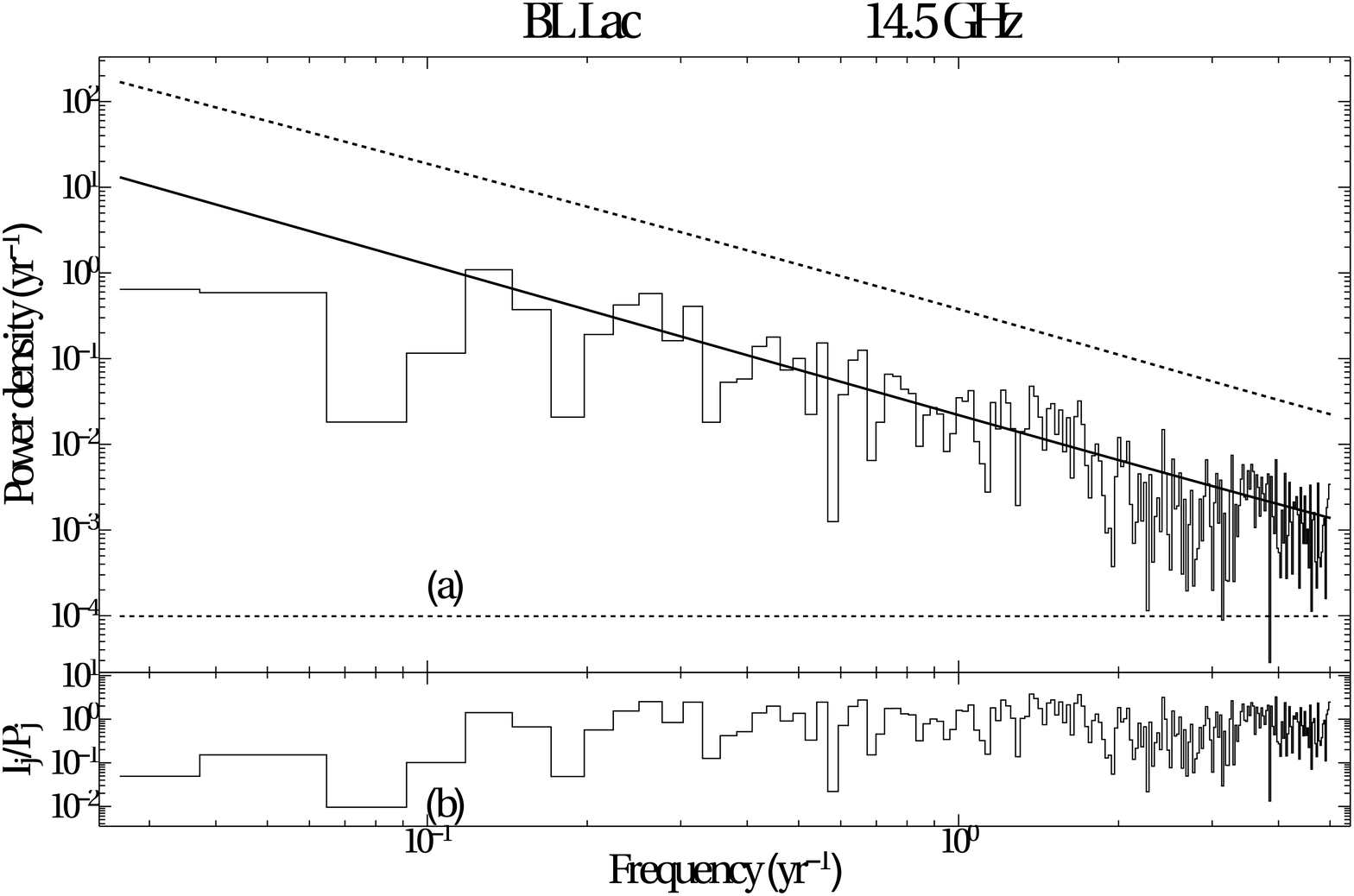}}
\centerline{\includegraphics[scale=0.155]{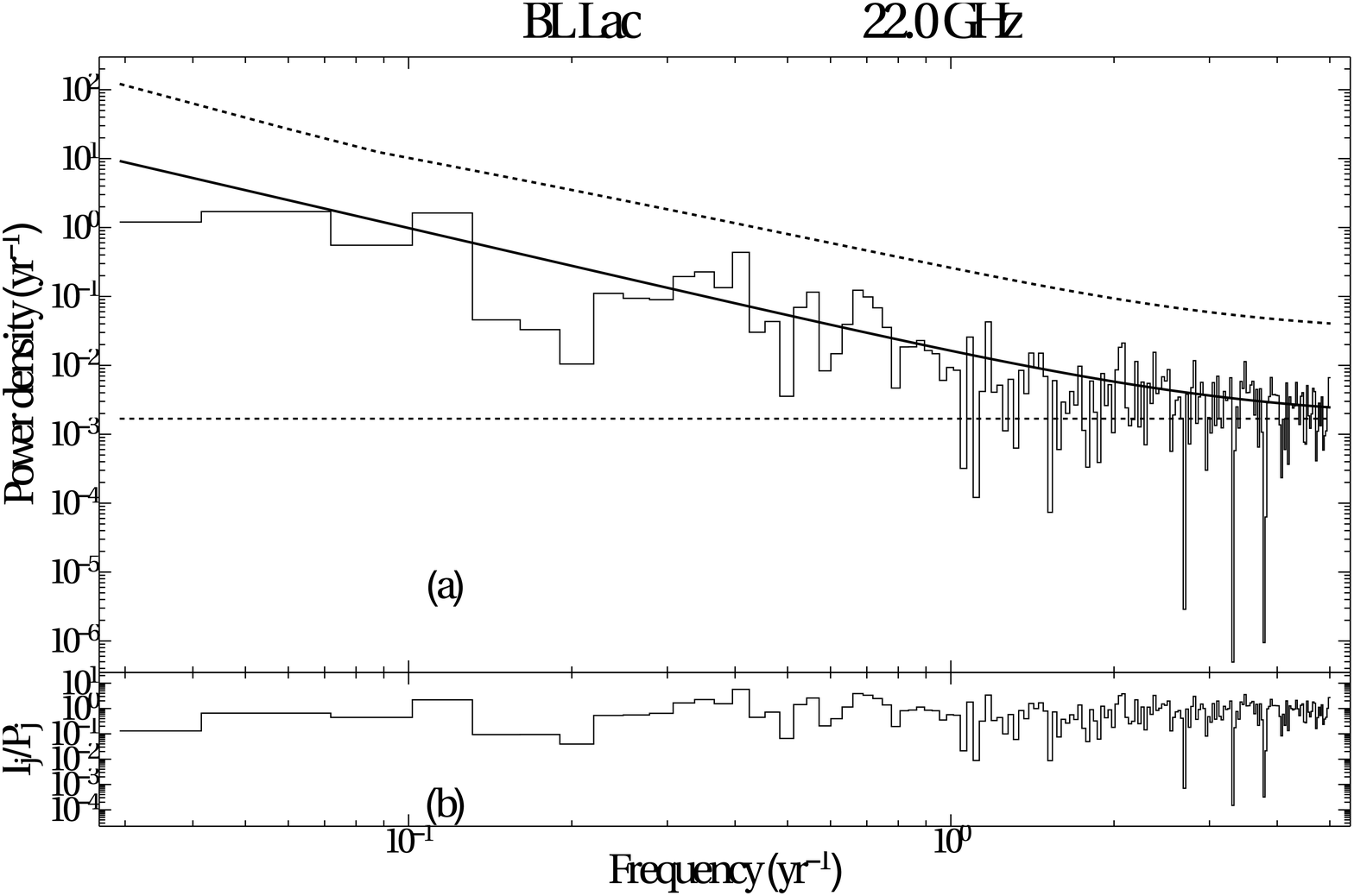}}
\centerline{\includegraphics[scale=0.155]{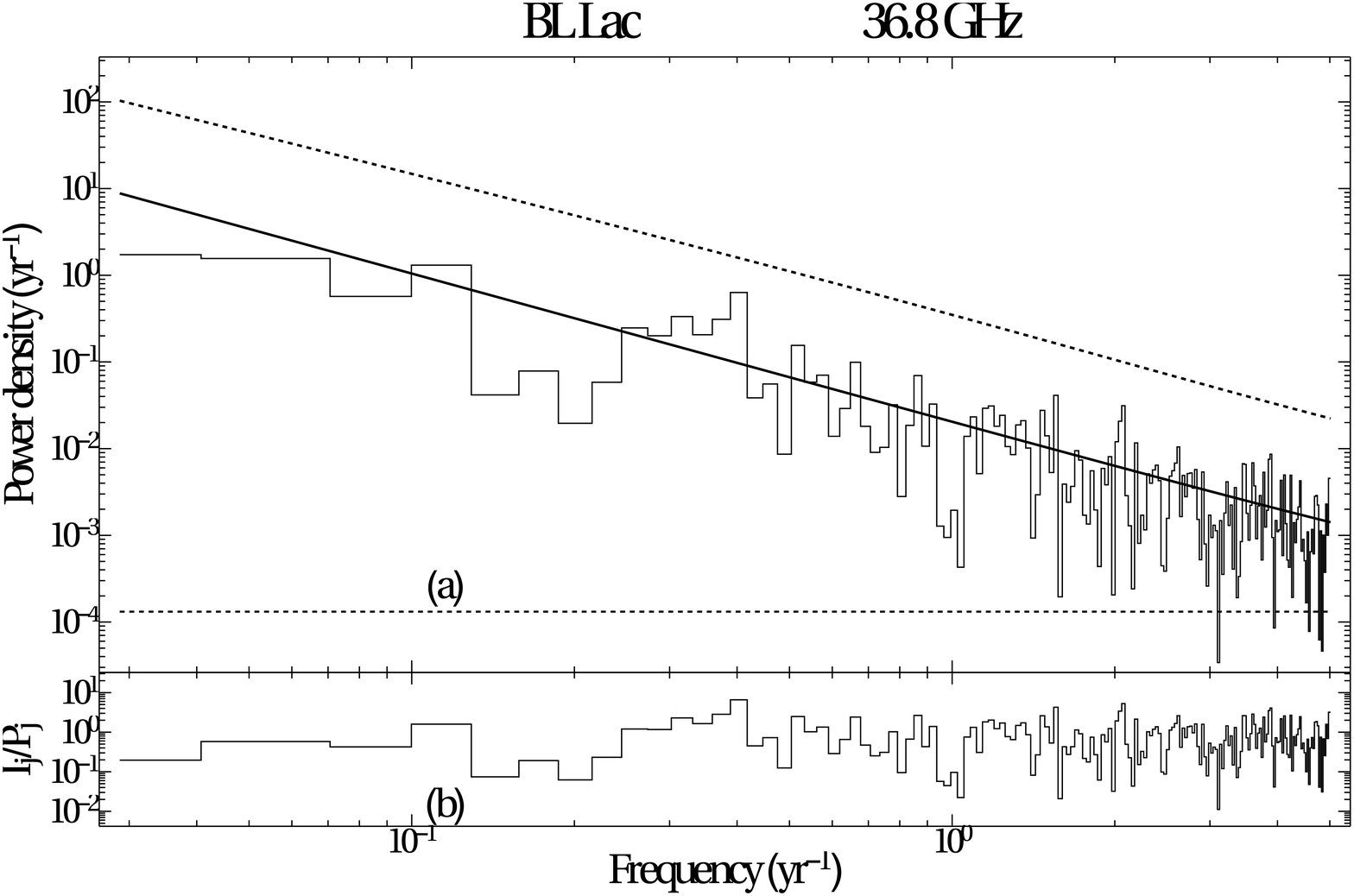}}
\caption{Periodogram analysis of BL Lacertae.}
\label{segpsda}
\end{figure}

\begin{table}
{\bf Table 11.} Results from the parametric PSD models fit to the periodogram of the 4.8 - 36.8 GHz light curves of 3C 279.
Column details same as given in the Table 10.
\centering
\scalebox{.73}{
\begin{tabular}{lllllll}
\hline
Observation & PSD       & \multicolumn{3}{c}{PSD Fit parameters}    & AIC & Model \\
Frequency \&& model     & \multicolumn{3}{l}{} &     & likelihood\\
Segment     &           & log(A) & $\alpha$ & log(f$_b$)            &     & \\ \hline
4.8 GHz: Seg. 1 & {\bf PL} & -3.72 $\pm$ 0.16 & -2.3 $\pm$ 0.5 & & 1403.06 & 1.00 \\
 & BPL& -1.99 $\pm$ 0.18 & -2.4 $\pm$ 0.4 & -1.21 $\pm$ 0.08 & 1405.44 & 0.30 \\ 
4.8 GHz: Seg. 2 & {\bf PL} & -3.75 $\pm$ 0.18 & -3.0 $\pm$ 0.5 & & 1211.84 & 1.00 \\
 & BPL & -1.00 $\pm$ 0.18 & -3.4 $\pm$ 0.7 & -1.20 $\pm$ 0.08 & 1214.72 & 0.24 \\ \hline
8.0 GHz: Seg. 1 & {\bf PL} & -3.25 $\pm$ 0.18 & -2.4 $\pm$ 0.5 & & 1145.07 & 1.00 \\
 & BPL & -1.50 $\pm$ 0.18 & -2.5 $\pm$ 0.5 & -1.19 $\pm$ 0.08 & 1147.82 & 0.25 \\ 
8.0 GHz: Seg. 2 & PL & -3.14 $\pm$ 0.18 & -2.6 $\pm$ 0.5 & & 1113.81 & 1.00 \\
 & BPL & -1.04 $\pm$ 0.18 & -2.7 $\pm$ 0.5 & -1.22 $\pm$ 0.08 & 1115.75 & 0.38 \\ \hline
14.5 GHz: Seg. 1 & {\bf PL} & -2.87 $\pm$ 0.17 & -2.3 $\pm$ 0.5 & & 1005.63 & 1.00 \\
 & BPL& -1.12 $\pm$ 0.18 & -2.5 $\pm$ 0.5 & -1.19 $\pm$ 0.08 & 1008.34 & 0.26 \\ 
14.5 GHz: Seg. 2 & PL & -2.46 $\pm$ 0.15 & -1.8 $\pm$ 0.3 & & 949.16 & 1.00 \\
 & BPL & -1.25 $\pm$ 0.18 & -2.0 $\pm$ 0.3 & -1.22 $\pm$ 0.08 & 950.74 & 0.45 \\ \hline
22.2 GHz: Seg. 1 & {\bf PL} & -2.11 $\pm$ 0.15 & -2.0 $\pm$ 0.3 & & 8939.95 & 1.00 \\
 & BPL& -0.80 $\pm$ 0.18 & -2.0 $\pm$ 0.3 & -1.21 $\pm$ 0.08 & 842.86 & 0.23 \\ 
22.2 GHz: Seg. 2 & {\bf PL} & -2.48 $\pm$ 0.15 & -1.9 $\pm$ 0.4 & & 996.01 & 1.00 \\
 & BPL & -1.19 $\pm$ 0.18 & -2.0 $\pm$ 0.4 & -1.26 $\pm$ 0.08 & 998.30 & 0.32 \\ \hline
36.8 GHz: Seg. 1 & {\bf PL} & -2.24 $\pm$ 0.16 & -2.0 $\pm$ 0.4 & & 890.74 & 1.00 \\
 & BPL& -0.76 $\pm$ 0.18 & -2.2 $\pm$ 0.4 & -1.25 $\pm$ 0.08 & 892.99 & 0.33 \\ 
36.8 GHz: Seg. 2 & {\bf PL} & -2.35 $\pm$ 0.16 & -1.9 $\pm$ 0.4 & & 895.54 & 1.00 \\
 & BPL & -1.03 $\pm$ 0.18 & -2.0 $\pm$ 0.4 & -1.24 $\pm$ 0.08 & 898.43 & 0.24 \\ \hline
\end{tabular}}
\label{3c279tab}
\end{table}
 
\section{Conclusions}    
    
A prominent constituent of radio loud AGNs is their core-jet morphology. We studied the core shift effect in the blazars S5 0716+714, 3C 279 and BL
Lac using their 4.8 GHz - 36.8 GHz radio light curves obtained from more than three decades of continuous monitoring.

\begin{enumerate}
    
\item
From the time lag and frequency fits, we found the weighted mean $k_r$ values
as 1.03 for S5 0716+714, 1.84 for segment 1 of 3C 279, 1.02 for segment 2 of 3C 279, while for BL Lacertae it is 1.06
thus being consistent with the equipartition between the magnetic field and particle energy densities.

\item In case of the BL Lacertae, S5 0716+714, we found $\Omega_{r\nu}$ = $19.81 \pm 10.49$ pcGHz for $\mu = 1.4$ mas/yr.
For $\mu = 0.35$ mas/yr in case of 3C 279 we calculated $\Omega_{r\nu}$ = $6.69 \pm 2.77$ pcGHz for Segment 1 and
$\Omega_{r\nu}$ = $8.24 \pm 2.93$ pcGHz for Segment 2.
Using $\mu = 1.08$ mas/yr in BL Lacertae we found the resultant $\Omega_{r\nu}$ to be $6.29 \pm 0.45$ pcGHz.

\item For S5 0716+714 we found weighted mean $B$ = 0.22 $\pm$ 0.36 G and $B_{core} = 15 \pm 5 mG$.
In case of the FSRQ 3C 279, We infer weighted mean $B$ = 0.0005 $\pm$ 0.0004 G and $B_{core} = 0.014 \pm 0.03 mG$ for segment 1 while $B$ = 0.35 $\pm$ 0.45 G and
$B_{core} = 17 \pm 6 mG$ for segment 2.
For BL Lacertae, we estimated $B$ = 0.02 $\pm$ 0.06 G and $B_{core} = 8 \pm 2 mG$. 

\item The light curves of all three sources were analyzed to infer the PSD shape. No statistically significant QPO was detected in either of
the sources. The power law PSD shape is the best fit model for our sources with mean slope of $-1.3$ for S5 0716+714, $-2.3$ for 3C 279 and $-1.8$
for BL Lacertae. 

\item We have estimated the spin from the derived field strength based on reasonable fiducial values and are 0.16 for BL Lacertae,
0.17 for S5 0716+714 and 0.9 for 3C 279. Independent estimates of spin from the Iron line or other techniques if available, can greatly help in reducing uncertainties. Using more accurate measurements of the field strength, bolometric and the jet luminosity, one can explore other electrodynamical and hybrid jet models and place constraints on the spin and mass of the black hole.  
\item The methodology can be applied to a large sample of quasars to estimate black hole spin, compare with estimates
from independent methods, study the distribution of inferred spins, and relate this to source properties including radio
loudness, jet luminosity, redshift, amongst others.
\end{enumerate}
These studies give the evidence of frequency dependent core shifts and also encourage us further to calculate jet parameters in the
region unresolved by VLBI technique.
The linear polarization based electric vector position angle (EVPA) follows a bimodal distribution with a tendency of polarization
orthogonal to the jet in quasars and along the jet in BL Lac objects \cite[e.g.][]{1993ApJ...416..519C,2003MNRAS.338..312G}. The shock
mechanism produces magnetic fields compressed in the plane of the shock, resulting in polarization along the jet for a transverse shock.
As shocks are transient events, the resultant jet polarization orientation may not be retained over long durations. Further, as internal
shocks can be oblique, with a range of orientations, a natural bimodal distribution of the relative EVPAs cannot be expected.
The observation of a bimodal distribution has been inferred in the computation of the linear polarization and the EVPA variability in the
context of a relativistic jet flow carrying large-scale helical magnetic fields \cite[e.g.][]{2005MNRAS.360..869L}. The measurement of the field
strength and its variation along the length of the pc-scale jet using the core shift method can aid detailed polarization models. A particular
model (Mangalam, 2017; in preparation) invokes the assumption of helical magnetic structure and special relativistic kinematics of a blob
transiting the jet to compute the instantaneous Doppler factor and hence predict the varying flux density, linear polarization and EVPA.
Radio continuum surveys with the SKA can provide AGN imaging and timing (flux density, polarization) data across a range of
frequencies simultaneously, ideally suited for the application of the time delay method. The SKA-VLBI synergy can provide key and
unique contributions to the current study and additionally, in the context of studies of magnetic fields in AGN
jets \citep{Roy2016}. Some of these include (1) verification of the SSA model using high resolution sub--pc-scale images,
as the SSA opacity is related to the magnetic field strength and core distance \cite[e.g.][]{1998A&A...330...79L},
(2) offering a sizeable sample for statistical and comparative studies of pc-scale jet kinematics and energetics which can
serve as crucial inputs to fully relativistic radiation magneto-hydrodynamic simulations, and (3) exploiting the higher
sensitivity to understand evolution of magnetic fields in these AGNs over cosmic time.

\begin{table}
{\bf Table 12.} PSD results for BL Lacertae. Column details same as given in the Table 11.
\centering
\scalebox{.73}{
\begin{tabular}{lllllll}
\hline
Observation & PSD       & \multicolumn{3}{c}{PSD Fit parameters}    & AIC & Model \\
Frequency   & model     & \multicolumn{3}{l}{} &     & likelihood\\
            &           & log(A) & $\alpha$ & log(f$_b$)            &     & \\ \hline
4.8 GHz & {\bf PL} & -2.00 $\pm$ 0.10 & -1.9 $\pm$ 0.2 & & 1668.5 & 1.00 \\
 & BPL & -0.50 $\pm$ 0.18 & -2.0 $\pm$ 0.1 & -1.54 $\pm$ 0.08 & 1672.32 & 0.15 \\ \hline
8.0 GHz & {\bf PL} & -1.87 $\pm$ 0.09 & -1.8 $\pm$ 0.2 & & 1954.28 & 1.00 \\
 & BPL & -0.44 $\pm$ 0.18 & -1.9 $\pm$ 0.2 & -1.65 $\pm$ 0.08 & 1957.80 & 0.17 \\ \hline
14.5 GHz & {\bf PL} & -1.66 $\pm$ 0.10 & -1.8 $\pm$ 0.2 & & 1490.63 & 1.00 \\
 & BPL & -0.35 $\pm$ 0.18 & -1.8 $\pm$ 0.2 & -1.58 $\pm$ 0.08 & 1494.79 & 0.13 \\ \hline
22.2 GHz & {\bf PL} & -1.83 $\pm$ 0.15 & -1.8 $\pm$ 0.4 & & 1329.95 & 1.00 \\
 & BPL & -0.40 $\pm$ 0.18 & -1.9 $\pm$ 0.3 & -1.53 $\pm$ 0.08 & 1333.18 & 0.20 \\ \hline
36.8 GHz & {\bf PL} & -1.69 $\pm$ 0.10 & -1.7 $\pm$ 0.2 & & 1377.88 & 1.00 \\
 & BPL & -0.42 $\pm$ 0.18 & -1.8 $\pm$ 0.2 & -1.54 $\pm$ 0.08 & 1381.02 & 0.21 \\ \hline
\end{tabular}}
\label{bllactab}
\end{table}

\section*{Acknowledgments}

We thank the referee for a constructive discussion and suggestions
which have improved the manuscript. PM is supported by the Chinese Academy of Sciences President’s International
Fellowship Initiative (CAS-PIFI; 
grant no. 2016PM024) post-doctoral fellowship and the National Natural Science Foundation of China (NSFC) 
Research Fund for International Young Scientists (grant no. 11650110438). ACG is partially supported by CAS 
President’s International Fellowship Initiative (PIFI; grant no. 2016VMB073). The work at UMRAO was supported 
in part by a series of grants from the NSF and NASA, most recently AST-0607523 and NASA Fermi GI grants 
NNX09AU16G, NNX10AP16G, NNX11AO13G, and NNX13AP18G. MFG is supported by the National Science Foundation of China 
(grants 11473054 and U1531245) and by the Science and Technology Commission of Shanghai Municipality (grant 14ZR1447100).

\label{lastpage}
\end{document}